\documentclass[0trackchanges,resetfootnote,twocolumn]{aastex7}

\newcommand{\shortname}{RACS~J0320$-$35}
\newcommand\Tstrut{\rule{0pt}{2.6ex}}         
\newcommand\Bstrut{\rule[-0.9ex]{0pt}{0pt}}   
\newcommand\BBstrut{\rule[-1.3ex]{0pt}{0pt}}   

\begin{document}

\title{X-ray investigation of possible super-Eddington accretion in a radio-loud quasar at $z=6.13$}

\author[orcid=0000-0003-1516-9450,gname=Luca,sname=Ighina]{Luca Ighina}
\affiliation{Center for Astrophysics | Harvard \& Smithsonian, 60 Garden St., Cambridge, MA 02138, USA}
\affiliation{INAF, Osservatorio Astronomico di Brera, via Brera 28, 20121, Milano, Italy}
\email[show]{luca.ighina@cfa.harvard.edu}  

\author[orcid=0000-0002-2339-8264,gname=Alessandro, sname=Caccianiga]{Alessandro Caccianiga} 
\affiliation{INAF, Osservatorio Astronomico di Brera, via Brera 28, 20121, Milano, Italy}
\email{alessandro.caccianiga@inaf.it}  

\author[orcid=0000-0002-7898-7664,gname=Thomas,sname=Connor]{Thomas Connor}
\affiliation{Center for Astrophysics | Harvard \& Smithsonian, 60 Garden St., Cambridge, MA 02138, USA}
\email{thomas.connor@cfa.harvard.edu}  

\author[0000-0002-9770-0315,gname=Alberto,sname=Moretti]{Alberto Moretti}
\affiliation{INAF, Osservatorio Astronomico di Brera, via Brera 28, 20121, Milano, Italy}
\email{alberto.moretti@inaf.it}  

\author[orcid=0000-0001-9879-7780,gname=Fabio,sname=Pacucci]{Fabio Pacucci}
\affiliation{Center for Astrophysics | Harvard \& Smithsonian, 60 Garden St., Cambridge, MA 02138, USA}
\affiliation{Black Hole Initiative, Harvard University, 20 Garden St., Cambridge, MA 02138, USA}
\email{fabio.pacucci@cfa.harvard.edu}  

\author[0000-0002-8978-0626,gname=Cormac,sname=Reynolds]{Cormac Reynolds}
\affiliation{SKA Observatory, Science Operations Centre, CSIRO ARRC, 26 Dick Perry Avenue, Kensington, WA 6151, Australia}
\email{Cormac.Reynolds@csiro.au}  

\author[0000-0002-9149-2973,gname=José,sname=Afonso]{José Afonso}
\affiliation{Instituto de Astrofísica e Ciências do Espaço, Universidade de Lisboa, OAL, Tapada da Ajuda, Lisboa, Portugal}
\affiliation{Departamento de Física, Faculdade de Ciências, Universidade de Lisboa, Lisbon, Portugal}
\email{Cormac.Reynolds@csiro.au}  

\author[0000-0001-8166-6602,gname=Bruno,sname=Arsioli]{Bruno Arsioli}
\affiliation{Instituto de Astrofísica e Ciências do Espaço, Universidade de Lisboa, OAL, Tapada da Ajuda, Lisboa, Portugal}
\affiliation{Departamento de Física, Faculdade de Ciências, Universidade de Lisboa, Lisbon, Portugal}
\email{bruno.arsioli@gmail.com}  

\author[0000-0003-4747-4484,gname=Silvia,sname=Belladitta]{Silvia Belladitta}
\affiliation{Max Planck Institut für Astronomie, Königstuhl 17, D-69117 Heidelberg, Germany}
\affiliation{INAF – Osservatorio di Astrofisica e Scienza dello Spazio di Bologna, Via Gobetti 93/3, I-40129 Bologna, Italy}
\email{belladitta@mpia.de}  

\author[0000-0002-2239-6099,gname=Jess,sname=Broderick]{Jess W. Broderick}
\affiliation{SKA Observatory, Science Operations Centre, CSIRO ARRC, 26 Dick Perry Avenue, Kensington, WA 6151, Australia}
\affiliation{International Centre for Radio Astronomy Research, Curtin University, 1 Turner Avenue, Bentley, WA, 6102, Australia}
\email{jess.broderick@skao.int}  

\author[0000-0003-1246-6492,gname=Daniele,sname=Dallacasa]{Daniele Dallacasa}
\affiliation{Dipartimento di Fisica e Astronomia ‘Augusto Righi’, Università degli Studi di Bologna, Via Gobetti 93/2, 40129 Bologna, Italy}
\affiliation{INAF - Istituto di Radioastronomia, Via Gobetti 101, I-40129 Bologna, Italy}
\email{ddallaca@ira.inaf.it}  

\author[0000-0001-7551-2252,gname=Roberto,sname=Della Ceca]{Roberto Della Ceca}
\affiliation{INAF, Osservatorio Astronomico di Brera, via Brera 28, 20121, Milano, Italy}
\email{roberto.dellaceca@inaf.it}

\author[0000-0003-3291-3704,gname=Francesco,sname=Haardt]{Francesco Haardt}
\affiliation{Dipartimento di Scienza e Alta Tecnologia, Università degli Studi dell’Insubria, via Valleggio 11, I-22100 Como, Italy}
\affiliation{INAF, Osservatorio Astronomico di Brera, Via E. Bianchi 46, I-23807 Merate, Italy}
\affiliation{INFN, Sezione Milano-Bicocca, P.za della Scienza 3, I-20126 Milano, Italy}
\email{Francesco.Haardt@uninsubria.it}  

\author[0000-0003-3216-7190,gname=Erini,sname=Lambrides]{Erini Lambrides}
\affiliation{NASA Goddard Space Flight Center, Code 662, Greenbelt, 20771, MD, USA}
\email{elambrid@gmail.com}

\author[0000-0002-9415-3766,gname=James,sname=Leung]{James K. Leung}
\affiliation{David A. Dunlap Department of Astronomy and Astrophysics, University of Toronto, 50 St. George Street, Toronto, ON M5S 3H4, Canada}
\affiliation{Dunlap Institute for Astronomy and Astrophysics, University of Toronto, 50 St. George Street, Toronto, ON M5S 3H4, Canada}
\affiliation{Racah Institute of Physics, The Hebrew University of Jerusalem, Jerusalem 91904, Israel}
\email{jamesk.leung@utoronto.ca}  

\author[0000-0003-3291-3704,gname=Alessandro,sname=Lupi]{Alessandro Lupi}
\affiliation{Como Lake Center for Astrophysics, DiSAT, Universit\`a degli Studi dell'Insubria,  via Valleggio 11, I-22100, Como, Italy}
\affiliation{INFN, Sezione Milano-Bicocca, P.za della Scienza 3, I-20126 Milano, Italy}
\affiliation{INAF – Osservatorio di Astrofisica e Scienza dello Spazio di Bologna, Via Gobetti 93/3, I-40129 Bologna, Italy}
\email{alessandro.lupi@uninsubria.it}  

\author[0000-0003-1177-3896,gname=Israel,sname=Matute]{Israel Matute}
\affiliation{Instituto de Astrofísica e Ciências do Espaço, Universidade de Lisboa, OAL, Tapada da Ajuda, Lisboa, Portugal}
\affiliation{Departamento de Física, Faculdade de Ciências, Universidade de Lisboa, Lisbon, Portugal}
\email{imatute@fc.ul.pt}  

\author[0000-0001-7880-8825,gname=Fabio,sname=Rigamonti]{Fabio Rigamonti}
\affiliation{INAF, Osservatorio Astronomico di Brera, Via E. Bianchi 46, I-23807 Merate, Italy}
\affiliation{INFN, Sezione Milano-Bicocca, P.za della Scienza 3, I-20126 Milano, Italy}
\affiliation{Como Lake Center for Astrophysics, DiSAT, Universit\`a degli Studi dell'Insubria,  via Valleggio 11, I-22100, Como, Italy}
\email{fabio.rigamonti@inaf.it}  

\author[0000-0001-5619-5896,gname=Paola,sname=Severgnini]{Paola Severgnini}
\affiliation{INAF, Osservatorio Astronomico di Brera, via Brera 28, 20121, Milano, Italy}
\email{paola.severgnini@inaf.it}  

\author[0000-0003-3506-5536,gname=Nick,sname=Seymour]{Nick Seymour}
\affiliation{International Centre for Radio Astronomy Research, Curtin University, 1 Turner Avenue, Bentley, WA, 6102, Australia}
\email{nick.seymour@curtin.edu.au}  

\author[0000-0003-0256-0995,gname=Fabrizio,sname=Tavecchio]{Fabrizio Tavecchio}
\affiliation{INAF, Osservatorio Astronomico di Brera, Via E. Bianchi 46, I-23807 Merate, Italy}
\email{fabrizio.tavecchio@inaf.it}  

\author[0000-0002-8853-9611,gname=Cristian,sname=Vignali]{Cristian Vignali}
\affiliation{INAF – Osservatorio di Astrofisica e Scienza dello Spazio di Bologna, Via Gobetti 93/3, I-40129 Bologna, Italy}
\affiliation{Dipartimento di Fisica e Astronomia ‘Augusto Righi’, Università degli Studi di Bologna, Via Gobetti 93/2, 40129 Bologna, Italy}
\email{cristian.vignali@unibo.it}

\begin{abstract}

We present radio and X-ray observations of the recently discovered $z=6.13$ radio-powerful quasar RACS~J032021.44$-$352104.1 using uGMRT, ATCA, LBA, and \textit{Chandra}.
The observed radio properties are in line with what is typically observed in high-$z$ radio quasars ($\alpha_{\rm r}=0.72\pm 0.02$ and L$_{\rm 1.4GHz}=5.8 \pm 0.9 \times 10^{26}$~W~Hz$^{-1}$). Despite the relatively low X-ray flux observed $F_{\rm 0.5-7.0~keV}=2.3\pm0.5 \times 10^{-14}$~erg~sec$^{-1}$~cm$^{-2}$, the intrinsic luminosity in the 2--10~keV rest frame is markedly high,  $L_{\rm 2-10~keV}=1.8^{+1.1}_{-0.7} \times 10^{46}$~erg~sec$^{-1}$, making RACS~J032021.44$-$352104.1 one of the most luminous quasars currently known at $z>5.5$. The high X-ray luminosity is largely driven by an extrapolation to energies below the observable X-ray window with {\it Chandra} and the slope derived in the 0.5-7~keV band (or 3.5--50~keV in the rest-frame; $\Gamma_{\rm X}=3.3\pm0.4$).
By analysing the overall spectral energy distribution of the quasar we found that the remarkably soft X-ray emission: (1) cannot be produced by relativistic jets, even when relativistic boosting is considered; and (2) is consistent with expectations for a super-Eddington accreting SMBH.
If such a high accretion rate was confirmed, this source would be a unique laboratory to study high accretion in the early Universe and could help resolve some challenges inherent in early black hole growth paradigms. 
 
\end{abstract}



\section{Introduction}
\label{sec:intro}

The observation of supermassive black holes (SMBHs; M$_{\rm BH}\gtrsim10^{8}$~M$_\odot$) hosted in $z\gtrsim6$ quasars \citep[e.g.,][]{Wang2021} is one of the most important ways to constrain the initial mass of their seed BH. While there are several theoretical models predicting the formation of the first seed BHs \citep[e.g.,][]{Volonteri2010,Volonteri2021}, in order to explain the SMBHs we observe at $z\gtrsim6$, models producing very massive seeds (M$_{\rm seed}\gtrsim10^4$~M$_\odot$) at $z\sim15$ and accreting close to their Eddington limit\footnote{
The Eddington limit for the accretion onto a BH of mass $M_{\rm BH}$ with radiative accretion efficiency $\epsilon$ is $\dot{M}_{\mathrm{Edd}} \approx 2.2  M_{\rm BH} \epsilon^{-1} \times 10^{-9}\ {\rm yr}^{-1}$. Super-Eddington accretion is when $\lambda_{\rm Edd} = \dot{M}/\dot{M}_{\mathrm{Edd}} > 1$.
}
are favoured (e.g. \citealt{Bogdan2024}). However, these models typically require very rare and unique physical conditions \citep{Schauer2017,Lupi2021,Latif2022}, and struggle to explain the large abundances of $\sim10^9$~M$_\odot$ high-$z$ quasars observed \citep{Shen2020,Fan2023}. Another way to reconcile theory with observations is to consider accretion above the Eddington limit, as shown in many simulations (\citealt{Bhowmick2022,Massonneau2023,Lupi2024}). At the same time, it has been recently suggested that super-Eddington accretion can also explain the observational properties (namely the X-ray non-detection; e.g., \citealt{Maiolino2024b, Yue_2024_LRD}) of the new population of active galactic nuclei (AGN) revealed by the James Webb Space Telescope (JWST; see, e.g., \citealt{Pacucci2024,Lambrides2024,Madau2024}). Nevertheless, observational evidence for super-Eddington accretion has only been found for a handful of sources at high redshift \citep[e.g.][]{Wolf2023,Abuter2024} and suggested for some quasars in the epoch of reionization \citep{Yang2021,Zappacosta2023,Belladitta2025}. For example, \cite{Suh2024} recently reported the detection of an X-ray luminous $\sim7\times10^6$~M$_\odot$ BH at redshift $z\sim4$. Based on the large X-ray and bolometric luminosity (L$_{\rm bol}\sim10^{46.7}$~erg~sec$^{-1}$), the authors suggest that this system is accreting $\sim40$ times its Eddington limit.

However, the vast majority of $z>6$ UV-bright quasar population has accretion rates consistent with being Eddington limited \citep[e.g.][]{Shen2019,Farina2022,Mazzucchelli2023}. We must stress that these values are based on somewhat uncertain black hole mass estimates (typically based on single epoch, SE, measurements). Indeed, it has been shown how the BH mass of these very high-redshift systems can be overestimated, by up to an order of magnitude, which in turn underestimates the accretion rates and implies that the need for very massive seed BHs is less severe \citep[see, e.g.,][]{Lupi2024b,King2024,Lambrides2024}.

In this letter, we present radio and X-ray observations of RACS~J032021.44$-$352104.1 (\shortname hereafter) which suggest that the SMBH hosted in this system is accreting at a super-Eddington rate. This radio quasar belongs to a larger sample (see \citealt{Ighina2025}) selected from the combination of the first data release of the Rapid ASKAP continuum Survey (RACS; \citealt{McConnell2020, Hale2021}) together with the Dark Energy Survey (DES; \citealt{Abbott2021}). \shortname \, was then confirmed to be at $z=6.13$ (based on the Lyman break) with the Gemini-South telescope \citep{Ighina2023}. 

Throughout the paper, we report spectral power-law slopes following the convention $S_\nu \propto \nu^{-\alpha}$, and all errors are reported at a 68 percent confidence level, unless otherwise specified. Throughout this work, we adopt a flat $\Lambda$CDM cosmology of $H_0=70$ km s$^{-1}$, $\Omega_{\rm M}=0.3$, and $\Omega_{\rm \Lambda}=0.7$.

\section{Radio and X-ray observations}
\label{sec:observations}

Dedicated radio observations for \shortname \, were obtained with the upgraded Giant Metrewave Radio Telescope (uGMRT; at 400 and 650~MHz), the Australia Telescope Compact Array (ATCA; at 2.1, 5.5 and 9.0~GHz), and the Australian Large Baseline Array (LBA; 2.4 and 8.4~GHz). We report in Appendix \ref{sec:app_radio_data} a description of the observations, the steps of the data reduction and the final images and fluxes obtained. Together with public data available from radio surveys (see Appendix \ref{sec:app_radio_data}), we were able to constrain the radio emission of the quasar over the $\sim$0.1--10~GHz observed frame (or 0.7--70~GHz in the rest frame) on arcseconds scales. At the same time, the LBA observations provide an estimate and an upper-limit on the milli-arcsec emission at at 2.3 and 8.4~GHz, respectively.

\begin{figure*}
 	\includegraphics[width=0.5\hsize]{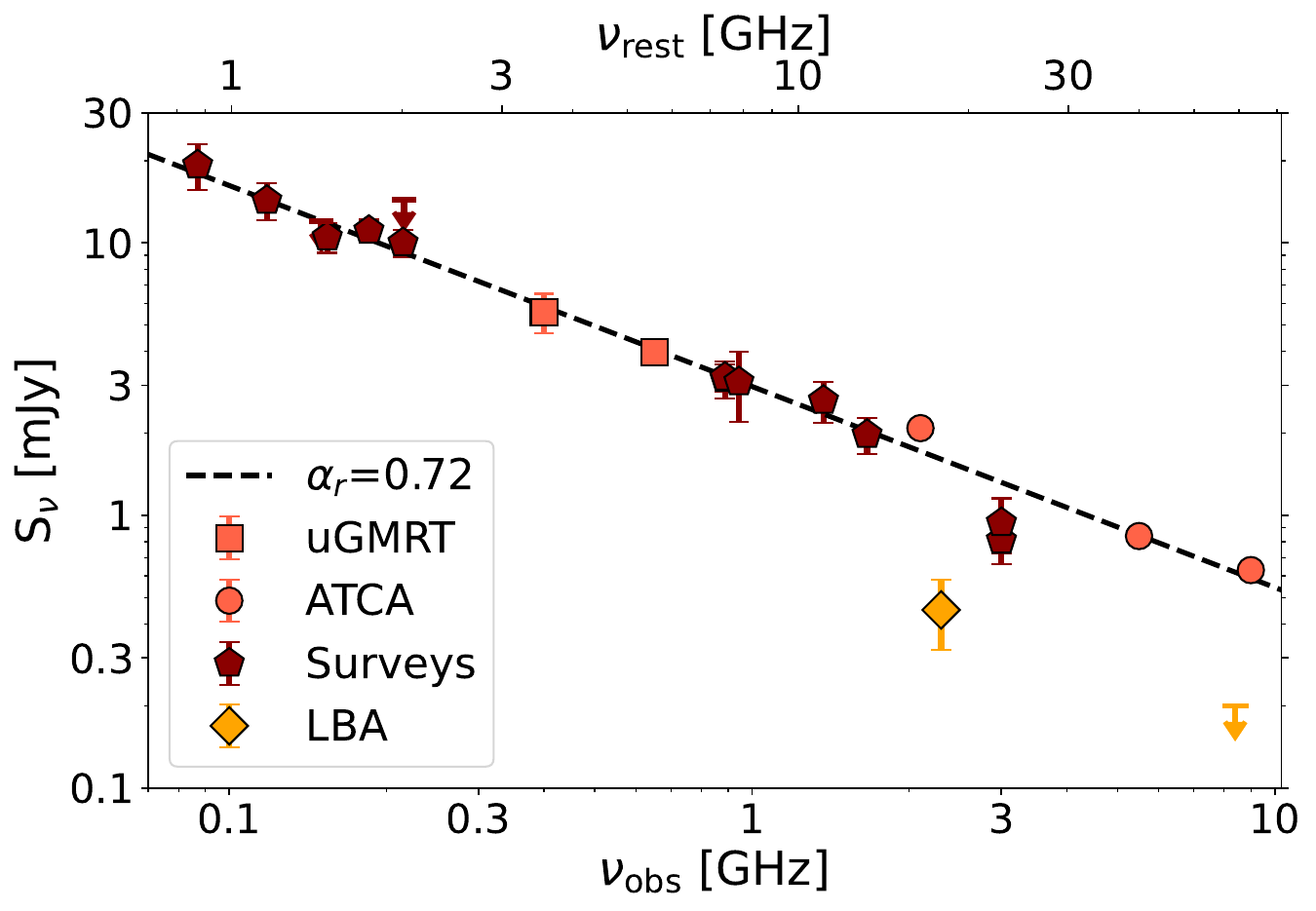}
     \includegraphics[width=0.47\hsize]{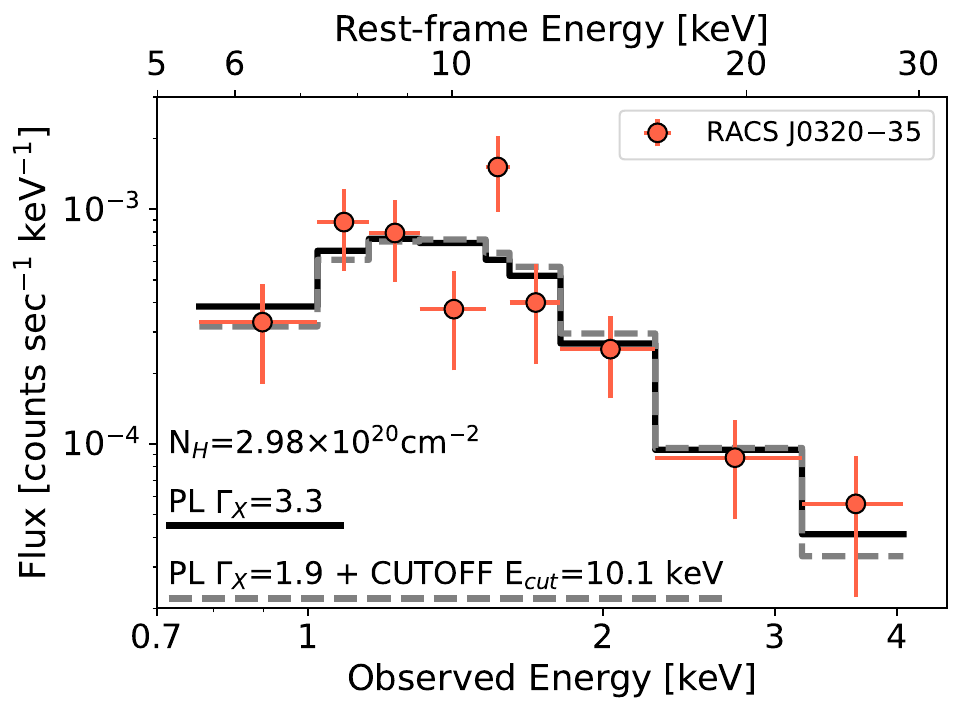}

 \caption{{\bf Left:} Radio spectrum of \shortname\ obtained from dedicated observations (uGMRT+ATCA; light red squares and circles) as well as data available from public surveys (dark red pentagons), as described in the text.Measurements from the VAST survey are shown as a single data-point at the median value (3.2~mJy~beam$^{-1}$) and the standard deviation (0.5~mJy~beam$^{-1}$) as uncertainty. The best fit power law with is shown as a black dashed line. We also show the measurement and upper limit from the VLBI-LBA observations (yellow diamond); as these observations are of significantly higher angular resolution, they were not used during the fit.
 {\bf Right:} X-ray spectra of RACS~J0320$-$35 obtained with {\it Chandra}. The best-fit power law with Galactic absorption is shown as a black solid line, while the best-fit power law ($\Gamma_{\rm X}=1.9$, fixed) with an exponential cutoff and Galactic absorption is shown as a grey dashed line. Photon counts have been re-binned to $5\sigma$ for plotting purposes only.}
    \label{fig:X_and_R_spec}
\end{figure*}

We show in the left panel of Fig. \ref{fig:X_and_R_spec} the radio spectrum of \shortname. The spectral index obtained from a single power law fit to the flux density measurements on arcsec scales is $\alpha_{\rm r}=0.72\pm0.02$. Although observations are separated across six years (2019--2025), we do not find evidence for variation. Indeed, the reduced $\chi^2$ parameter derived from the 28 measurements from ASKAP as part of the Variables and Slow Transients \citep[VAST;][]{Murphy2013,Murphy2021} over more than two years (2023 July -- 2025 April) survey at 888~MHz with a constant model is $\chi^2_{\rm red}=0.83$. Similarly, the median peak flux densities in these images ($S_{\rm 888MHz}=3.2$~mJy~beam$^{-1}$) is consistent with the estimate from the first scan of the RACS-low survey ($S_{\rm 888MHz}=3.2$~mJy~beam$^{-1}$), despite being taken about six years apart in the observed frame.
Moreover, fitting a radio power law solely in individual epochs (GLEAM-X, October 2020; uGMRT, April 2022; ATCA, September 2022), the slope and normalization remain consistent with the single power-law fit. These findings suggest that, if present, variability does not significantly affect the overall shape of the spectrum. However, monitoring on larger time window is needed to constrain the variation on timescales of years in the rest frame.

The target appears point-like on arcsecond scales, based on a 2D Gaussian fit to the uGMRT and ATCA images. However, the flux recovered at 2.3~GHz on milliarcsecond scales with the LBA (0.45$\pm0.13$~mJy) is only $\sim$20-30\% of the total flux recovered at larger scales (2.09~mJy at 2.1~GHz measured with ATCA or 1.61~mJy at 2.3~GHz from the best-fit power law). Similarly, at 8.4~GHz the emission produced on milliarcsecond scales is $<30$\% of the total flux density recovered on arcsecond scales. Although no significant variation was found on arcsecond scales, variability might still be present on the smaller scales.

In the X-rays, we observed \shortname\ for 60~ks with the Advanced CCD Imaging Spectrometer (ACIS; \citealt{Garmire2003}) on \textit{Chandra} as part of Proposal 24700061 (PI:Ighina). A description of the observations and the data reduction is reported in Appendix \ref{sec:chandra}. 
In Fig. \ref{fig:X_and_R_spec}, right panel, we show the best-fit model obtained from the fit of a power law absorbed by the Galactic column density \citep[N$_{\rm H} =2.98\times 10^{20}$~cm$^{-2}$;][]{HI4PI2016}.

To fit the observed emission, we adopted two models, both of which include absorption by the Galactic column density \citep[N$_{\rm H} =2.98\times 10^{20}$~cm$^{-2}$;][]{HI4PI2016}: single power, as typically adopted in the literature \citep[e.g.][]{Li2021,Connor2021}, and a power law with an exponential cutoff, expected at low energies for highly accreting BHs (see discussion in Sec. \ref{sec:superEdd} and e.g. \citealt{Tortosa2022}).
By considering a single power, we derived a very soft and luminous X-ray emission ($\Gamma_{\rm X}=3.3\pm0.7$ and $L_{\rm 2-10~keV}=1.8^{+2.2}_{-0.9}\times10^{46}$~erg~sec$^{-1}$, errors at 90\% confidence level) even in comparison to the optical ($\alpha_{\rm ox}=0.97\pm0.07$ errors at 90\% confidence level, see \citealt{Ighina2023} for the 2500\AA\, luminosity)\footnote{where $\alpha_{\rm ox}=0.384\rm log(L_{\rm 2keV}/L_{2500\AA})$.}. The corresponding observed flux is f$_\mathrm{0.5-7.0keV}=2.3_{-0.5}^{+1.1}\times 10^{-14}$~erg~sec$^{-1}$~cm$^{-2}$ (errors at 90\% confidence level).

If we consider a power law with a high energy exponential cutoff where the photon index of the power law and the energy of the cutoff are both free to vary, the fit does not converge. Therefore, following the approach of \cite{Tortosa2024}, we fixed the value of the photon index to $\Gamma_{\rm X}=1.9$ and estimated a cutoff energy of $E_{\rm cut}=10.1_{-3.7}^{+10.7}$~keV (errors at 90\% confidence level). The corresponding luminosity and flux are $L_{\rm 2-10~keV}=0.9\pm0.3\times10^{46}$~erg~sec$^{-1}$ and f$_\mathrm{0.5-7.0keV}=1.8_{-0.2}^{+0.3}\times 10^{-14}$~erg~sec$^{-1}$~cm$^{-2}$, respectively (errors at 90\% confidence level). In the case of a power law with a photon index fixed to $\Gamma_{\rm X}=2.2$, the corresponding energy cutoff, luminosity and flux are $E_{\rm cut}=13.0_{-5.5}^{+24.8}$~keV, $L_{\rm 2-10~keV}=1.0\pm0.3\times10^{46}$~erg~sec$^{-1}$ and f$_\mathrm{0.5-7.0keV}=1.9{\pm0.3}\times 10^{-14}$~erg~sec$^{-1}$~cm$^{-2}$, respectively (errors at 90\% confidence level).
In the following, we consider the best-fit single power law for the comparison of \shortname\ to other high-$z$ quasars, since most of the luminosities in the literature were computed with this model.

Finally, as a reference, we also report the values of flux and luminosity by assuming $\Gamma_{\rm X}=2$, which is often adopted in the literature for sources without meaningful constraints on the photon index (see e.g. \citealt{Zuo2024}). In this way we obtained a luminosity which is about an order of magnitude fainter compared to the best-fit photon index, with $L^{\Gamma_{\rm X}=2}_{\rm 2-10~keV}=1.6\times10^{45}$~erg~sec$^{-1}$, even though the corresponding observed flux is only a factor $\sim3$ lower (f$^{\Gamma_{\rm X}=2}_\mathrm{0.5-7.0keV}=7.5\times 10^{-15}$~erg~sec$^{-1}$~cm$^{-2}$). This difference is mainly due to the extrapolation to energies not directly sampled with {\it Chandra}, which make the luminosity estimate highly dependant on the photon index value. As highlighted in the next section, many more objects at high redshift covered with relatively shallow observations might have a similar soft X-ray emission and the assumption of $\Gamma_{\rm X}=2$ can systematically underestimate their luminosities.

\section{Comparison with \lowercase{z}~$>5.5$ quasars}

Based on the X-ray analysis of the {\it Chandra} observations of \shortname, its X-ray luminosity in the canonical 2--10~keV energy range is among the brightest ones observed in high-$z$ quasars, even when considering sources with jets aligned close to our line of sight (i.e., blazars; see \citealt{Ighina2019}).
In the left panel of Fig. \ref{fig:Lum_comp} we compare the rest-frame 2--10~keV X-ray luminosity of \shortname \, to other known $z>5.5$ quasars with dedicated {\it Chandra} or {\it XMM-Newton} X-ray observations from the literature. As clear from the plot, \shortname \, is among the X-ray-brightest sources in this redshift range, with an X-ray luminosity about $\sim$15 times higher than the median of all the X-ray-detected quasars, $\bar{L}_{\rm 2-10~keV}\sim1.3\times10^{45}$~erg~sec$^{-1}$. The only two sources with a comparable luminosity at $z>6$ are PSO~J030947+271757108 at $z=6.1$ \citep[blue pentagons][]{Belladitta2020,Moretti2021} and CFHQS~J142952+544717 at $z=6.19$ \citep[yellow triangles][]{Willott2010,Migliori2023,Marcotulli2025}, while a third radio quasar with a similar luminosity ($L_{\rm 2-10keV}=1.7-3.6~\times 10^{46}$~erg~s$^{-1}$) is at $z=5.47$ \citep{Khorunzhev2021}.

Interestingly, even though both these $z>6$ quasars and \shortname\ are radio loud, suggesting they all host powerful relativistic jets (see \citealt{Frey2011,Spingola2020}), their multi-wavelength properties are quite different. PSO~J030947+271757108 was classified as a flat-spectrum radio quasar (i.e., a blazar) based on its bright radio and X-ray luminosity (see \citealt{Belladitta2020,Spingola2020}). This source presented a statistically significant variability in the soft X-rays, with an increase of the 0.5--2~keV flux of a factor $\sim3$ in only $\sim300$~sec in the rest frame \citep{Moretti2021}.
Similarly, based on recent {\it NuSTAR} observations, CFHQS~J142952+544717 also showed signs of variability, increasing its total flux by a factor $\sim2.6$ over a timescale of $\sim110$ days in the observed frame ($\sim15$~days at $z=6.19$; see \citealt{Marcotulli2025}) with respect to previous e-ROSITA, XMM-{\it Newton} and {\it Chandra} estimates \citep{Medvedev2020,Medvedev2021,Migliori2023}.
In both cases, the authors argue that the variability (and therefore the X-ray emission) is produced by the relativistic jets. Indeed, to explain the short timescales of these variations relativistic effects are needed, since they increase the time in the source's rest frame.
However, in the case of \shortname, we do not detect any significant variability on a rest-frame timescale of $\sim30$ days (similar to CFHQS~J142952+544717), as detailed in Appendix \ref{sec:variab_X}, albeit further monitoring on larger timescales is needed to rule out the presence of X-ray variability in the source.
Moreover, in both cases the slope of the X-ray emission is significantly harder ($\Gamma_{\rm X}=1.7\pm0.2$ for PSO~J030947+271757108 and $\Gamma_{\rm X}=2.2\pm0.2$ for CFHQS~J142952+544717, errors at 90\% confidence level; \citealt{Ighina2022a,Migliori2023}) compared to \shortname\ ($\Gamma_{\rm X}=3.3\pm0.4$). Indeed, not only is \shortname \, one of the most X-ray luminous quasars currently known at $z>5.5$, but it also presents a much larger photon index value compared to what is typically inferred for optically bright quasars -- both radio-loud and radio-quiet-- where $\Gamma_{\rm X}$ generally ranges from $\sim1.8-2.0$ (e.g. \citealt{Shemmer2005,Vignali2005,Shemmer2014,Moretti2014,Nanni2017,Ai2017,Banados2018c,Shaban2022}). 
The only $z>6$ source with a similar X-ray shape is HSC~J092120+000722, a radio-quiet quasar at $z=6.56$ \citep{Matsuoka2018a}. Based on dedicated {\it Chandra} observations, \cite{Wolf2023} derived an X-ray luminosity of $L_{\rm 2-10keV}=3.7^{+4.0}_{-1.9}~\times 10^{45}$~erg~s$^{-1}$, consistent with other quasars at similar redshift (see pink diamond in Fig. \ref{fig:Lum_comp}), but with an unusually steep photon index value, $\Gamma_{\rm X}=3.2^{+0.7}_{-0.6}$. Interestingly, using the MgII broad emission line, \cite{Wolf2023} concluded that this system is hosting a $\sim2.5\times10^8$~M$_\odot$ SMBH potentially accreting at a super-Eddington rate ($\lambda_{Edd}\approx2.3$). 
As discussed in the next section, the large value of the photon index derived for this object (as for \shortname) is likely related to the high accretion rate.

However, we also stress that most of the X-ray observations currently available for $z>5.5$ quasars are relatively shallow and, due to the small number counts, they can only be used to estimate the observed flux. Whereas, the slope of the X-ray emission remains unconstrained (see e.g. \citealt{Vito2019,Ighina2024b}). In these cases a value of $\Gamma_X=2$ is typically assumed when computing the rest-frame properties which, however, would underestimate the L$_{\rm 2-10keV}$ luminosity if the emission was softer, as shown in the previous section for \shortname. 
This also means that there might be many more quasars at high redshift with a similarly steep X-ray emission, potentially indicating that super-Eddington accretion is more common in the early Universe.


\section{Origin of the X-ray emission}

\begin{figure*}
 	\includegraphics[width=0.48\hsize]{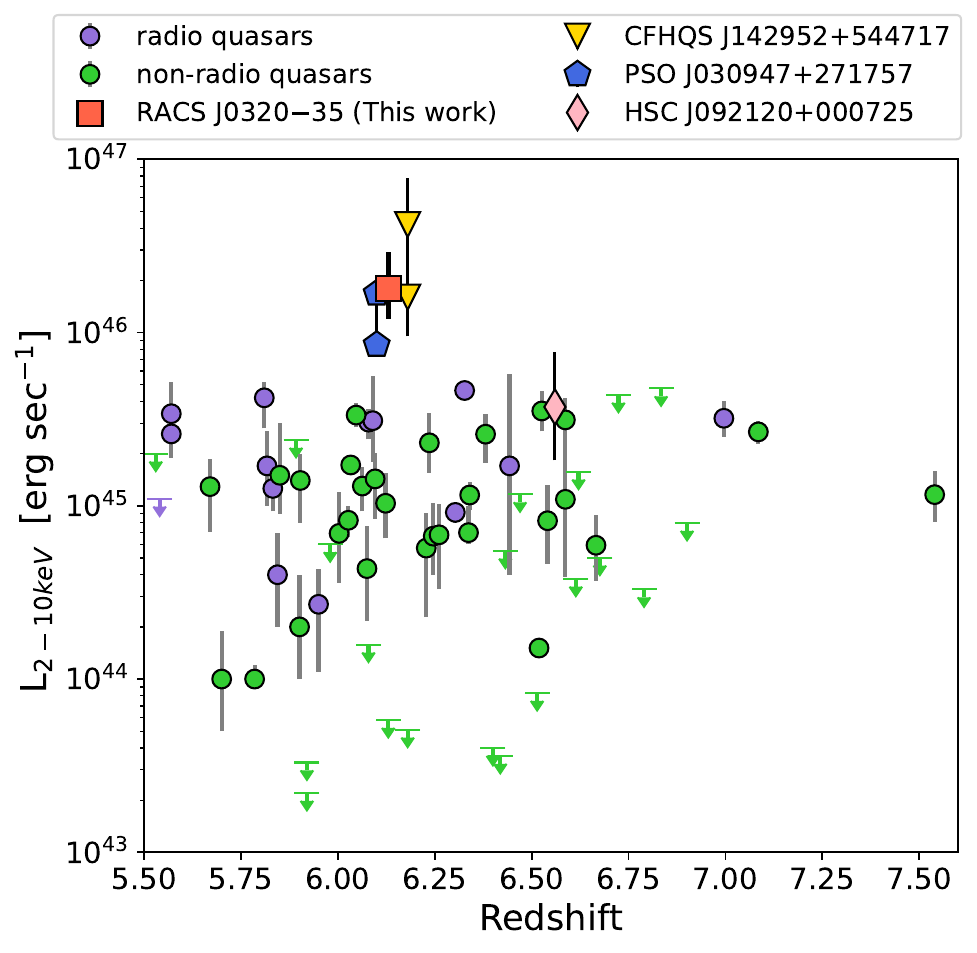}
 	\includegraphics[width=0.51\hsize]{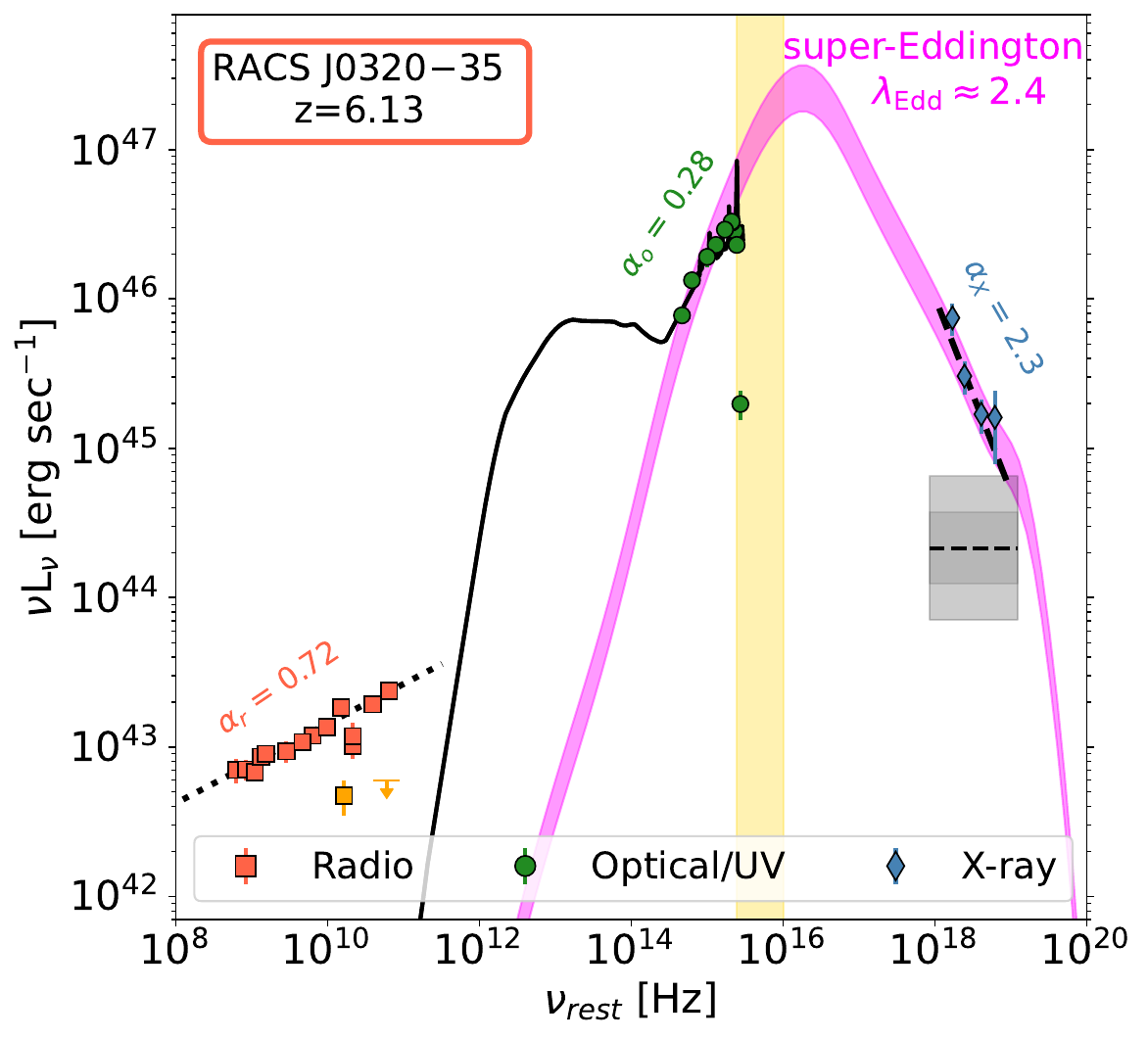}

 \caption{{\bf Left panel:} X-ray luminosity in the 2--10~keV energy band (rest frame) as a function of redshift for the $z>5.5$ quasars with X-ray observations from either {\it Chandra} or {\it XMM-Newton} available in the literature. We highlight the following objects: \shortname \,(red square; this work), HSC~J092120.56+000722.9 (pink diamond; \citealt{Wolf2023}), CFHQS~J142952+544717 (yellow triangles; \citealt{Migliori2023,Marcotulli2025}) and PSO~J030947+271757 (purple pentagon; \citealt{Moretti2021}). Since these last two sources present variable/flaring emission, we show two data-points representing the quiescent and the flaring state respectively. All the other quasars are reported with different colours based if they are also detected in the radio band (purple) or not (green). {\bf Right panel:} Rest-frame, multi-wavelength spectral energy distribution of \shortname. The X-ray weak super-Eddington SED from \cite{Pacucci2024} is shown in magenta. The solid black line is a quasar template \citep{Polletta2007} matched to the optical-UV data-points. The dashed black line is the X-ray emission expected from the UV-X-ray relation derived by \cite{Lusso2016} and assuming $\Gamma_{\rm X}=2.0$. The gray regions show the 1,2$\sigma$ dispersion of the relation. The vertical yellow region indicates frequencies heavily affected by the absorption of the intergalactic medium.}
    \label{fig:Lum_comp}
\end{figure*}

As discussed in the previous sub-section, \shortname \, stands out in terms of the intensity and shape of the X-ray emission when compared to the general quasar population, even at high redshift. For this reason, the origin of its high-energy emission is not easy to determine. Here we discuss different scenarios which could, in principle, reproduce the observed multi-wavelength SED of \shortname \, (shown in Fig. \ref{fig:Lum_comp}, right panel).\\

\subsection{Considering the emission from the relativistic jets}
Given the radio nature of \shortname, the X-ray luminosity of this source can be interpreted as due to the relativistically boosted radiation produced by jets oriented close to our line of sight---that is, the source is a blazar. 
Currently known high-$z$ blazars belong to the flat-spectrum radio quasar (FSRQ) population and present both broad emission lines in the rest-frame UV spectrum as well as a strong and `hard' X-ray emission (or `flat' photon index, $\Gamma_{\rm X}\lesssim1.7$; \citealt{Ighina2019,Moretti2021,Banados2024}). In these systems, the high energy emission is normally interpreted as originating via inverse Compton interaction of the electrons within the jets with external seed photons produced by the accretion disc, the BLR and/or the dusty torus (see e.g. \citealt{Ghisellini2009}).  While \shortname \, also shows strong X-ray emission, the shape of its X-ray spectrum is not consistent with the one expected from the inverse Compton (IC) interaction with external photons. At the same time, the IC interaction of the electrons with the low-frequency photons produced by the same electrons through Synchrotron (Synchrotron-Self Compton; SSC) would require a radio emission orders of magnitudes larger than what observed in \shortname. 

The absence of the Ly$\alpha$ broad emission line (see \citealt{Ighina2023}) and the soft high-energy radiation could still be consistent with a BL Lac blazar nature, making it the most distant currently known in this class (e.g. \citealt{Paliya2020}). These systems are characterised by a very low accretion dominated by the advection of gas into the BH. This process makes the accretion less luminous compared to a typical geometrically thin and optically thick disc, resulting in lower amount of ionizing photons and, consequently, of broad emission lines intensity. At the same time, the continuum emission in the radio, optical, UV and X-ray bands is dominated by the radiation produced by the jets after being amplified by relativistic boosting.

Based on the SED shape of \shortname, increasing in the optical/UV\footnote{based on data-points from DES \citep{Abbott2021}, the VISTA Kilo-degree Infrared Galaxy \citep{Edge2013} and the Wide-field Infrared Survey Explorer catalogue \citep{Eisenhardt2020}.} and decreasing in the X-rays as a function of frequency ($\alpha_\nu^{\rm o}=0.28\pm0.03$; see \citealt{Ighina2023}), this source would be classified as a high synchrotron peaked blazar (HBL). In this case, the observed X-ray emission would be produced through a synchrotron process, as opposed to IC in the case of FSRQs (e.g. \citealt{Ghisellini2013}). The best-fit photon index derived for \shortname \, ($\Gamma_{\rm X}=3.3\pm0.4$) would also be consistent with the ones typically observed in this HBL class ($\Gamma_{\rm X} \sim 2-3$; see e.g. \citealt{Middei2022}). Moreover, the X-ray-to-radio ratio\footnote{quantified by $\alpha_{\rm xr}=0.13\times log(F_{\rm 1keV}/F_{\rm 5GHz})$, where the monochromatic fluxes are defined in the rest-frame \citep{Fossati1997}.}  derived for \shortname\ would be consistent with the typical values derived for the HBL population \citep[$\alpha_{\rm xr}=0.51\pm0.03$; see, e.g.][]{Donato2001}, even though HBL sources are significantly fainter in both the radio and X-ray band ($log(\bar{L}_{\rm 5GHz}/{\rm erg/s})\sim41.5$ and $log(\bar{L}_{\rm 1keV}/{\rm erg/s})\sim44.6$; \citealt{Donato2001}) with respect to to \shortname\ ($log({L}_{\rm 5GHz}/{\rm erg/s})\sim43.0$ and $log({L}_{\rm 1keV}/{\rm erg/s})\sim46.8$).
Despite some similarities with HBL objects, there are several pieces of information that would rule out this nature. The radio properties of \shortname---namely a relatively steep spectral index ($\alpha_{\rm r}=0.72\pm0.02$) and a faint core component on VLBI scales---suggest that its low-frequency emission is not dominated by relativistic beaming as expected for HBL \citep[e.g.][]{Wu2007,Liuzzo2013}. Indeed, even if partially resolved, in HBL sources we would still expect the core emission to dominate the flux density on milliarcsecond scales at the rest-frame frequencies sampled with the LBA ($\sim15-60$~GHz; see, e.g. \citealt{Piner2010}). Given the faint core emission, the overall radio emission observed in \shortname\ is dominated by the extended regions of the relativistic jets, which are resolved in the LBA image (i.e. on scales $\gtrsim50$~mas or $\gtrsim300$~pc). This means the quasar is not a compact steep-spectrum or a peaked-spectrum radio source, that is, it is not a young radio object.

Moreover, its overall radio+optical/UV+X-ray SED cannot be reproduced by a single component (typically log parabola, e.g., \citealt{MassaroE2004}), as expected if they are all associated with the synchrotron emission of the relativistic jet \citep[e.g.][]{Giommi2021}. We also checked $\gamma$-ray observations from the FERMI Large Area Telescope (LAT), but did not find any significant high-energy emission associated with the relativistic jets at the position of the quasar (see Appendix \ref{sec:fermi}). Finally, the absence of significant variability in the radio band on timescales of about one year in the rest frame also disfavours a HBL nature for \shortname \, \citep[see e.g.][]{Hovatta2014}, although a monitoring on a larger time window is needed to properly characterise the presence/absence of variability. Similarly, multi-epoch monitoring in the optical-UV-X-ray rest-frame bands, where we expect HBL to be the highly variable on timescales from days to years \citep[e.g.][]{Zhang2017,Wierzcholska2025}, will also be crucial to further rule out the BL Lac nature of \shortname. We stress that, if the HBL nature of \shortname\ were confirmed, it would be, by far, the highest redshift BL Lac blazar currently known \citep[e.g.][]{Paliya2020b}.

\subsection{Super-Eddington accretion} 
\label{sec:superEdd}
Another possibility to explain the unique X-ray slope of \shortname \ is the presence of a SMBH accreting above its Eddington limit.

Recently, there has been renewed interest in studying the X-ray properties of fast accretors because of a new population of high-$z$ AGN uncovered by JWST at $z > 4$ that are not detected in the X-rays (see, e.g, \citealt{Maiolino2024, Yue_2024_LRD}). Several studies have investigated how super-Eddington accretion can naturally lead to an apparent X-ray weakness at high redshift, both from a theoretical perspective \citep{Madau2024,Lambrides2024,Inayoshi2024} and through detailed simulations (e.g., \citealt{Pacucci2024}). In both cases the X-ray non detections are explained as due to a very steep X-ray emission that, when evaluated in the observed 0.5--7~keV energies (i.e. $\gtrsim3-40$~keV in the rest-frame for $z\gtrsim4$ sources), falls below the detection limit of current facilities.

From an observational point of view, evidence for a correlation between $\Gamma_{\rm X}$ and $\lambda_{\rm Edd}$ was found by several studies at low redshift \citep[e.g.,][]{Shemmer2008,Fanali2013,Brightman2013}. By analysing the X-ray properties of highly accreting ($\lambda_{\rm Edd}>0.8$) sources at $z\sim0.1-0.6$, \cite{Huang2020} and \cite{Liu2021} found that the median photon index of these objects is larger ($\bar{\Gamma}_X=2.2\pm0.2$) compared to sources with slower accretion ($\lambda_{\rm Edd}<0.3$; $\bar{\Gamma}_X=1.8\pm0.2$).
Similarly, several studies targeting $z>6$ UV-bright quasars revealed that they have a steeper photon index compared to $z<6$ sources (see \citealt{Vito2019,Wang2021b,Zappacosta2023}). From the analysis of 18 highly accreting $z>6$ quasars, \cite{Zappacosta2023} found the median photon index of these sources to be $\bar{\Gamma}_{\rm X}=2.4\pm0.1$, again significantly different from the $\Gamma_{\rm X}=1.8-2.0$ normally found at lower redshift. While some works interpreted this trend as an increase of the typical accretion rate of bright quasars as a function of redshift (e.g. \citealt{Risaliti2009,Wang2021}), other works did not confirm the presence of a $\Gamma_{\rm X}-\lambda_{\rm Edd.}$ relation \citep[see e.g.][]{Laurenti2022, Tortosa2024}.

In this context, the photon index value derived for \shortname \, is larger, at more than a 95\% confidence level, than all the super-Eddington sources analysed by \cite{Liu2021} as well as than the median value derived by \cite{Zappacosta2023}. At the same time, the best-fit value derived for \shortname, $\Gamma_{\rm X}=3.3\pm0.4$, is fully consistent with the typical values derived from the simulation of $\lambda_{\rm Edd}\sim2-4$ AGN described in \citeauthor{Pacucci2024} (\citeyear{Pacucci2024}; median photon index of 3.1) as well as with the model proposed by \citeauthor{Madau2024} (\citeyear{Madau2024}; expected photon index values $\gtrsim2.8$). We stress that here the photon index value is simply a measurement of the soft-to-hard X-ray emission in this system, without the assumption of a physically motivated model. Indeed, in the case of super-Eddington accretion we can expect the presence of a very low-energy cutoff ($E_{\rm cut}\sim30-40$~keV; \citealt{Kara2017,Madau2024,Inayoshi2024}), which would also be consistent with the best-fit estimates derived in Sec. \ref{sec:observations} ($E_{\rm cut}=10.1^{+4.8}_{-2.6}$~keV). Indeed, the combination of a very low energy value for the cutoff and the high luminosity would put \shortname\ in a unique place of the parameter space, compared to the general AGN population (see, e.g, fig. 4 in \citealt{Bertola2022}).

By making conservative assumptions on the bolometric luminosity of \shortname, we can have a rough estimate of the physical parameters of the accretion process. Given the optical and X-ray luminosity of \shortname, we can expect the corresponding bolometric luminosity of $L\gtrsim3\times 10^{46}$~erg~s$^{-1}$. The corresponding X-ray bolometric correction ($K_{\rm X}=L_{\rm bol}/L_{\rm X}$) is $K_{\rm X}>20$, based on the $1\sigma$ dispersion of the relation derived by \cite{Duras2020}. At the same time, we can also assume that BH hosted in \shortname\ has mass $\lesssim10^9$~M$_\odot$, based on SMBHs with similar optical/UV properties \citep[e.g.][]{Mazzucchelli2023}. The corresponding Eddington ratio would be $\lambda_{\rm Edd}>2.8$ (or $\lambda_{\rm Edd}>1.4$, for the exponential cutoff model), supporting the super-Eddington accretion scenario. We note that these estimates do not take potential trends of the bolometric correction as a function of redshift \citep[e.g.][]{Maiolino2024} or Eddington ratio \citep[e.g.][]{Gupta2024} into account, which would further increase the derived value of $\lambda_{\rm Edd}$. Conversely, if we considered a BH accreting sub-Eddington,  $\lambda_{\rm Edd}\sim0.5$, the inferred BH mass would be $M_{\rm BH}\gtrsim6 \times 10^9\ {\rm M}_\odot$---comparable to the most massive $z>6$ SMBH known yet \citep[e.g.][]{Wu2015} and with roughly three times the X-ray luminosity \citep{Connor2021b}.



Several models also predict different optical/UV properties in case of super-Eddington accretion, both in terms of continuum \citep[e.g.][]{Pognan2020} and emission lines \citep[e.g.][]{Lambrides2024}. In particular, in the scenario proposed by \cite{Madau2025} the strength of high-ionisation emission lines depends on the viewing angle with respect to the rotation axis of the disc (see their Fig. 7). Indeed, by assuming a disc-like geometry for the broad-line region \citep[e.g.][]{Gaskell2009,Savic2024,Rigamonti2025}, the total UV continuum perceived by the BLR clouds and an observer with a different orientation can be very different. For the specific class type 1 AGN---that is, observed close to the axis---the continuum emission produced by the disc outshines the line emission, resulting in apparent weak broad emission lines.
As mentioned before, in the optical spectrum of \shortname\ \citep{Ighina2023} the Ly$\alpha$ line was not detected, broadly consistent with the scenario described in \cite{Madau2025}. However, the Ly$\alpha$ profile can also be affected by the absorption of neutral hydrogen; therefore further NIR observations covering additional emission lines are needed to test these predictions.

To visually compare the SED expected from a super-Eddington quasar and the observations obtained for \shortname, we considered the SEDs presented in \cite{Pacucci2024}. In particular, these authors used General Relativistic Radiation Magneto Hydrodynamic (GRRMHD, see \citealt{Pacucci2024} for details about the codes used) simulations of BHs with mass ${\rm M}_{\rm BH} = 10^7 \, {\rm M}_\odot$ accreting at super-Eddington rates up to $\lambda_{\mathrm{Edd}} = 13.4$ to compute the resulting SEDs. The study finds that the observed X-ray emission for mildly super-Eddington accretion ($1.4 < \lambda_{\mathrm{Edd}} < 4$) can be extremely steep, with a median photon index $\Gamma_{\rm X} = 3.1$ and a mode of $\Gamma_{\rm X} = 4.4$, especially in slowly spinning or non-spinning black holes ($a \sim 0$)\footnote{where $a$ is the non-dimensional spin parameter which varies between 0, i.e., non-spinning, to 1, i.e., maximally spinning.} observed at inclination angles greater than $30^\circ$ from the polar axis.
Based on these simulations, the non-detection of the $\sim 10^7 \, {\rm M}_\odot$ SMBHs observed with JWST is due to the steep X-ray emission resulting in a faint emission at high energies, which are redshifted to the observed band. 


We show in the right panel of Fig. \ref{fig:Lum_comp} the expected optical-UV-X-ray SED based on the results \cite{Pacucci2024}. This SED is appropriate for a non-spinning ($a = 0$) SMBH accreting with $\lambda_{\rm Edd} = 2.4$; the emission is observed from an inclination of $i=10^\circ$ from the polar axis (see fig. 4, top-left panel in \citealt{Pacucci2024}). We stress that the simulations presented in \cite{Pacucci2024} focused on $\sim10^7$~M$_\odot$ BHs and result in lower UV and X-ray luminosities. For this reason, we rescaled the SED obtained from the simulations to match the measurements of RACS~J0320$-$35 (i.e. by a factor $\approx$100). This normalization can be explained, by a first order approximation, by a larger mass SMBH ($\sim10^9$~M$_\odot$, or even larger for $i>10^\circ$), as expected for \shortname.
As shown in Fig. \ref{fig:Lum_comp}, there is an excellent agreement between the expected emission and the observed data across the different wavelength bands. In particular, the quantitative properties in terms of relative UV-to-X-ray emission and shape of the simulated SED in \cite{Pacucci2024} considered here match closely those measured for \shortname: ${\alpha}_{\rm ox} = 0.99$ and $\Gamma_{\rm X} = 3.14$. 

We note that the presence of radio-bright jets in \shortname\ would suggest larger values of the spin parameter compared to the one considered for the SED shown in Fig. \ref{fig:Lum_comp}. While the comparison in Fig. \ref{fig:Lum_comp} is only meant to show the broad consistency between the quasars and the simulations of \cite{Pacucci2024}, we also note that most of the radio emission is produced by extended regions of the jets (see Sec. \ref{sec:observations}), that is, related to previous activity of the SMBH while the X-ray emission is measuring the more recent accretion process. Therefore, the SMBH could have slowed down due to the energy extracted by the relativistic jets themselves and decreased the value of $a$. Based on eq. 7 in \cite{Meier2002}, the e-folding spindown timescale for a BH accreting at its Eddington limit and with a jet duty cycle of 10\% is $\approx2\times10^5$~yr. By assuming that the radio jet in \shortname\ expanded with a 0.1$c$ velocity \citep[e.g.][]{An2012} up to $\approx0.3-1$~arcsec (or $\approx1.7-5.6$~kpc; i.e. in between the ATCA and LBA resolutions), the associated jet activity timescale is $\approx1-10\times10^5$~yr (for an inclination of 10--30$^\circ$), that is, enough for the BH to significantly spin down.
Using GRRMHD simulations, \cite{Ricarte2023} showed that for 10--10$^{4}$~M$_\odot$ BHs the equilibrium spin parameter in the presence of relativistic jets and mildly super-Eddington accretion lies between $a\sim0.3-0.6$. Detailed simulations, similar to the ones presented in \cite{Pacucci2024}, targeting a higher mass range $\sim10^{8-9}$~M$_\odot$ for different spin values will be crucial to accurately constrain the properties of the SMBH hosted in \shortname.

\section{Conclusions}

In this work we presented the multi-wavelength properties of a recently discovered $z=6.13$ radio-bright quasar, \shortname\ \citep{Ighina2023,Ighina2025}. 

Based on dedicated uGMRT+ATCA observations, we concluded that the source presents a radio emission typical of the general radio-loud population ($\alpha_{\rm r}=0.72\pm0.02$; e.g. \citealt{Calistro2017}) and does not show signs of strong variability on timescales of $\sim6$ years in the observed frame. Dedicated VLBI-LBA observations revealed that the core radio emission (on $\sim20$~mas scales) is faint, a factor $\sim3-5$ lower compared to the one observed on arcsec scales. Finally the overall radio+optical/NIR+X-ray SED cannot be reproduced by a single log-parabola component. Together, these observations suggest that the emission observed in \shortname \, is not dominated by relativistic boosting---that is, \shortname\ is not a blazar, albeit further multi-epoch and multi-wavelength observations are needed to fully discard this scenario.
 
While the low-energy properties of \shortname \, are consistent with the typical radio-loud population of quasars, high-energy observations with the {\it Chandra} telescope uncovered an unusually strong and soft X-ray emission. In particular, \shortname, is, within errors, the most X-ray luminous quasar at $z>5.5$ ($L_{\rm 2-10kev}=1.8_{-0.7}^{+1.1}\times 10^{46}$~erg~s$^{-1}$) and the one with the steepest photon index ($\Gamma_{\rm X}=3.3\pm0.4$). 
At the same time, we also note that an accurate estimate of the photon index is not available for most of the high-$z$ quasars currently detected in the X-rays. Therefore, deeper X-ray observations of a large sample of $z\gtrsim6$ sources are necessary to test whether such soft X-ray emission is more common in the early Universe.

The most likely scenario that could explain the steep X-ray spectrum in \shortname\ is that its SMBH is accreting above its Eddington limit. Indeed, the observed X-ray slope (or potential cutoff, depending on the model) is fully consistent with theoretical predictions describing super-Eddington accretion \citep[e.g.][]{Madau2024,Inayoshi2024}. Moreover, we showed that the optical+UV+X-ray SED of \shortname \, can be well reproduced by the presence of a SMBH with $M_{\rm BH}\approx10^{9}$~M$_\odot$ and $\lambda_{\rm Edd}=2.4$, based on the simulations and models presented in \cite{Pacucci2024}.

The overall multi-wavelength properties of \shortname \, make it one of the most promising candidates for super-Eddington accretion in the early Universe. Many recent works on super-Eddington accretion focused on the AGN population uncovered by JWST. However, due to their low BH mass ($\sim10^{7}$~M$_\odot$), these systems are much fainter, and only upper-limits are available in the X-rays \citep[e.g.][] {Maiolino2024}. Given its high luminosity across the entire electromagnetic spectrum, \shortname\ offers the perfect laboratory where we can directly test predictions of theoretical models and simulations in the context of super-Eddington accretion. To this end, future IR spectroscopic observations covering broad emission lines (e.g. CIV, MgII, H$_\alpha$ and H$_\beta$) will be essential to constrain the mass and accretion properties of the SMBH hosted in \shortname\ and confirm or discard the presence of super-Eddingotn accretion. At the same time, deeper X-ray observations will also be crucial to better constrain the X-ray intensity and shape, especially in the soft part of the spectrum, where we expect the majority of the X-ray emission. 

Finally, the radio-bright nature of this unique quasar might imply an accretion--jets relation. For example, in the scenario proposed by \cite{Jolley2008}, jets can enhance the accretion onto the central SMBH by converting part of the accreting mass-energy into jet kinetic power instead of radiation, which otherwise generally limits further accretion. As relativistic jets carry away a substantial amount of kinetic power but very little mass, jetted quasars could, in principle, accrete more material than quasars with a similar UV luminosity but without relativistic jets (see e.g. \citealt{Connor2024}).

\begin{acknowledgments}
 We want to thank the anonymous referee for their comments, which improved the quality of the paper.
L.I., A.C. and A.M. acknowledge financial support from INAF under the projects ``Quasar jets in the early Universe'' (Ricerca Fondamentale 2022) and ``Testing the obscuration in the early Universe'' (Ricerca Fondamentale 2023).
Support for L.I.'s  work was provided by the National Aeronautics and Space Administration through Chandra Award Number GO3-24069X issued by the Chandra X-ray Observatory Center, which is operated by the Smithsonian Astrophysical Observatory for and on behalf of the National Aeronautics Space Administration under contract NAS8-03060.
T.C. acknowledges support from NASA Contract NAS8-03060 to the \textit{Chandra} X-ray Center.
AL acknowledges support by the PRIN MUR “2022935STW" funded by European Union-Next Generation EU, Missione 4 Componente 2, CUP C53D23000950006. 
FR acknowledges the support from the Next Generation EU funds within the National Recovery and Resilience Plan (PNRR), Mission 4 - Education and Research, Component 2 - From Research to Business (M4C2), Investment Line 3.1 - Strengthening and creation of Research Infrastructures, Project IR0000012 – “CTA+ - Cherenkov Telescope Array Plus.
JA, IM and BA acknowledge financial support from the Science and Technology Foundation (FCT, Portugal) through research grants UIDB/04434/2020 (DOI: 10.54499/UIDB/04434/2020), UIDP/04434/2020 (DOI: 10.54499/UIDP/04434/2020) and  UID/04434/2025 (DOI: 10.54499/UID/04434/2020).

The Australia Telescope Compact Array is part of the Australia Telescope National Facility (\url{https://ror.org/05qajvd42}) which is funded by the Australian Government for operation as a National Facility managed by CSIRO.
The Long Baseline Array is part of the Australia Telescope National Facility which is funded by the Australian Government for operation as a National Facility managed by CSIRO.
We acknowledge the Gomeroi, Gamilaroi and Wiradjuri people as the Traditional Owners of the ATCA, Mopra and Parkes Observatory site, respectively.
This work was supported by resources provided by the Pawsey Supercomputing Research Centre with funding from the Australian Government and the Government of Western Australia.
We thank the staff of the GMRT who have made these observations possible. The GMRT is run by the National Centre for Radio Astrophysics of the Tata Institute of Fundamental Research.
The scientific results reported in this article are based on observations made by the \textit{Chandra X-ray Observatory} contained in the \textit{Chandra} Data Collection (CDC) 318~\href{https://doi.org/10.25574/cdc.318}{doi:10.25574/cdc.318}.
This research has made use of software provided by the \textit{Chandra} X-ray Center (CXC) in the application package CIAO. 

\end{acknowledgments}





%
\facilities{Chandra, ATCA, LBA, uGMRT}

\software{astropy \citep{astropy2013,astropy:2018,astropy:2022},  
          CIAO \citep{Fruscinone2006}, 
          XSPEC \citep{Arnaud1996}
          CASA \citep{Mcmullin2007}
          MIRIAD \citep{Sault1995}
          AIPS \citep{Wells1985}
          CAPTRUE \citep{Kale2021}
          }


\appendix

\section{Radio observations and analysis}
\label{sec:app_radio_data}

In \cite{Ighina2023} we were only able to loosely constrain the radio emission observed in RACS~J0320$-$35 based on the measurements and upper-limits available from public surveys. To better characterise the spectral shape and the emission produced at different scales in this system, we performed dedicated observations with the uGMRT, the ATCA, and the LBA arrays. In this section, we describe these observations and the corresponding data reduction.

\subsection{Survey observations}

\shortname\  was covered as part of several public radio surveys in addition to the observations reported in \citeauthor{Ighina2023} (\citeyear{Ighina2023}; see their table 3), namely with the Murchinson Widefield Array (MWA) and with the Australian Square Kilometre Array Pathfinder (ASKAP). In particular, RACS~J0320$-$35 is detected in the second data release of the GaLactic and Extra-Galactic All-sky MWA extended (GLEAM-X; \citealt{Ross2024}) survey, covering the 70--231~MHz frequency range. Here we consider the flux density measurements from the five wide bands of the GLEAM-X survey as reported in the catalogue: 19.3$\pm$3.5 at 87~MHz, 14.3$\pm$2.1 at 118~MHz, 10.5$\pm$1.2 at 154~MHz, 11.1$\pm$0.9 at 185~MHz and 10.0$\pm$1.0 at 215~MHz.
In Fig. \ref{fig:radio_images} we show the GLEAM-X image in the very wide band (170--231~MHz) centred on the optical position of RACS~J0320$-$35. The quasar also been detected in several scans of the RACS survey, including the 3rd scan at low frequency ($S_{\rm944MHz}=3.09\pm0.88$~mJy) and the first scans at mid ($S_{\rm1.37GHz}=2.64\pm0.43$~mJy; \citealt{Duchesne2023,Duchesne2024}) and high ($S_{\rm1.66GHz}=1.98\pm0.28$~mJy; \citealt{Duchesne2025}) frequencies. Finally, RACS~J0320$-$35 belongs to the High Declination field of the VAST \citep{Murphy2013,Murphy2021} survey performed with ASKAP. \shortname\ was covered for a total of 28 times at 888~MHz between 2023 July and 2025 April. 
We estimated the peak flux density of \shortname\ directly from the images, we did not apply corrections to account for small differences in the flux scales of each epoch (e.g. \citealt{McConnell2020}). In Fig. \ref{fig:VAST_light_curve} we show the flux density of \shortname\ at 888~MHz as a function of time. The average peak flux density of the source is $S_{\rm 888MHz}=3.2$~mJy~beam$^{-1}$ and the standard deviation is 0.5~mJy~beam$^{-1}$, consistent with the median RMS (0.4~mJy~beam$^{-1}$) and the typical uncertainty including epoch-to-epoch flux scale variations ($\sim$0.5~mJy~beam$^{-1}$; see e.q. 7 in \citealt{McConnell2020}). To check for any potential sings of variability we computed the modulation index parameter ($V$) and the reduced chi squared ($\chi^2_{\rm red}$) with respect to a constant model (see eq. 1 and 2 in \citealt{Murphy2021}). For \shortname\ we obtained: $V=0.16$ and $\chi^2_{\rm red}=0.83$. Both these values suggests that the radio emission in the quasar is not significantly variable on timescales up to $\sim2$ years. Moreover, we also note that the median values derived from VAST are fully consistent with the peak flux density measured from the first scan of RACS-low (performed on 2019 July) at 888~MHz ($S_{\rm 888MHz}=3.2\pm0.3$~mJy~beam$^{-1}$). Once again, these measurements suggest no significant variation on timescales of $\sim$6 years in the observed frame, or $\sim1$ in the rest frame, without accounting for potential beaming effects.

\subsection{uGMRT observations}
uGMRT observations of RACS~J0320$-$35 were performed on 2022 April 23 (project 42\_001; PI:Ighina) using the GMRT wideband backend (GWB; \citealt{Reddy2017} with a bandwidth of 200~MHz) in band-3 (centred at 400~MHz) and in band-4 (centred at 650~MHz). During the run, we observed 3C48 as primary calibrator and 0409$-$179 as secondary calibrator. The data reduction was performed using the CAsa Pipeline-cum-Toolkit for Upgraded GMRT data REduction (\texttt{CAPTURE}; \citealt{Kale2021}) code by applying further flagging depending on the specific observation. For the imaging analysis, we adopted a robust parameter of 0.5, that is, a compromise between resolution and sensitivity. We report the images obtained in Fig. \ref{fig:radio_images} and the results of a 2D Gaussian fit using the Common Astronomy Software Applications package (\texttt{CASA}, \citealt{Mcmullin2007}) in Tab. \ref{tab:radio_fit}. When fitting the radio spectrum we also considered a conservative 10\% (added in quadrature) in the errors to account for the uncertainty related to the calibration process.

\subsection{ATCA observations}
\begin{figure*}
\centering
	\includegraphics[width=0.312\hsize]{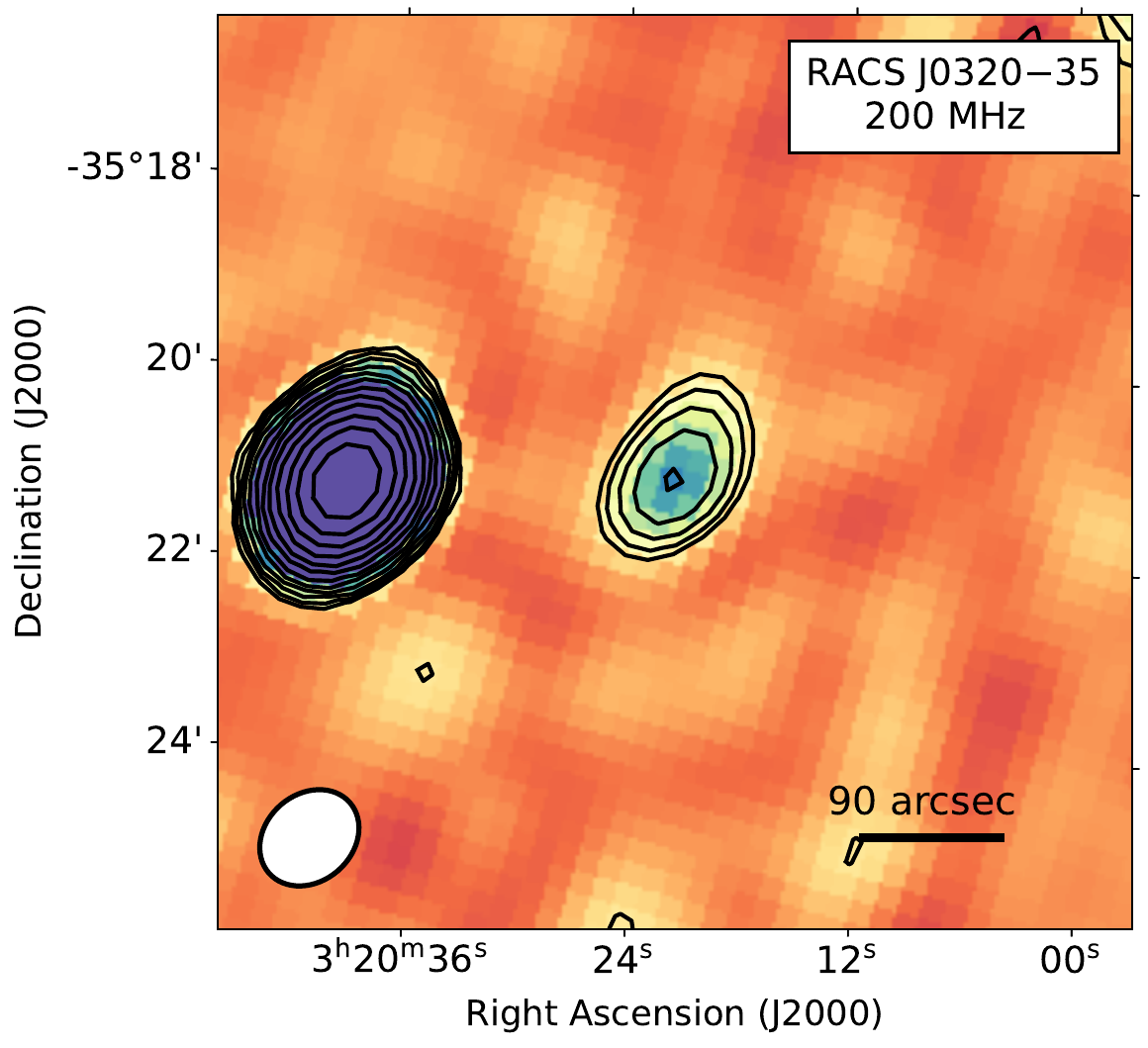}
	\includegraphics[width=0.33\hsize]{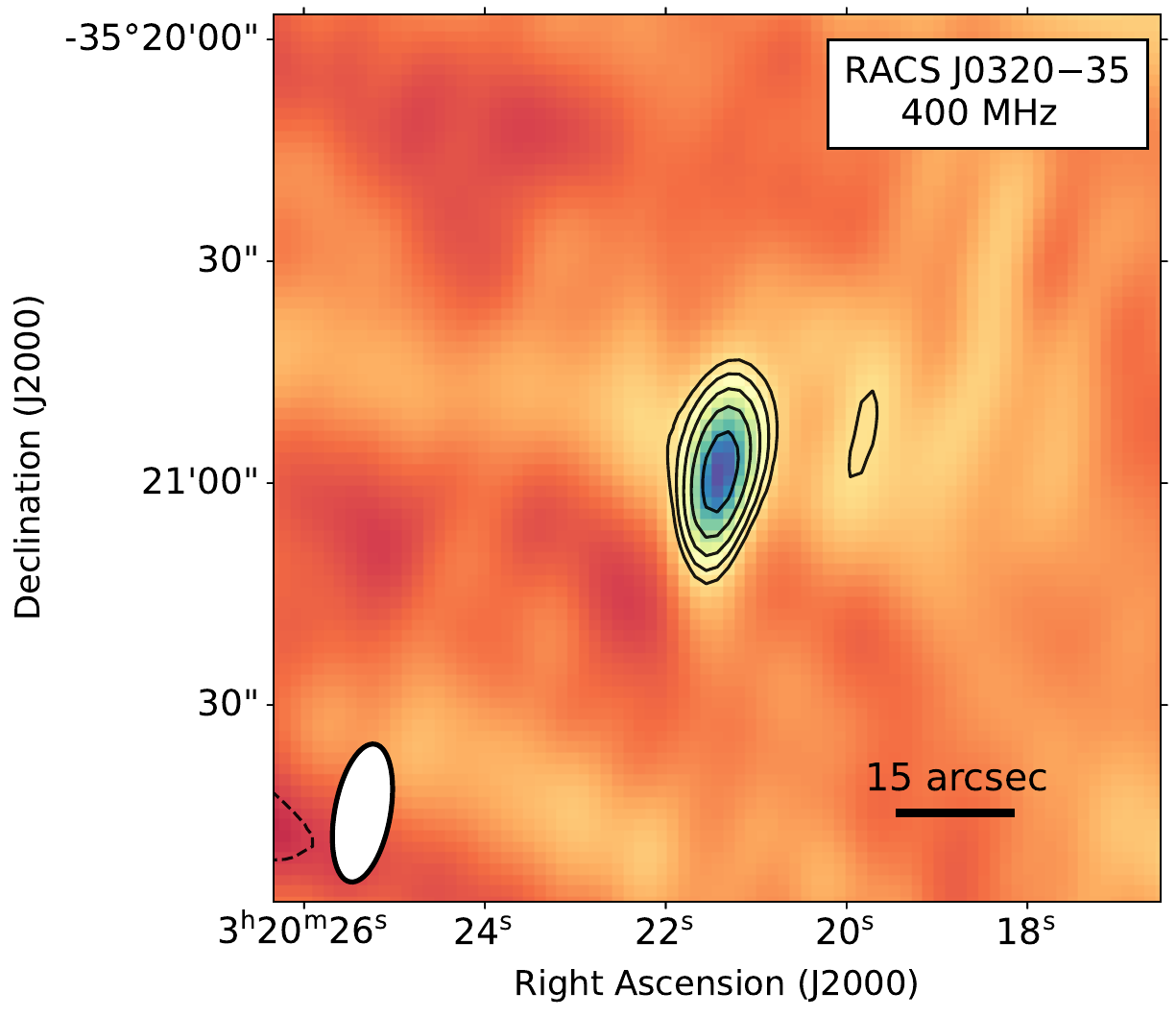}
	\includegraphics[width=0.33\hsize]{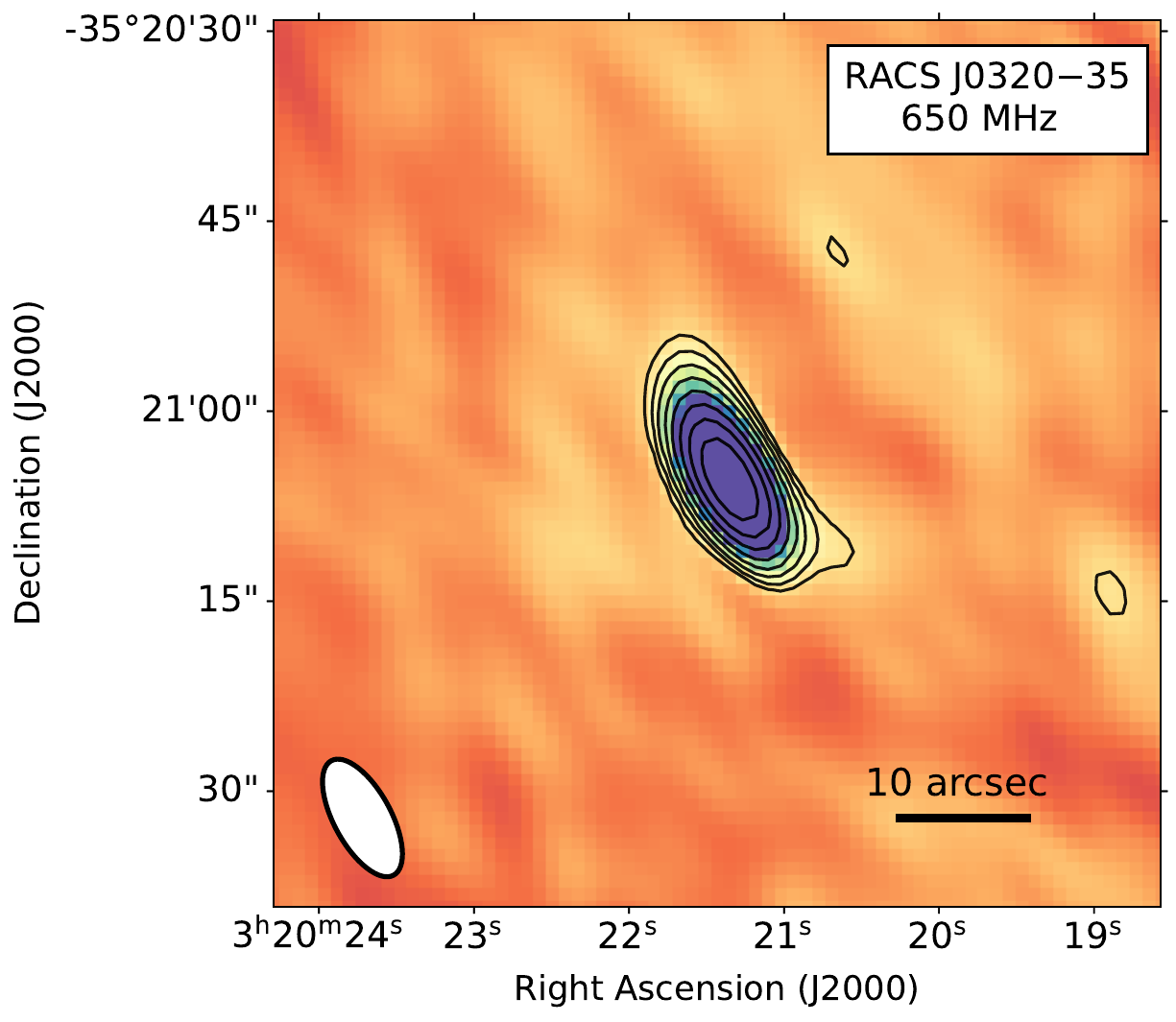}\\
 	\includegraphics[width=0.325\hsize]{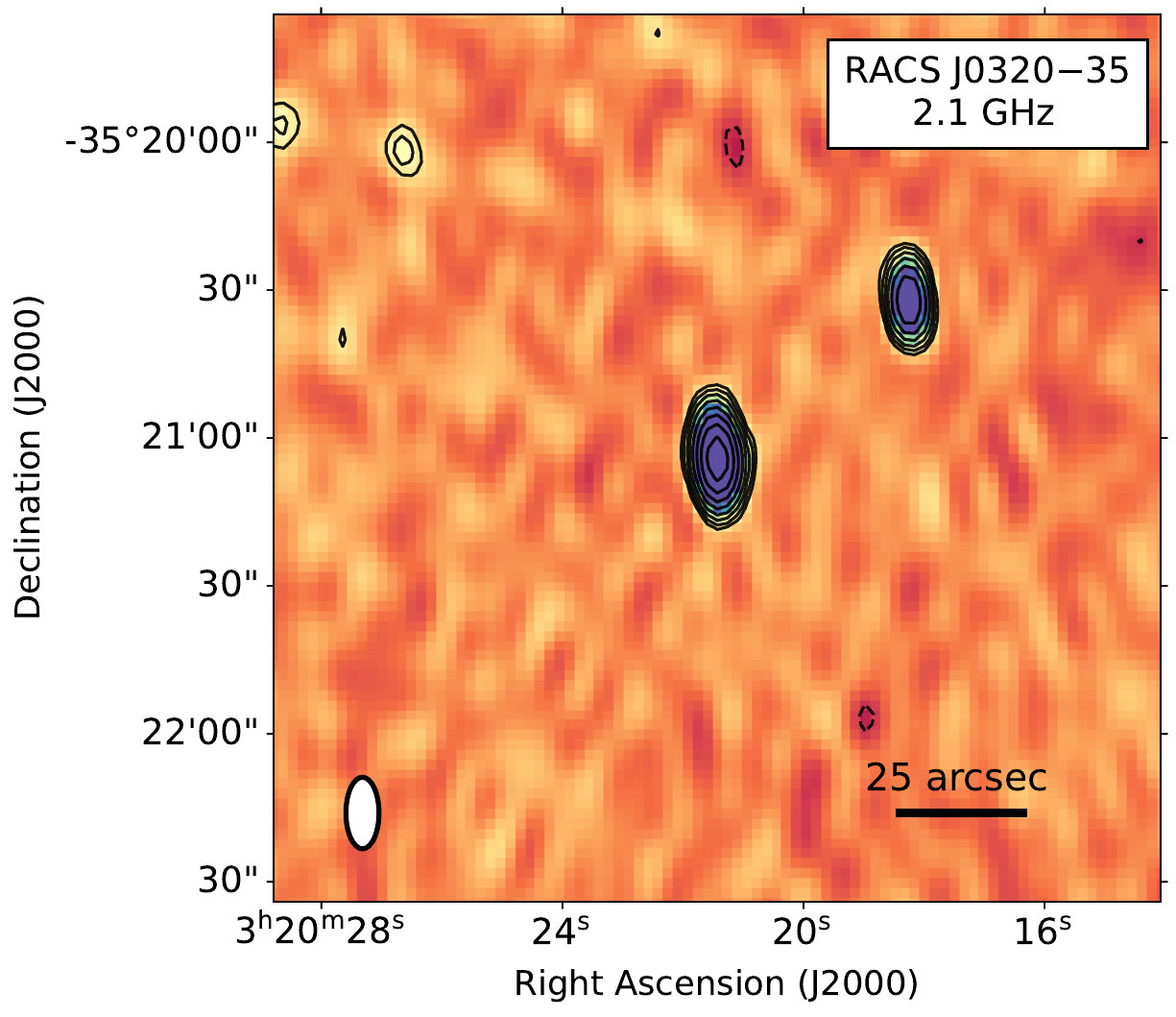}
	\includegraphics[width=0.325\hsize]{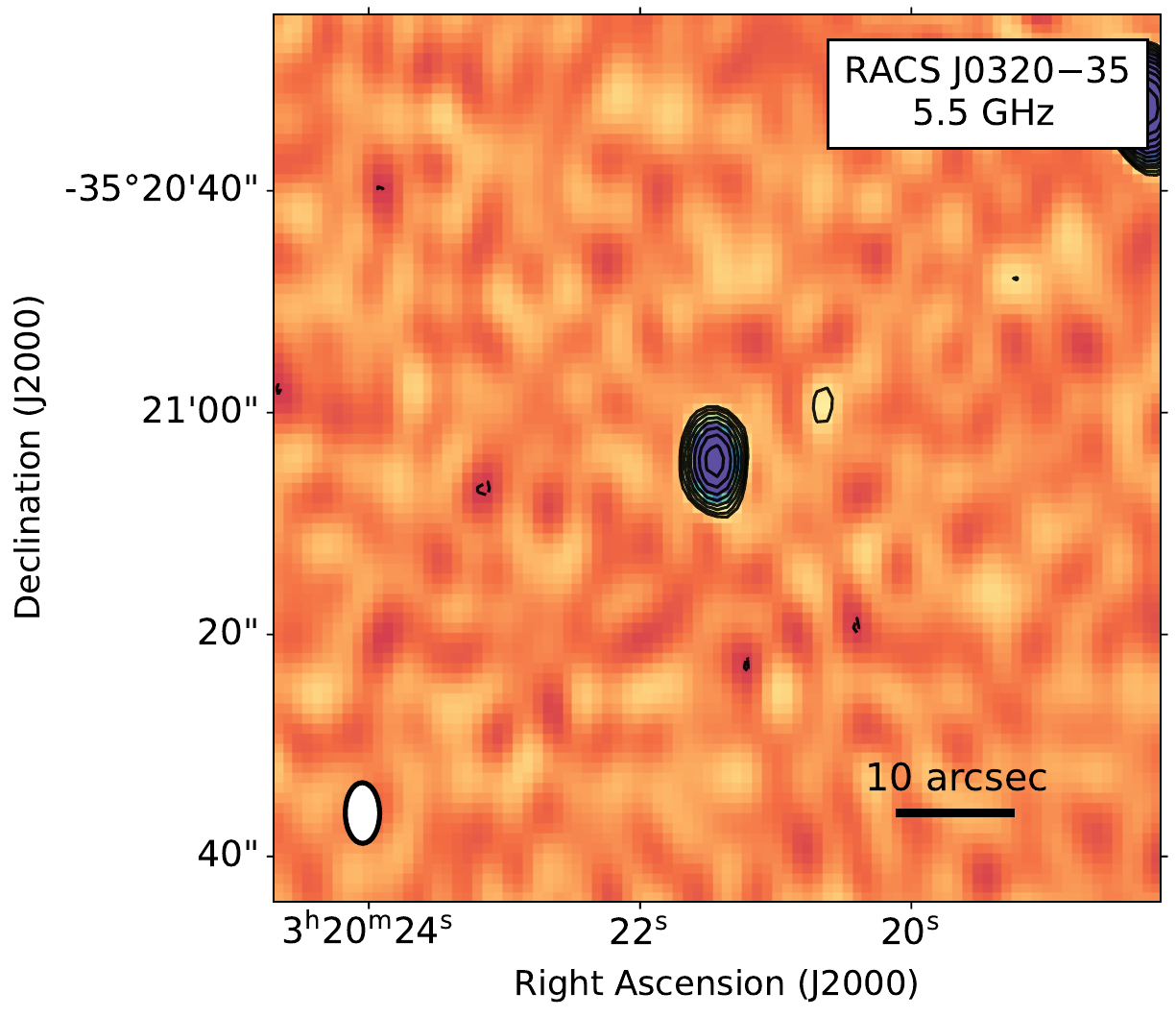}
	\includegraphics[width=0.325\hsize]{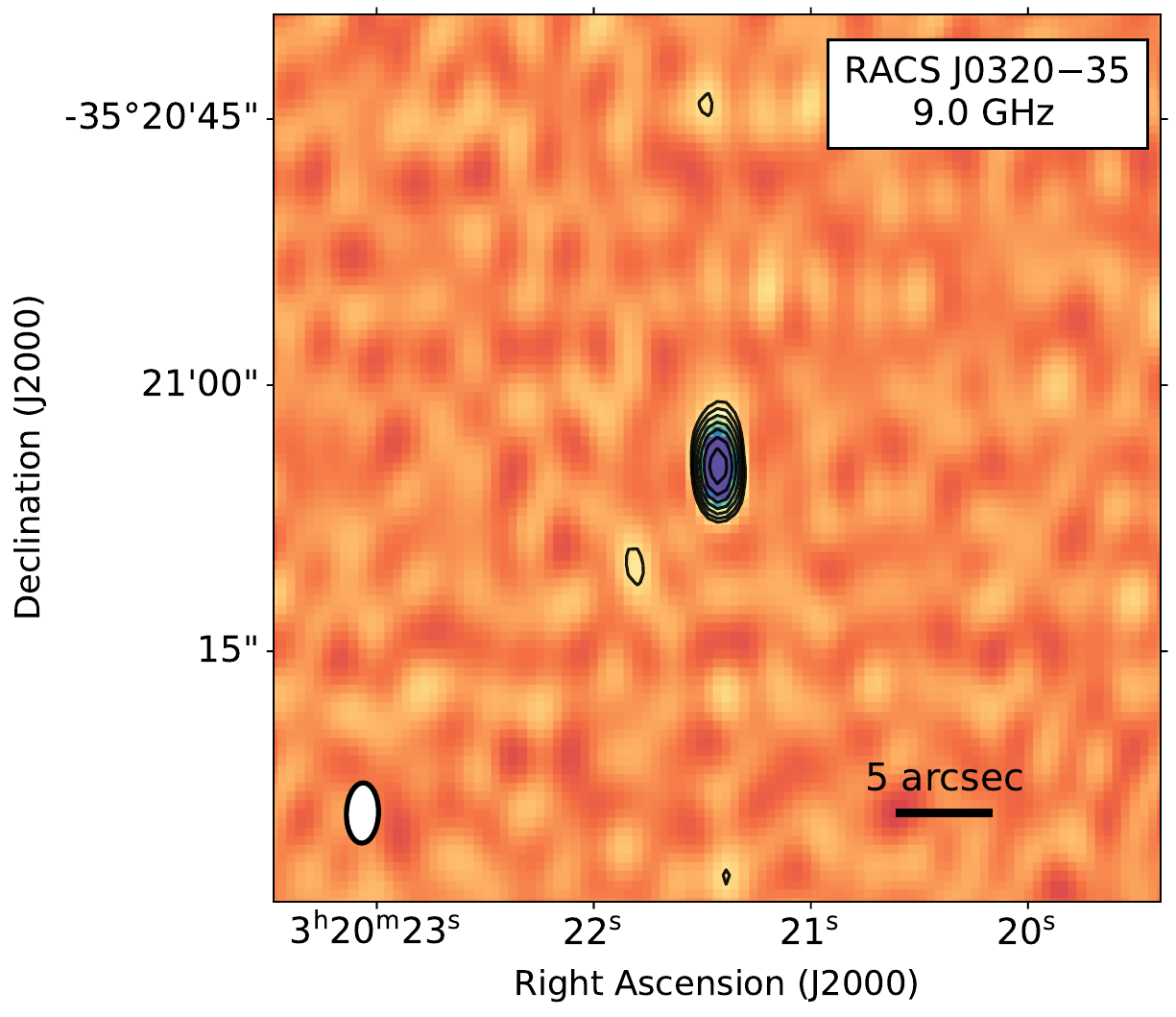}\\
 \caption{MWA (200~MHz), uGMRT (400 and 650~MHz) and ATCA (2.1, 5.5 and 9~GHz) images centred on the optical position of RACS~J0320$-$35. Contours start at $\pm$3$\times$RMS and increase by factors of $\sqrt{2}$.}
    \label{fig:radio_images}
\end{figure*}

\begin{figure*}
\centering
	\includegraphics[width=0.6\hsize]{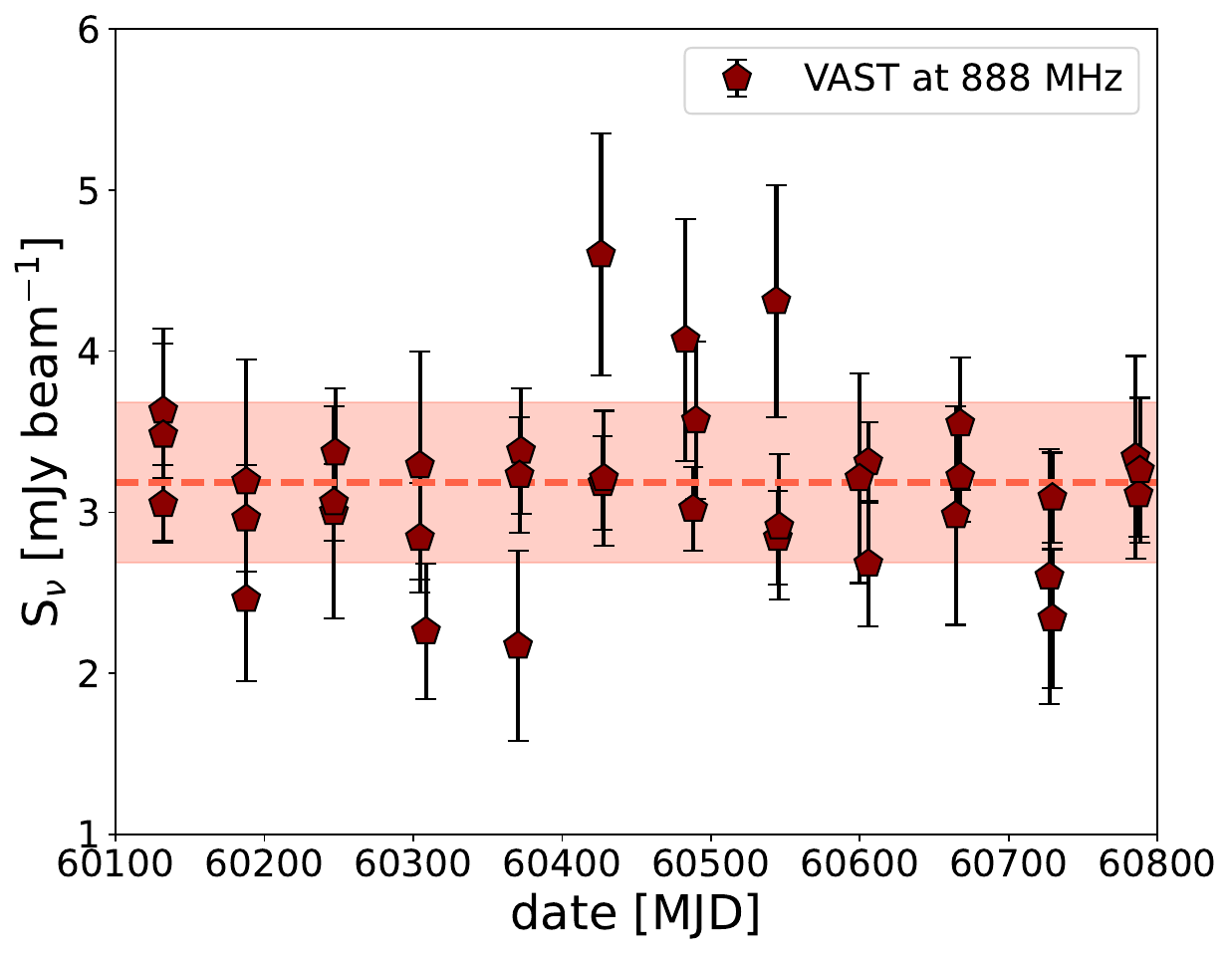}
 \caption{Light curve at 888~MHz, observed frame, of \shortname\ based on the data from the VAST survey. Measurements span a range of $\sim$2 years in the observed frame, form July 2023 to April 2025. The dashed line is the weighted average, while the shaded area represents the 1$\sigma$ standard deviation (=0.5~mJy~beam$^{-1}$), comparable to the median uncertainty of the measurements (=0.4~mJy~beam$^{-1}$).}
    \label{fig:VAST_light_curve}
\end{figure*}

Dedicated ATCA observations of \shortname\ were performed on 2022 September 03 with the most extended, 6D configuration under the project C3477 (PI:Ighina). However, antenna CA06 was not available, reducing the maximum baseline to $\sim$2500~m. Observations were carried out at 2.1, 5.5 and 9~GHz using the Compact Array Broadband Bandwidth (CABB; \citealt{Wilson2011}), with a nominal bandwidth of 2048~MHz (divided into channels of 1~MHz). We used the source PKS~B1934$-$638 as the standard primary calibrator \citep{Reynolds1994}, observed at the beginning of the session. To calibrate the phases throughout the session we periodically observed 0402$-$362. The final on-source time (after flagging) for RACS~J0320$-$35 is 1.7, 2.4 and 2.2~h at 2.1, 5.5 and 9~GHz respectively. To process the data (calibration and imaging), we used the \texttt{MIRIAD} data-reduction package \citep{Sault1995} following a standard reduction together with two cycles of phase self-calibration. For imaging, we adopted a standard robust parameter of 0.5 and the results are shown in Fig. \ref{fig:radio_images}. Finally, we performed a 2D Gaussian fit on the target using the CASA software. The source is clearly detected and appears point-like in all the observations. We report the results of the fit in Tab. \ref{tab:radio_fit}. When fitting the radio spectrum we also considered a conservative 10\% (added in quadrature) in the errors to account for the uncertainty related to the calibration process.

\subsection{VLBI-LBA observations}
VLBI observations of RACS~J0320$-$35 were performed with the Australian The Long Baseline Array (LBA; project v619, PI:Ighina) on 2022 October 
21 (at 2.3~GHz) and 23 (at 8.4~GHz). Each observing run was 12~h and included another target (as well as their calibrators and slewing time) plus 1~h of setup and fringe finding. During the observations we used 0332-375 for the phase referencing. For both sessions the ATCA, Mopra, Parkes, Ceduna, Katherine, Yarragadee and Warkworth antennae were used. During the 8.4~GHz session, the Hobart antenna was also added to the array. However, all the Yarragadee and the Warkworth measurements were flagged at 2.3~GHz, whereas all of the Katherine measurements were flagged for the 8.4~GHz observations. The final longest baselines were: 2611 and 5362~km at 2.3 and 8.4~GHz respectively.

\begin{figure*}
\centering
 	\includegraphics[width=0.4\hsize]{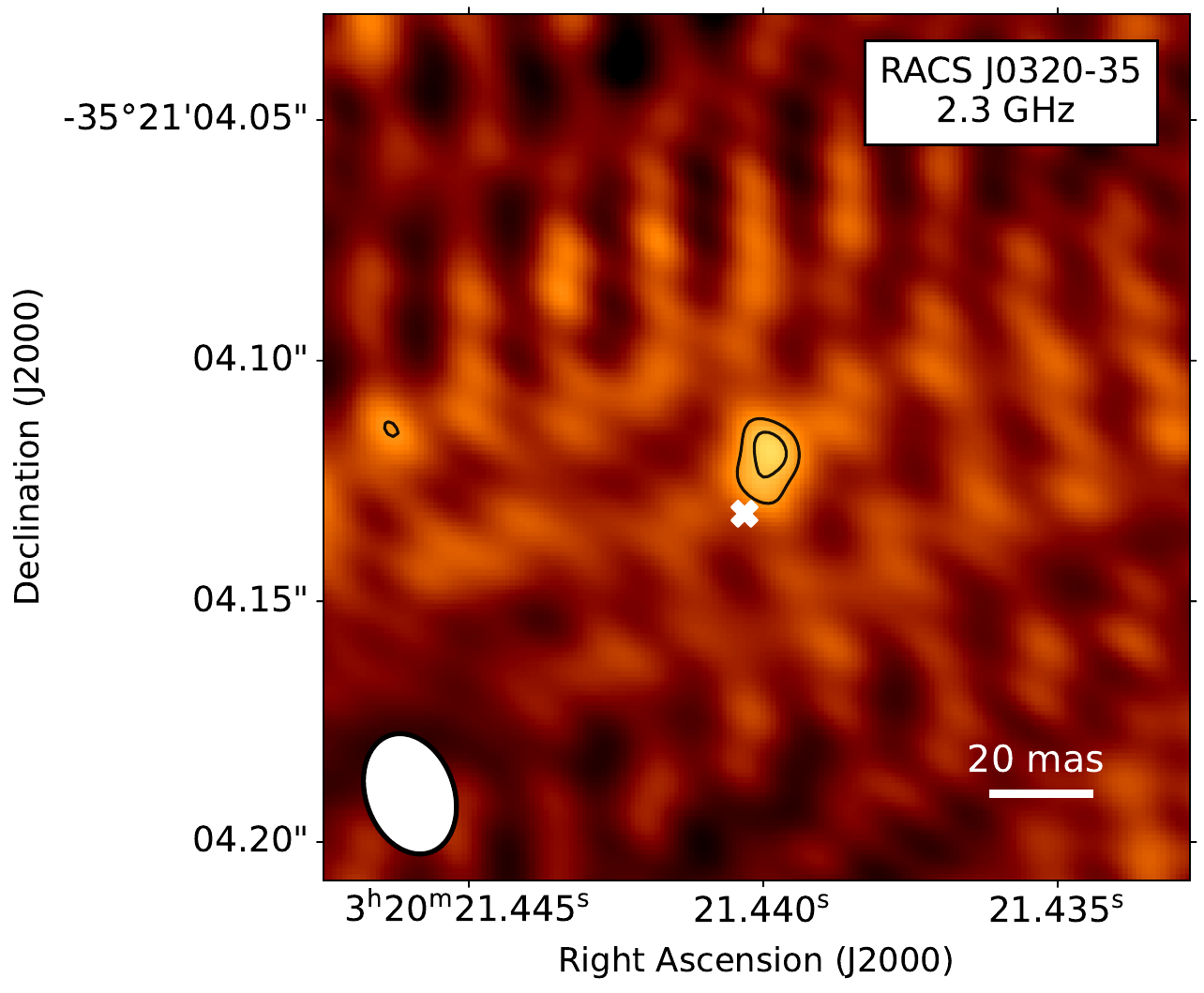}
	\includegraphics[width=0.418\hsize]{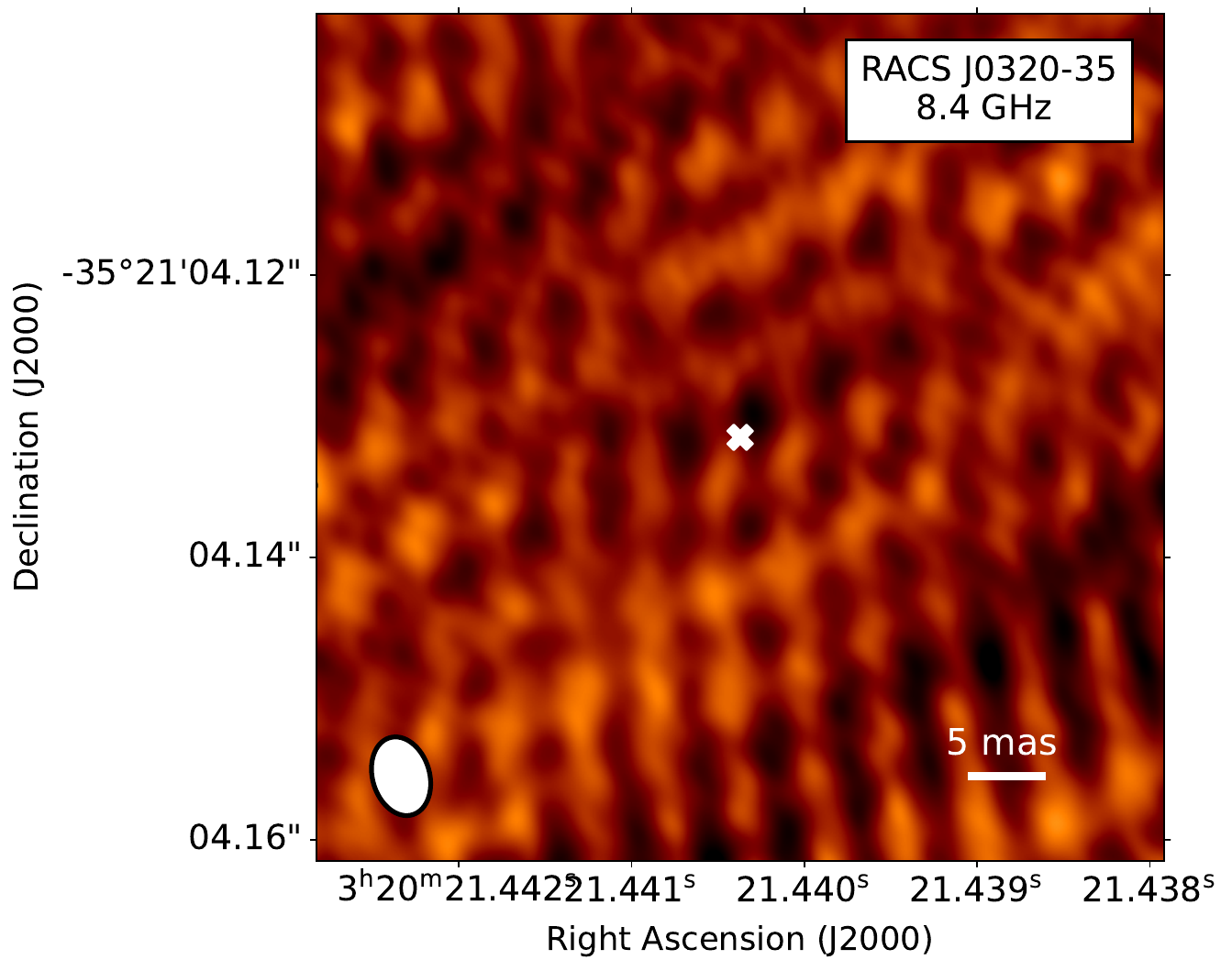}\\
 \caption{LBA-VLBI images of RACS~J0320$-$35 centred at 2.3 (left) and 8.4~GHz (right). The white cross shows the optical position of RACS~J0320$-$35 from the DES catalogue. Contours start at $\pm$3$\times$RMS and increase by factors of $\sqrt{2}$. At these scales, most of the radio emission observed in the ATCA images is resolved out.}
    \label{fig:LBA_images}
\end{figure*}

Data were processed using the NRAO’s Astronomical Imaging Processing System (\texttt{AIPS}; \citealt{Wells1985}), where the calibration and flagging followed the general procedure outlined in the AIPS cookbook. The accuracy of the flux scale is $\sim$20\%. We considered this uncertainty, added in quadrature, when computing the errors on the fluxes measured from these images. 
For imaging, we adopted a weighting robust parameter equal to 2 in order to enhance the sensitivity at the expense of the angular resolution. The restored images obtained after cleaning (using the \texttt{imagr} task in \texttt{AIPS}) are reported in Fig. \ref{fig:LBA_images}. 
The final images reach an RMS of $\sim95$~\textmu Jy~beam$^{-1}$ and  $\sim55$~\textmu Jy~beam$^{-1}$ at 2.3 and 8.4~GHz, respectively. 

At 2.3~GHz, a 5$\sigma$ radio signal is present $\sim16$~mas away from the optical position of the quasar reported in the DES catalogue (white cross in Fig. \ref{fig:LBA_images}). Even though the S/N of this radio emission alone is relatively low (S/N$\sim$5), the close position to the expected core is an indication that this emission is true and likely produced by the inner-most regions of the jets, close to the accretion disc. Therefore, we considered its peak flux density (S$_{\rm2.3GHz}^{\rm peak}=450\pm130$~\textmu Jy~beam$^{-1}$) as an estimate of the radio emission produced within the core of RACS~J0320$-$35. At 8.4~GHz, there is no significant radio signal at the target position, nor within $\sim$200~mas radius around it. Therefore, using a 3$\sigma$ upper limit at 8.4~GHz (S$_{\rm8.4GHz}<170$~\textmu Jy~beam$^{-1}$), we can conclude that the radio core of the source RACS~J0320$-$35 has a spectral index $\alpha_{\rm r}^{\rm core}\gtrsim0.5$.

\begin{table*}
	\centering
\caption{Best-fit values obtained from a 2D Gaussian fit to the uGMRT and ATCA images of RACS~J0320$-$35.}
\begin{tabular}{ccccccccc}
\hline
\hline
Array & Frequency & Int. flux & Peak surf. Brightness & Beam sizes & P.A. & Off-source RMS & Obs. Date\Tstrut\\
& (GHz) & (mJy) & (mJy~beam$^{-1}$) &  maj$\times$min & (deg) & (\textmu Jy~beam$^{-1}$) & Y--M--D\Bstrut\\
\hline
uGMRT & 0.40 & 5.58$\pm$0.88    & 5.17$\pm$0.39  & 18.9$''\times$7.5$''$  &$-$100.3  & 370 & 2022--04--23\Tstrut\\
uGMRT & 0.65 & 3.97$\pm$0.17    & 3.76$\pm$0.08  &  10.3$''\times$4.5$''$ &$-$61.4   & 80  & 2022--04--23\\
ATCA  & 2.1   & 2.09$\pm$0.09    & 1.91$\pm$0.05  & 14.5$''\times$6.7$''$  & 0.1      & 45  & 2022--09--03\\
ATCA & 5.5    & 0.84$\pm$0.03    & 0.87$\pm$0.02  & 5.5$''\times$3.1$''$   & 0.5      & 19  & 2022--09--03\\
ATCA & 9.0    & 0.63$\pm$0.03    & 0.63$\pm$0.02  & 3.4$''\times$1.8$''$   & $-$1.9   & 19  & 2022--09--03\Bstrut\\

\hline

LBA  & 2.3   & -- & 0.45$\pm$0.13  & 25.7mas$\times$18.3mas & 11.6 & 95 & 2022--10--21 \Tstrut\\
LBA  & 8.4   & -- & $<0.17$        & 5.7mas$\times$4.0mas   & 16.6 & 55 & 2022--10--23 \Bstrut\\
\hline

\end{tabular}
\tablecomments{Col. (1) Array used for a given observation; col. (2) central frequency of the observation; col. (3) integrated flux density; col. (4) peak surface brightness; col. (5) size of the synthesised beam in arcseconds. The LBA beams are in units of milliarcseconds; col. (6) position angle (P.A.), east of north, of the synthesised beam; col. (7) RMS of the image nearby the source; col. (8) date observations were performed.}
\label{tab:radio_fit}

\end{table*}

\section{High-energy observations and analysis}
\label{sec:variab_X}

\subsection{X-ray observations with Chandra}
\label{sec:chandra}

\begin{figure}
\centering
 	\includegraphics[width=0.478\hsize]{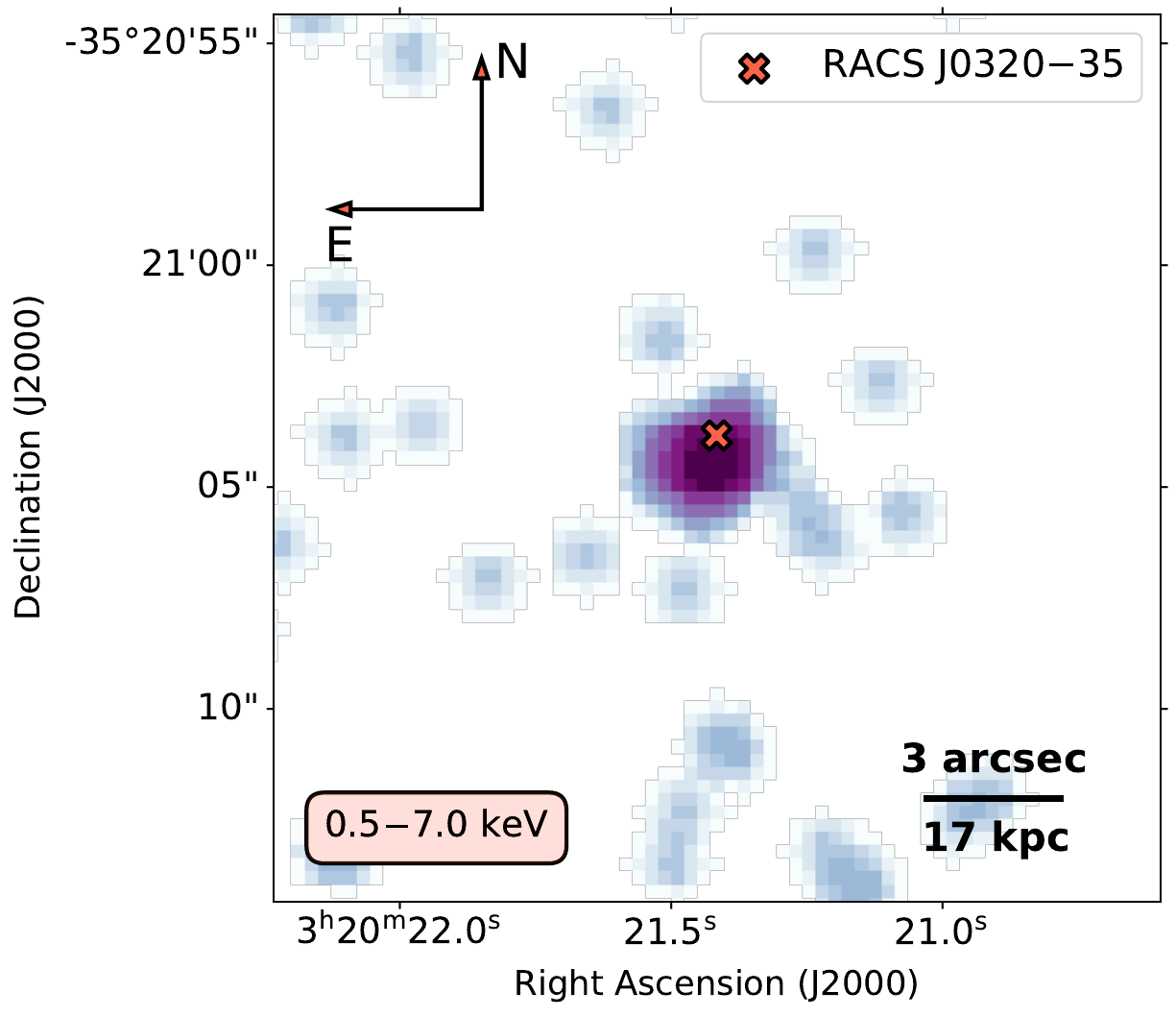}
 	\includegraphics[width=0.512\hsize]{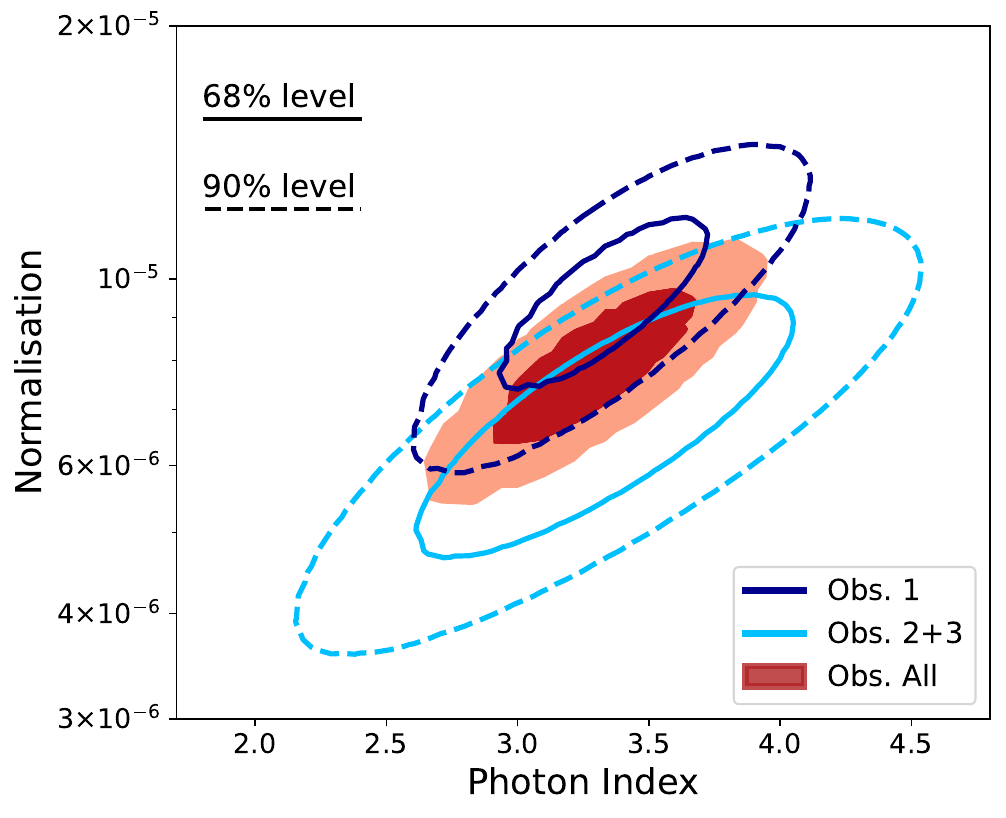}
 \caption{{\bf Left:} {\it Chandra} images (20$''\times$20$''$) of RACS~J0320$-$35 in the energy band 0.5--7~keV. The red cross indicates the optical position of the quasar, consistent with the X-ray source detected in the {\it Chandra} image. {\bf Rgiht:} Contour levels of the photon index and normalisation parameter derived from the fit of the {\it Chandra} observations. Different colours indicate different observation segments, with the filled ellipses showing the contours from all the observations combined. Dashed (solid) lines indicate the 90\% (68\%) confidence regions.}
    \label{fig:X_cont}
\end{figure}

In this section we describe the {\it Chandra} X-ray observations of \shortname. The overall exposure time, 60~ksec, was split into two groupings, with one observation (27112; 29.68~ks) conducted in 2023 July and two observations (26709 and 29162; 15.87 and 14.74~ks, respectively) in 2023 December; all observations are contained in the \textit{Chandra} Data Collection (CDC) 318~\href{https://doi.org/10.25574/cdc.318}{doi:10.25574/cdc.318}. Events were recorded in the Very Faint telemetry format and Timed Exposure mode, with \shortname\ positioned on the back-illuminated S3 chip. 
Data reduction was performed using the software \texttt{CIAO} (v. 4.16; \citealt{Fruscinone2006}) with CALDB (v4.11.2). We show in Fig. \ref{fig:X_cont}, left panel, the 0.5--7.0~keV image obtained from the combination of the different exposures. A relatively strong X-ray source is detected at the optical position of the quasar.

We used \texttt{specextract} to extract events from a 2\arcsec \, source region and from a background annulus of 10\textendash $30^{\prime\prime}$, both  centred on the optical/NIR position of the quasar. 
The target is detected with 52 net counts over a background of $\lesssim$1 photon in the 0.5--7.0~keV energy band.
We then analysed the extracted spectra using \texttt{XSPEC} v 12.11.1 \citep{Arnaud1996} and performed a fit minimizing the modified C-statistic \citep{Cash1979,Wachter1979}. We binned the spectrum to one net count per energy bin and we limited the fit to the energy range of 0.5--7.0~keV. 
We considered two models: a simple power law absorbed by the Galactic column density along the line of sight (\texttt{tbabs}$\times$\texttt{pow}) and a power low with an exponential cutoff at higher energies and Galactic absorption (\texttt{tbabs}$\times$\texttt{zcutoffpl}). In both cases we fixed the column density to N$_{\rm H} =2.98\times 10^{20}$~cm$^{-2}$ \citep{HI4PI2016}. For the power law with an exponential cutoff model we fixed the value of the power law photon index to $\Gamma_{\rm X}=1.9$ and 2.2 \citep[e.g.][]{Tortosa2024}, since the fit would not converge with three free parameters. We report in Table \ref{tab:Gem_x-ray_values} the best-fit parameters derived for each model.


In order to check for potential variability between the two observing epochs, we analysed the 27112 and 26709+29162 (taken less than a day apart) observations separately. For this analysis we adopted a single power law fit.
We show in Fig. \ref{fig:X_cont}, right panel, the contours of the best fits of these observations. Contours represent the 68\% and 90\% confidence limits on the modified C Statistic. Within uncertainties, both epochs are consistent with each other, indicating that there was no significant variation and that we can jointly fit the entire 60~ksec.

\subsection{$\gamma$-ray upper-limits from FERMI-LAT}
\label{sec:fermi}

Since BL Lac, and blazars in general, are the main population of extragalactic gamma-ray emitters \citep[e.g.][]{Konigl1981}, we also considered the data available from the Fermi-LAT in the MeV-GeV energy regime.
For the analysis, we followed the same setup described in \citep{Arsioli2025}, integrating over 17 years of observations with LAT, but considering a broader energy range, between 600 MeV to 800 GeV.
No significant $\gamma$-ray emission was found, that is, the likelihood analysis resulted in test statistics approximately zero, setting an upper limit flux of $\sim$2e-13~erg~cm$^{-2}$~s$^{-1}$ at 1~GeV\footnote{Considering Fermi-LAT sensitivity at high galactic latitude ($|$b$|>45^\circ$) and at the energy of 1~GeV; as described in \url{https://www.slac.stanford.edu/exp/glast/groups/canda/lat_Performance.htm}.}.  
A complementary light-curve search with time-bins of 100~days, aimed at unveiling flares that could be diluted in the long 17 years exposure, returned no significant excess. The most pronounced interval reaches only TS$\sim$6, well below the 3$\sigma$ detection threshold (TS$\geq$12) for this type of analysis.
We therefore found no evidence, either persistent or transient, for $\gamma$-ray emission in \shortname.


\begin{table}
	\centering
	\caption{Best-fit values obtained from the fit of the X-ray spectrum of RACS~J0320$-$35.}
	\label{tab:Gem_x-ray_values}
\centering
\begin{tabular}{cccccccc}
\\
\hline
\hline
Model & $\Gamma_{\rm X}$    & E$_{\rm cut}$ &  f$_\mathrm{0.5-7.0keV}$ &  
L$_\mathrm{2-10keV}$   &   ${\alpha}_{\rm ox}$   & cstat/d.o.f.\Tstrut\BBstrut\\
\hline
PL & 3.3$\pm0.4$  & -- &	2.3$_{-0.3}^{+0.5}$  &   1.8$_{-0.7}^{+1.1}$  & 0.97${\pm0.05}$  & 35 / 43 \\
PL+CUTOFF & 1.9$^*$ & 10.1$_{-2.6}^{+4.8}$ & 	1.8$\pm0.1$  &   0.9$\pm0.2$  & 1.19$_{-0.05}^{+0.06}$  & 36 / 43 \\
PL+CUTOFF & 2.2$^*$ & 13.0$_{-4.0}^{+8.9}$ & 	1.9$^{+0.1}_{-0.2}$  &   1.0$^{+0.3}_{-0.2}$  & 1.14$_{-0.05}^{+0.06}$  &  36 / 43\BBstrut\Tstrut\\
\hline
\hline
\end{tabular}

\tablecomments{During the fit we assumed a simple power law with Galactic absorption and the errors are reported at a 68 percent confidence level. Col. (1) photon index of the power law. `*' indicates a fixed value; col. (2) energy of the exponential cutoff in keV; (3) un-absorbed flux in the energy band 0.5--7~keV in units of 10$^{-14}$ erg sec$^{-1}$ cm$^{-2}$; col. (4) rest-frame luminosity in the energy range 2--10~keV in units of 10$^{46}$ erg sec$^{-1}$; col. (5) ${\alpha}_{\rm ox}$ parameter; col. (6) C-statistic and degrees of freedom of the fit.}

\end{table}


\bibliography{referenze}{}

\begin{thebibliography}{}
\expandafter\ifx\csname natexlab\endcsname\relax\def\natexlab#1{#1}\fi
\providecommand{\url}[1]{\href{#1}{#1}}
\providecommand{\dodoi}[1]{doi:~\href{http://doi.org/#1}{\nolinkurl{#1}}}
\providecommand{\doeprint}[1]{\href{http://ascl.net/#1}{\nolinkurl{http://ascl.net/#1}}}
\providecommand{\doarXiv}[1]{\href{https://arxiv.org/abs/#1}{\nolinkurl{https://arxiv.org/abs/#1}}}

\bibitem[{T.~M.~C. {Abbott} {et~al.}(2021){Abbott}, {Adam{\'o}w}, {Aguena}, {Allam}, {Amon}, {Annis}, {Avila}, {Bacon}, {Banerji}, {Bechtol}, {Becker}, {Bernstein}, {Bertin}, {Bhargava}, {Bridle}, {Brooks}, {Burke}, {Carnero Rosell}, {Carrasco Kind}, {Carretero}, {Castander}, {Cawthon}, {Chang}, {Choi}, {Conselice}, {Costanzi}, {Crocce}, {da Costa}, {Davis}, {De Vicente}, {DeRose}, {Desai}, {Diehl}, {Dietrich}, {Drlica-Wagner}, {Eckert}, {Elvin-Poole}, {Everett}, {Evrard}, {Ferrero}, {Fert{\'e}}, {Flaugher}, {Fosalba}, {Friedel}, {Frieman}, {Garc{\'\i}a-Bellido}, {Gaztanaga}, {Gelman}, {Gerdes}, {Giannantonio}, {Gill}, {Gruen}, {Gruendl}, {Gschwend}, {Gutierrez}, {Hartley}, {Hinton}, {Hollowood}, {Honscheid}, {Huterer}, {James}, {Jeltema}, {Johnson}, {Kent}, {Kron}, {Kuehn}, {Kuropatkin}, {Lahav}, {Li}, {Lidman}, {Lin}, {MacCrann}, {Maia}, {Manning}, {Maloney}, {March}, {Marshall}, {Martini}, {Melchior}, {Menanteau}, {Miquel}, {Morgan}, {Myles}, {Neilsen}, {Ogando}, {Palmese}, {Paz-Chinch{\'o}n}, {Petravick},
  {Pieres}, {Plazas}, {Pond}, {Rodriguez-Monroy}, {Romer}, {Roodman}, {Rykoff}, {Sako}, {Sanchez}, {Santiago}, {Scarpine}, {Serrano}, {Sevilla-Noarbe}, {Smith}, {Smith}, {Soares-Santos}, {Suchyta}, {Swanson}, {Tarle}, {Thomas}, {To}, {Tremblay}, {Troxel}, {Tucker}, {Turner}, {Varga}, {Walker}, {Wechsler}, {Weller}, {Wester}, {Wilkinson}, {Yanny}, {Zhang}, {Nikutta}, {Fitzpatrick}, {Jacques}, {Scott}, {Olsen}, {Huang}, {Herrera}, {Juneau}, {Nidever}, {Weaver}, {Adean}, {Correia}, {de Freitas}, {Freitas}, {Singulani}, {Vila-Verde}, \& {Linea Science Server}}]{Abbott2021}
{Abbott}, T.~M.~C., {Adam{\'o}w}, M., {Aguena}, M., {et~al.} 2021, \bibinfo{title}{{The Dark Energy Survey Data Release 2},} \apjs, 255, 20, \dodoi{10.3847/1538-4365/ac00b3}

\bibitem[{R. {Abuter} {et~al.}(2024){Abuter}, {Allouche}, {Amorim}, {Bailet}, {Berdeu}, {Berger}, {Berio}, {Bigioli}, {Boebion}, {Bolzer}, {Bonnet}, {Bourdarot}, {Bourget}, {Brandner}, {Cao}, {Conzelmann}, {Comin}, {Cl{\'e}net}, {Courtney-Barrer}, {Davies}, {Defr{\`e}re}, {Delboulb{\'e}}, {Delplancke-Str{\"o}bele}, {Dembet}, {Dexter}, {de Zeeuw}, {Drescher}, {Eckart}, {{\'E}douard}, {Eisenhauer}, {Fabricius}, {Feuchtgruber}, {Finger}, {F{\"o}rster Schreiber}, {Garcia}, {Garcia Lopez}, {Gao}, {Gendron}, {Genzel}, {Gil}, {Gillessen}, {Gomes}, {Gont{\'e}}, {Gouvret}, {Guajardo}, {Guieu}, {Hackenberg}, {Haddad}, {Hartl}, {Haubois}, {Hau{\ss}mann}, {Hei{\ss}el}, {Henning}, {Hippler}, {H{\"o}nig}, {Horrobin}, {Hubin}, {Jacqmart}, {Jocou}, {Kaufer}, {Kervella}, {Kolb}, {Korhonen}, {Lacour}, {Lagarde}, {Lai}, {Lapeyr{\`e}re}, {Laugier}, {Le Bouquin}, {Leftley}, {L{\'e}na}, {Lewis}, {Liu}, {Lopez}, {Lutz}, {Magnard}, {Mang}, {Marcotto}, {Maurel}, {M{\'e}rand}, {Millour}, {More}, {Netzer}, {Nowacki}, {Nowak}, {Oberti},
  {Ott}, {Pallanca}, {Paumard}, {Perraut}, {Perrin}, {Petrov}, {Pfuhl}, {Pourr{\'e}}, {Rabien}, {Rau}, {Riquelme}, {Robbe-Dubois}, {Rochat}, {Salman}, {Sanchez-Bermudez}, {Santos}, {Scheithauer}, {Sch{\"o}ller}, {Schubert}, {Schuhler}, {Shangguan}, {Shchekaturov}, {Shimizu}, {Sevin}, {Soulez}, {Spang}, {Stadler}, {Sternberg}, {Straubmeier}, {Sturm}, {Sykes}, {Tacconi}, {Tristram}, {Vincent}, {von Fellenberg}, {Uysal}, {Widmann}, {Wieprecht}, {Wiezorrek}, {Woillez}, \& {Zins}}]{Abuter2024}
{Abuter}, R., {Allouche}, F., {Amorim}, A., {et~al.} 2024, \bibinfo{title}{{A dynamical measure of the black hole mass in a quasar 11 billion years ago},} \nat, 627, 281, \dodoi{10.1038/s41586-024-07053-4}

\bibitem[{Y. {Ai} {et~al.}(2017){Ai}, {Fabian}, {Fan}, {Walker}, {Ghisellini}, {Sbarrato}, {Dou}, {Wang}, {Wu}, \& {Feng}}]{Ai2017}
{Ai}, Y., {Fabian}, A.~C., {Fan}, X., {et~al.} 2017, \bibinfo{title}{{XMM-Newton observation of the ultraluminous quasar SDSS J010013.02+280225.8 at redshift 6.326},} \mnras, 470, 1587, \dodoi{10.1093/mnras/stx1231}

\bibitem[{T. {An} \& W.~A. {Baan}(2012){An} \& {Baan}}]{An2012}
{An}, T., \& {Baan}, W.~A. 2012, \bibinfo{title}{{The Dynamic Evolution of Young Extragalactic Radio Sources},} \apj, 760, 77, \dodoi{10.1088/0004-637X/760/1/77}

\bibitem[{K.~A. {Arnaud}(1996){Arnaud}}]{Arnaud1996}
{Arnaud}, K.~A. 1996, \bibinfo{title}{{XSPEC: The First Ten Years},} 101, 17

\bibitem[{B. {Arsioli} {et~al.}(2025){Arsioli}, {Chang}, \& {Ighina}}]{Arsioli2025}
{Arsioli}, B., {Chang}, Y.-L., \& {Ighina}, L. 2025, \bibinfo{title}{{Mapping the cosmic gamma-ray horizon: the 1CGH catalogue of Fermi-LAT detections above 10 GeV},} \mnras, 539, 1458, \dodoi{10.1093/mnras/staf329}

\bibitem[{ {Astropy Collaboration} {et~al.}(2013){Astropy Collaboration}, {Robitaille}, {Tollerud}, {Greenfield}, {Droettboom}, {Bray}, {Aldcroft}, {Davis}, {Ginsburg}, {Price-Whelan}, {Kerzendorf}, {Conley}, {Crighton}, {Barbary}, {Muna}, {Ferguson}, {Grollier}, {Parikh}, {Nair}, {Unther}, {Deil}, {Woillez}, {Conseil}, {Kramer}, {Turner}, {Singer}, {Fox}, {Weaver}, {Zabalza}, {Edwards}, {Azalee Bostroem}, {Burke}, {Casey}, {Crawford}, {Dencheva}, {Ely}, {Jenness}, {Labrie}, {Lim}, {Pierfederici}, {Pontzen}, {Ptak}, {Refsdal}, {Servillat}, \& {Streicher}}]{astropy2013}
{Astropy Collaboration}, {Robitaille}, T.~P., {Tollerud}, E.~J., {et~al.} 2013, \bibinfo{title}{{Astropy: A community Python package for astronomy},} \aap, 558, A33, \dodoi{10.1051/0004-6361/201322068}

\bibitem[{ {Astropy Collaboration} {et~al.}(2018){Astropy Collaboration}, {Price-Whelan}, {Sip{\H{o}}cz}, {G{\"u}nther}, {Lim}, {Crawford}, {Conseil}, {Shupe}, {Craig}, {Dencheva}, {Ginsburg}, {Vand erPlas}, {Bradley}, {P{\'e}rez-Su{\'a}rez}, {de Val-Borro}, {Aldcroft}, {Cruz}, {Robitaille}, {Tollerud}, {Ardelean}, {Babej}, {Bach}, {Bachetti}, {Bakanov}, {Bamford}, {Barentsen}, {Barmby}, {Baumbach}, {Berry}, {Biscani}, {Boquien}, {Bostroem}, {Bouma}, {Brammer}, {Bray}, {Breytenbach}, {Buddelmeijer}, {Burke}, {Calderone}, {Cano Rodr{\'\i}guez}, {Cara}, {Cardoso}, {Cheedella}, {Copin}, {Corrales}, {Crichton}, {D'Avella}, {Deil}, {Depagne}, {Dietrich}, {Donath}, {Droettboom}, {Earl}, {Erben}, {Fabbro}, {Ferreira}, {Finethy}, {Fox}, {Garrison}, {Gibbons}, {Goldstein}, {Gommers}, {Greco}, {Greenfield}, {Groener}, {Grollier}, {Hagen}, {Hirst}, {Homeier}, {Horton}, {Hosseinzadeh}, {Hu}, {Hunkeler}, {Ivezi{\'c}}, {Jain}, {Jenness}, {Kanarek}, {Kendrew}, {Kern}, {Kerzendorf}, {Khvalko}, {King}, {Kirkby}, {Kulkarni},
  {Kumar}, {Lee}, {Lenz}, {Littlefair}, {Ma}, {Macleod}, {Mastropietro}, {McCully}, {Montagnac}, {Morris}, {Mueller}, {Mumford}, {Muna}, {Murphy}, {Nelson}, {Nguyen}, {Ninan}, {N{\"o}the}, {Ogaz}, {Oh}, {Parejko}, {Parley}, {Pascual}, {Patil}, {Patil}, {Plunkett}, {Prochaska}, {Rastogi}, {Reddy Janga}, {Sabater}, {Sakurikar}, {Seifert}, {Sherbert}, {Sherwood-Taylor}, {Shih}, {Sick}, {Silbiger}, {Singanamalla}, {Singer}, {Sladen}, {Sooley}, {Sornarajah}, {Streicher}, {Teuben}, {Thomas}, {Tremblay}, {Turner}, {Terr{\'o}n}, {van Kerkwijk}, {de la Vega}, {Watkins}, {Weaver}, {Whitmore}, {Woillez}, {Zabalza}, \& {Astropy Contributors}}]{astropy:2018}
{Astropy Collaboration}, {Price-Whelan}, A.~M., {Sip{\H{o}}cz}, B.~M., {et~al.} 2018, \bibinfo{title}{{The Astropy Project: Building an Open-science Project and Status of the v2.0 Core Package},} \aj, 156, 123, \dodoi{10.3847/1538-3881/aabc4f}

\bibitem[{ {Astropy Collaboration} {et~al.}(2022){Astropy Collaboration}, {Price-Whelan}, {Lim}, {Earl}, {Starkman}, {Bradley}, {Shupe}, {Patil}, {Corrales}, {Brasseur}, {N{"o}the}, {Donath}, {Tollerud}, {Morris}, {Ginsburg}, {Vaher}, {Weaver}, {Tocknell}, {Jamieson}, {van Kerkwijk}, {Robitaille}, {Merry}, {Bachetti}, {G{"u}nther}, {Aldcroft}, {Alvarado-Montes}, {Archibald}, {B{'o}di}, {Bapat}, {Barentsen}, {Baz{'a}n}, {Biswas}, {Boquien}, {Burke}, {Cara}, {Cara}, {Conroy}, {Conseil}, {Craig}, {Cross}, {Cruz}, {D'Eugenio}, {Dencheva}, {Devillepoix}, {Dietrich}, {Eigenbrot}, {Erben}, {Ferreira}, {Foreman-Mackey}, {Fox}, {Freij}, {Garg}, {Geda}, {Glattly}, {Gondhalekar}, {Gordon}, {Grant}, {Greenfield}, {Groener}, {Guest}, {Gurovich}, {Handberg}, {Hart}, {Hatfield-Dodds}, {Homeier}, {Hosseinzadeh}, {Jenness}, {Jones}, {Joseph}, {Kalmbach}, {Karamehmetoglu}, {Ka{l}uszy{'n}ski}, {Kelley}, {Kern}, {Kerzendorf}, {Koch}, {Kulumani}, {Lee}, {Ly}, {Ma}, {MacBride}, {Maljaars}, {Muna}, {Murphy}, {Norman}, {O'Steen},
  {Oman}, {Pacifici}, {Pascual}, {Pascual-Granado}, {Patil}, {Perren}, {Pickering}, {Rastogi}, {Roulston}, {Ryan}, {Rykoff}, {Sabater}, {Sakurikar}, {Salgado}, {Sanghi}, {Saunders}, {Savchenko}, {Schwardt}, {Seifert-Eckert}, {Shih}, {Jain}, {Shukla}, {Sick}, {Simpson}, {Singanamalla}, {Singer}, {Singhal}, {Sinha}, {Sip{H{o}}cz}, {Spitler}, {Stansby}, {Streicher}, {{{S}}umak}, {Swinbank}, {Taranu}, {Tewary}, {Tremblay}, {Val-Borro}, {Van Kooten}, {Vasovi{'c}}, {Verma}, {de Miranda Cardoso}, {Williams}, {Wilson}, {Winkel}, {Wood-Vasey}, {Xue}, {Yoachim}, {Zhang}, {Zonca}, \& {Astropy Project Contributors}}]{astropy:2022}
{Astropy Collaboration}, {Price-Whelan}, A.~M., {Lim}, P.~L., {et~al.} 2022, \bibinfo{title}{{The Astropy Project: Sustaining and Growing a Community-oriented Open-source Project and the Latest Major Release (v5.0) of the Core Package},} \apj, 935, 167, \dodoi{10.3847/1538-4357/ac7c74}

\bibitem[{E. {Ba{\~n}ados} {et~al.}(2018){Ba{\~n}ados}, {Connor}, {Stern}, {Mulchaey}, {Fan}, {Decarli}, {Farina}, {Mazzucchelli}, {Venemans}, {Walter}, {Wang}, \& {Yang}}]{Banados2018c}
{Ba{\~n}ados}, E., {Connor}, T., {Stern}, D., {et~al.} 2018, \bibinfo{title}{{Chandra X-Rays from the Redshift 7.54 Quasar ULAS J1342+0928},} \apjl, 856, L25, \dodoi{10.3847/2041-8213/aab61e}

\bibitem[{E. {Ba{\~n}ados} {et~al.}(2024){Ba{\~n}ados}, {Momjian}, {Connor}, {Belladitta}, {Decarli}, {Mazzucchelli}, {Venemans}, {Walter}, {Wang}, {Xie}, {Barth}, {Eilers}, {Fan}, {Khusanova}, {Schindler}, {Stern}, {Yang}, {Andika}, {Carilli}, {Farina}, {Fabian}, {Hennawi}, {Pensabene}, \& {Rojas-Ruiz}}]{Banados2024}
{Ba{\~n}ados}, E., {Momjian}, E., {Connor}, T., {et~al.} 2024, \bibinfo{title}{{A blazar in the epoch of reionization},} Nature Astronomy, \dodoi{10.1038/s41550-024-02431-4}

\bibitem[{S. {Belladitta} {et~al.}(2020){Belladitta}, {Moretti}, {Caccianiga}, {Spingola}, {Severgnini}, {Della Ceca}, {Ghisellini}, {Dallacasa}, {Sbarrato}, {Cicone}, {Cassar{\`a}}, \& {Pedani}}]{Belladitta2020}
{Belladitta}, S., {Moretti}, A., {Caccianiga}, A., {et~al.} 2020, \bibinfo{title}{{The first blazar observed at z \&gt; 6},} \aap, 635, L7, \dodoi{10.1051/0004-6361/201937395}

\bibitem[{S. {Belladitta} {et~al.}(2025){Belladitta}, {Ba{\~n}ados}, {Xie}, {Decarli}, {Onorato}, {Yang}, {Bischetti}, {Onoue}, {Loiacono}, {Mart{\'\i}nez-Ram{\'\i}rez}, {Mazzucchelli}, {Davies}, {Wolf}, {Schindler}, {Fan}, {Wang}, {Walter}, {Mkrtchyan}, {Stern}, {Farina}, \& {Venemans}}]{Belladitta2025}
{Belladitta}, S., {Ba{\~n}ados}, E., {Xie}, Z.-L., {et~al.} 2025, \bibinfo{title}{{Discovery and characterization of 25 new quasars at 4.6 < z < 6.9 from wide-field multi-band surveys},} arXiv e-prints, arXiv:2505.15923, \dodoi{10.48550/arXiv.2505.15923}

\bibitem[{E. {Bertola} {et~al.}(2022){Bertola}, {Vignali}, {Lanzuisi}, {Dadina}, {Cappi}, {Gilli}, {Matzeu}, {Chartas}, {Piconcelli}, \& {Comastri}}]{Bertola2022}
{Bertola}, E., {Vignali}, C., {Lanzuisi}, G., {et~al.} 2022, \bibinfo{title}{{The properties of the X-ray corona in the distant (z = 3.91) quasar APM 08279+5255},} \aap, 662, A98, \dodoi{10.1051/0004-6361/202142642}

\bibitem[{A.~K. {Bhowmick} {et~al.}(2022){Bhowmick}, {Blecha}, {Ni}, {Di Matteo}, {Torrey}, {Kelley}, {Vogelsberger}, {Weinberger}, \& {Hernquist}}]{Bhowmick2022}
{Bhowmick}, A.~K., {Blecha}, L., {Ni}, Y., {et~al.} 2022, \bibinfo{title}{{Probing the z {\ensuremath{\gtrsim}} 6 quasars in a universe with IllustrisTNG physics: impact of gas-based black hole seeding models},} \mnras, 516, 138, \dodoi{10.1093/mnras/stac2238}

\bibitem[{{\'A}. {Bogd{\'a}n} {et~al.}(2024){Bogd{\'a}n}, {Goulding}, {Natarajan}, {Kov{\'a}cs}, {Tremblay}, {Chadayammuri}, {Volonteri}, {Kraft}, {Forman}, {Jones}, {Churazov}, \& {Zhuravleva}}]{Bogdan2024}
{Bogd{\'a}n}, {\'A}., {Goulding}, A.~D., {Natarajan}, P., {et~al.} 2024, \bibinfo{title}{{Evidence for heavy-seed origin of early supermassive black holes from a z {\ensuremath{\approx}} 10 X-ray quasar},} Nature Astronomy, 8, 126, \dodoi{10.1038/s41550-023-02111-9}

\bibitem[{M. {Brightman} {et~al.}(2013){Brightman}, {Silverman}, {Mainieri}, {Ueda}, {Schramm}, {Matsuoka}, {Nagao}, {Steinhardt}, {Kartaltepe}, {Sanders}, {Treister}, {Shemmer}, {Brandt}, {Brusa}, {Comastri}, {Ho}, {Lanzuisi}, {Lusso}, {Nandra}, {Salvato}, {Zamorani}, {Akiyama}, {Alexander}, {Bongiorno}, {Capak}, {Civano}, {Del Moro}, {Doi}, {Elvis}, {Hasinger}, {Laird}, {Masters}, {Mignoli}, {Ohta}, {Schawinski}, \& {Taniguchi}}]{Brightman2013}
{Brightman}, M., {Silverman}, J.~D., {Mainieri}, V., {et~al.} 2013, \bibinfo{title}{{A statistical relation between the X-ray spectral index and Eddington ratio of active galactic nuclei in deep surveys},} \mnras, 433, 2485, \dodoi{10.1093/mnras/stt920}

\bibitem[{G. {Calistro Rivera} {et~al.}(2017){Calistro Rivera}, {Williams}, {Hardcastle}, {Duncan}, {R{\"o}ttgering}, {Best}, {Br{\"u}ggen}, {Chy{\.z}y}, {Conselice}, {de Gasperin}, {Engels}, {G{\"u}rkan}, {Intema}, {Jarvis}, {Mahony}, {Miley}, {Morabito}, {Prandoni}, {Sabater}, {Smith}, {Tasse}, {van der Werf}, \& {White}}]{Calistro2017}
{Calistro Rivera}, G., {Williams}, W.~L., {Hardcastle}, M.~J., {et~al.} 2017, \bibinfo{title}{{The LOFAR window on star-forming galaxies and AGNs - curved radio SEDs and IR-radio correlation at 0<z<2.5},} \mnras, 469, 3468, \dodoi{10.1093/mnras/stx1040}

\bibitem[{W. {Cash}(1979){Cash}}]{Cash1979}
{Cash}, W. 1979, \bibinfo{title}{{Parameter estimation in astronomy through application of the likelihood ratio.},} \apj, 228, 939, \dodoi{10.1086/156922}

\bibitem[{T. {Connor} {et~al.}(2024){Connor}, {Ba{\~n}ados}, {Cappelluti}, \& {Foord}}]{Connor2024}
{Connor}, T., {Ba{\~n}ados}, E., {Cappelluti}, N., \& {Foord}, A. 2024, \bibinfo{title}{{Uncovering the First AGN Jets with AXIS},} Universe, 10, 227, \dodoi{10.3390/universe10050227}

\bibitem[{T. {Connor} {et~al.}(2021{\natexlab{a}}){Connor}, {Stern}, {Ba{\~n}ados}, \& {Mazzucchelli}}]{Connor2021b}
{Connor}, T., {Stern}, D., {Ba{\~n}ados}, E., \& {Mazzucchelli}, C. 2021{\natexlab{a}}, \bibinfo{title}{{X-Ray Evidence Against the Hypothesis that the Hyperluminous z = 6.3 Quasar J0100+2802 is Lensed},} \apjl, 922, L24, \dodoi{10.3847/2041-8213/ac37b5}

\bibitem[{T. {Connor} {et~al.}(2021{\natexlab{b}}){Connor}, {Ba{\~n}ados}, {Stern}, {Carilli}, {Fabian}, {Momjian}, {Rojas-Ruiz}, {Decarli}, {Farina}, {Mazzucchelli}, \& {Earnshaw}}]{Connor2021}
{Connor}, T., {Ba{\~n}ados}, E., {Stern}, D., {et~al.} 2021{\natexlab{b}}, \bibinfo{title}{{Enhanced X-Ray Emission from the Most Radio-powerful Quasar in the Universe's First Billion Years},} \apj, 911, 120, \dodoi{10.3847/1538-4357/abe710}

\bibitem[{D. {Donato} {et~al.}(2001){Donato}, {Ghisellini}, {Tagliaferri}, \& {Fossati}}]{Donato2001}
{Donato}, D., {Ghisellini}, G., {Tagliaferri}, G., \& {Fossati}, G. 2001, \bibinfo{title}{{Hard X-ray properties of blazars},} \aap, 375, 739, \dodoi{10.1051/0004-6361:20010675}

\bibitem[{S.~W. {Duchesne} {et~al.}(2023){Duchesne}, {Thomson}, {Pritchard}, {Lenc}, {Moss}, {McConnell}, {Wieringa}, {Whiting}, {Wang}, {Wang}, {Rose}, {Raja}, {Murphy}, {Leung}, {Huynh}, {Hotan}, {Hodgson}, \& {Heald}}]{Duchesne2023}
{Duchesne}, S.~W., {Thomson}, A.~J.~M., {Pritchard}, J., {et~al.} 2023, \bibinfo{title}{{The Rapid ASKAP Continuum Survey IV: continuum imaging at 1367.5 MHz and the first data release of RACS-mid},} \pasa, 40, e034, \dodoi{10.1017/pasa.2023.31}

\bibitem[{S.~W. {Duchesne} {et~al.}(2024){Duchesne}, {Grundy}, {Heald}, {Lenc}, {Leung}, {McConnell}, {Murphy}, {Pritchard}, {Rose}, {Thomson}, {Wang}, {Wang}, \& {Whiting}}]{Duchesne2024}
{Duchesne}, S.~W., {Grundy}, J.~A., {Heald}, G.~H., {et~al.} 2024, \bibinfo{title}{{The Rapid ASKAP Continuum Survey V: Cataloguing the sky at 1 367.5 MHz and the second data release of RACS-mid},} \pasa, 41, e003, \dodoi{10.1017/pasa.2023.60}

\bibitem[{S.~W. {Duchesne} {et~al.}(2025){Duchesne}, {Ross}, {Thomson}, {Lenc}, {Murphy}, {Galvin}, {Hotan}, {Moss}, \& {Whiting}}]{Duchesne2025}
{Duchesne}, S.~W., {Ross}, K., {Thomson}, A.~J.~M., {et~al.} 2025, \bibinfo{title}{{The Rapid ASKAP Continuum Survey (RACS) VI: The RACS-high 1 655.5 MHz images and catalogue},} \pasa, 42, e038, \dodoi{10.1017/pasa.2025.2}

\bibitem[{F. {Duras} {et~al.}(2020){Duras}, {Bongiorno}, {Ricci}, {Piconcelli}, {Shankar}, {Lusso}, {Bianchi}, {Fiore}, {Maiolino}, {Marconi}, {Onori}, {Sani}, {Schneider}, {Vignali}, \& {La Franca}}]{Duras2020}
{Duras}, F., {Bongiorno}, A., {Ricci}, F., {et~al.} 2020, \bibinfo{title}{{Universal bolometric corrections for active galactic nuclei over seven luminosity decades},} \aap, 636, A73, \dodoi{10.1051/0004-6361/201936817}

\bibitem[{A. {Edge} {et~al.}(2013){Edge}, {Sutherland}, {Kuijken}, {Driver}, {McMahon}, {Eales}, \& {Emerson}}]{Edge2013}
{Edge}, A., {Sutherland}, W., {Kuijken}, K., {et~al.} 2013, \bibinfo{title}{{The VISTA Kilo-degree Infrared Galaxy (VIKING) Survey: Bridging the Gap between Low and High Redshift},} The Messenger, 154, 32

\bibitem[{P.~R.~M. {Eisenhardt} {et~al.}(2020){Eisenhardt}, {Marocco}, {Fowler}, {Meisner}, {Kirkpatrick}, {Garcia}, {Jarrett}, {Koontz}, {Marchese}, {Stanford}, {Caselden}, {Cushing}, {Cutri}, {Faherty}, {Gelino}, {Gonzalez}, {Mainzer}, {Mobasher}, {Schlegel}, {Stern}, {Teplitz}, \& {Wright}}]{Eisenhardt2020}
{Eisenhardt}, P. R.~M., {Marocco}, F., {Fowler}, J.~W., {et~al.} 2020, \bibinfo{title}{{The CatWISE Preliminary Catalog: Motions from WISE and NEOWISE Data},} \apjs, 247, 69, \dodoi{10.3847/1538-4365/ab7f2a}

\bibitem[{X. {Fan} {et~al.}(2023){Fan}, {Ba{\~n}ados}, \& {Simcoe}}]{Fan2023}
{Fan}, X., {Ba{\~n}ados}, E., \& {Simcoe}, R.~A. 2023, \bibinfo{title}{{Quasars and the Intergalactic Medium at Cosmic Dawn},} \araa, 61, 373, \dodoi{10.1146/annurev-astro-052920-102455}

\bibitem[{R. {Fanali} {et~al.}(2013){Fanali}, {Caccianiga}, {Severgnini}, {Della Ceca}, {Marchese}, {Carrera}, {Corral}, \& {Mateos}}]{Fanali2013}
{Fanali}, R., {Caccianiga}, A., {Severgnini}, P., {et~al.} 2013, \bibinfo{title}{{Studying the relationship between X-ray emission and accretion in AGN using the XMM-Newton Bright Serendipitous Survey},} \mnras, 433, 648, \dodoi{10.1093/mnras/stt757}

\bibitem[{E.~P. {Farina} {et~al.}(2022){Farina}, {Schindler}, {Walter}, {Ba{\~n}ados}, {Davies}, {Decarli}, {Eilers}, {Fan}, {Hennawi}, {Mazzucchelli}, {Meyer}, {Trakhtenbrot}, {Volonteri}, {Wang}, {Worseck}, {Yang}, {Gutcke}, {Venemans}, {Bosman}, {Costa}, {De Rosa}, {Drake}, \& {Onoue}}]{Farina2022}
{Farina}, E.~P., {Schindler}, J.-T., {Walter}, F., {et~al.} 2022, \bibinfo{title}{{The X-shooter/ALMA Sample of Quasars in the Epoch of Reionization. II. Black Hole Masses, Eddington Ratios, and the Formation of the First Quasars},} \apj, 941, 106, \dodoi{10.3847/1538-4357/ac9626}

\bibitem[{G. {Fossati} {et~al.}(1997){Fossati}, {Celotti}, {Ghisellini}, \& {Maraschi}}]{Fossati1997}
{Fossati}, G., {Celotti}, A., {Ghisellini}, G., \& {Maraschi}, L. 1997, \bibinfo{title}{{Unifying models for X-ray-selected and radio-selected BL Lac objects},} \mnras, 289, 136, \dodoi{10.1093/mnras/289.1.136}

\bibitem[{S. {Frey} {et~al.}(2011){Frey}, {Paragi}, {Gurvits}, {Gab{\'a}nyi}, \& {Cseh}}]{Frey2011}
{Frey}, S., {Paragi}, Z., {Gurvits}, L.~I., {Gab{\'a}nyi}, K.~{\'E}., \& {Cseh}, D. 2011, \bibinfo{title}{{Into the central 10 pc of the most distant known radio quasar. VLBI imaging observations of J1429+5447 at z = 6.21},} \aap, 531, L5, \dodoi{10.1051/0004-6361/201117341}

\bibitem[{A. {Fruscione} {et~al.}(2006){Fruscione}, {McDowell}, {Allen}, {Brickhouse}, {Burke}, {Davis}, {Durham}, {Elvis}, {Galle}, {Harris}, {Huenemoerder}, {Houck}, {Ishibashi}, {Karovska}, {Nicastro}, {Noble}, {Nowak}, {Primini}, {Siemiginowska}, {Smith}, \& {Wise}}]{Fruscinone2006}
{Fruscione}, A., {McDowell}, J.~C., {Allen}, G.~E., {et~al.} 2006, \bibinfo{title}{{CIAO: Chandra's data analysis system},} SPIE, 6270, 62701V, \dodoi{10.1117/12.671760}

\bibitem[{G.~P. {Garmire} {et~al.}(2003){Garmire}, {Bautz}, {Ford}, {Nousek}, \& {Ricker}}]{Garmire2003}
{Garmire}, G.~P., {Bautz}, M.~W., {Ford}, P.~G., {Nousek}, J.~A., \& {Ricker}, George~R., J. 2003, \bibinfo{title}{{Advanced CCD imaging spectrometer (ACIS) instrument on the Chandra X-ray Observatory},} 4851, 28, \dodoi{10.1117/12.461599}

\bibitem[{C.~M. {Gaskell}(2009){Gaskell}}]{Gaskell2009}
{Gaskell}, C.~M. 2009, \bibinfo{title}{{What broad emission lines tell us about how active galactic nuclei work},} \nar, 53, 140, \dodoi{10.1016/j.newar.2009.09.006}

\bibitem[{G. {Ghisellini} {et~al.}(2013){Ghisellini}, {Haardt}, {Della Ceca}, {Volonteri}, \& {Sbarrato}}]{Ghisellini2013}
{Ghisellini}, G., {Haardt}, F., {Della Ceca}, R., {Volonteri}, M., \& {Sbarrato}, T. 2013, \bibinfo{title}{{The role of relativistic jets in the heaviest and most active supermassive black holes at high redshift},} \mnras, 432, 2818, \dodoi{10.1093/mnras/stt637}

\bibitem[{G. Ghisellini \& F. Tavecchio(2009)Ghisellini \& Tavecchio}]{Ghisellini2009}
Ghisellini, G., \& Tavecchio, F. 2009, \bibinfo{title}{{Canonical high power blazars},} \mnras, 397, 985, \dodoi{10.1111/j.1365-2966.2009.15007.x}

\bibitem[{P. {Giommi} {et~al.}(2021){Giommi}, {Perri}, {Capalbi}, {D'Elia}, {Barres de Almeida}, {Brandt}, {Pollock}, {Arneodo}, {Di Giovanni}, {Chang}, {Civitarese}, {De Angelis}, {Leto}, {Verrecchia}, {Ricard}, {Di Pippo}, {Middei}, {Penacchioni}, {Ruffini}, {Sahakyan}, {Israyelyan}, \& {Turriziani}}]{Giommi2021}
{Giommi}, P., {Perri}, M., {Capalbi}, M., {et~al.} 2021, \bibinfo{title}{{X-ray spectra, light curves and SEDs of blazars frequently observed by Swift},} \mnras, 507, 5690, \dodoi{10.1093/mnras/stab2425}

\bibitem[{K.~K. {Gupta} {et~al.}(2024){Gupta}, {Ricci}, {Temple}, {Tortosa}, {Koss}, {Assef}, {Bauer}, {Mushotzy}, {Ricci}, {Ueda}, {Rojas}, {Trakhtenbrot}, {Chang}, {Oh}, {Li}, {Kawamuro}, {Diaz}, {Powell}, {Stern}, {Megan Urry}, {Harrison}, \& {Cenko}}]{Gupta2024}
{Gupta}, K.~K., {Ricci}, C., {Temple}, M.~J., {et~al.} 2024, \bibinfo{title}{{BASS: XLIII. Optical, UV, and X-ray emission properties of unobscured Swift/BAT active galactic nuclei},} \aap, 691, A203, \dodoi{10.1051/0004-6361/202450567}

\bibitem[{C.~L. {Hale} {et~al.}(2021){Hale}, {McConnell}, {Thomson}, {Lenc}, {Heald}, {Hotan}, {Leung}, {Moss}, {Murphy}, {Pritchard}, {Sadler}, {Stewart}, \& {Whiting}}]{Hale2021}
{Hale}, C.~L., {McConnell}, D., {Thomson}, A.~J.~M., {et~al.} 2021, \bibinfo{title}{{The Rapid ASKAP Continuum Survey Paper II: First Stokes I Source Catalogue Data Release},} \pasa, 38, e058, \dodoi{10.1017/pasa.2021.47}

\bibitem[{ {HI4PI Collaboration} {et~al.}(2016){HI4PI Collaboration}, {Ben Bekhti}, {Fl{\"o}er}, {Keller}, {Kerp}, {Lenz}, {Winkel}, {Bailin}, {Calabretta}, {Dedes}, {Ford}, {Gibson}, {Haud}, {Janowiecki}, {Kalberla}, {Lockman}, {McClure-Griffiths}, {Murphy}, {Nakanishi}, {Pisano}, \& {Staveley-Smith}}]{HI4PI2016}
{HI4PI Collaboration}, {Ben Bekhti}, N., {Fl{\"o}er}, L., {et~al.} 2016, \bibinfo{title}{{HI4PI: A full-sky H I survey based on EBHIS and GASS},} \aap, 594, A116, \dodoi{10.1051/0004-6361/201629178}

\bibitem[{T. {Hovatta} {et~al.}(2014){Hovatta}, {Pavlidou}, {King}, {Mahabal}, {Sesar}, {Dancikova}, {Djorgovski}, {Drake}, {Laher}, {Levitan}, {Max-Moerbeck}, {Ofek}, {Pearson}, {Prince}, {Readhead}, {Richards}, \& {Surace}}]{Hovatta2014}
{Hovatta}, T., {Pavlidou}, V., {King}, O.~G., {et~al.} 2014, \bibinfo{title}{{Connection between optical and {\ensuremath{\gamma}}-ray variability in blazars},} \mnras, 439, 690, \dodoi{10.1093/mnras/stt2494}

\bibitem[{J. {Huang} {et~al.}(2020){Huang}, {Luo}, {Du}, {Hu}, {Wang}, \& {Li}}]{Huang2020}
{Huang}, J., {Luo}, B., {Du}, P., {et~al.} 2020, \bibinfo{title}{{On the Relation between the Hard X-Ray Photon Index and Accretion Rate for Super-Eddington Accreting Quasars},} \apj, 895, 114, \dodoi{10.3847/1538-4357/ab9019}

\bibitem[{L. {Ighina} {et~al.}(2023){Ighina}, {Caccianiga}, {Moretti}, {Belladitta}, {Broderick}, {Drouart}, {Leung}, \& {Seymour}}]{Ighina2023}
{Ighina}, L., {Caccianiga}, A., {Moretti}, A., {et~al.} 2023, \bibinfo{title}{{New radio-loud QSOs at the end of the Re-ionization epoch},} \mnras, 519, 2060, \dodoi{10.1093/mnras/stac3668}

\bibitem[{L. {Ighina} {et~al.}(2019){Ighina}, {Caccianiga}, {Moretti}, {Belladitta}, {Della Ceca}, {Ballo}, \& {Dallacasa}}]{Ighina2019}
{Ighina}, L., {Caccianiga}, A., {Moretti}, A., {et~al.} 2019, \bibinfo{title}{{X-ray properties of z $>$ 4 blazars},} \mnras, 489, 2732, \dodoi{10.1093/mnras/stz2340}

\bibitem[{L. {Ighina} {et~al.}(2022){Ighina}, {Moretti}, {Tavecchio}, {Caccianiga}, {Belladitta}, {Dallacasa}, {Della Ceca}, {Sbarrato}, \& {Spingola}}]{Ighina2022a}
{Ighina}, L., {Moretti}, A., {Tavecchio}, F., {et~al.} 2022, \bibinfo{title}{{Direct observation of an extended X-ray jet at z = 6.1},} \aap, 659, A93, \dodoi{10.1051/0004-6361/202142676}

\bibitem[{L. {Ighina} {et~al.}(2024){Ighina}, {Caccianiga}, {Moretti}, {Broderick}, {Leung}, {Paterson}, {Rigamonti}, {Seymour}, {Belladitta}, {Drouart}, {Galvin}, \& {Hurley-Walker}}]{Ighina2024b}
{Ighina}, L., {Caccianiga}, A., {Moretti}, A., {et~al.} 2024, \bibinfo{title}{{Comprehensive view of a z {\ensuremath{\sim}} 6.5 radio-loud quasi-stellar object: From the radio to the optical/NIR to the X-ray band},} \aap, 687, A242, \dodoi{10.1051/0004-6361/202449369}

\bibitem[{L. {Ighina} {et~al.}(2025){Ighina}, {Caccianiga}, {Moretti}, {Broderick}, {Leung}, {Rigamonti}, {Seymour}, {Afonso}, {Connor}, {Vignali}, {Wang}, {An}, {Arsioli}, {Bisogni}, {Dallacasa}, {Della Ceca}, {Liu}, {L{\'o}pez-S{\'a}nchez}, {Matute}, {Reynolds}, {Rossi}, {Spingola}, {Severgnini}, \& {Tavecchio}}]{Ighina2025}
{Ighina}, L., {Caccianiga}, A., {Moretti}, A., {et~al.} 2025, \bibinfo{title}{{High-$z$ radio Quasars in RACS I: Selection, identification, and multi-wavelength properties},} arXiv e-prints, arXiv:2504.10573, \dodoi{10.48550/arXiv.2504.10573}

\bibitem[{K. {Inayoshi} {et~al.}(2024){Inayoshi}, {Kimura}, \& {Noda}}]{Inayoshi2024}
{Inayoshi}, K., {Kimura}, S.~S., \& {Noda}, H. 2024, \bibinfo{title}{{Weakness of X-rays and Variability in High-redshift AGNs with Super-Eddington Accretion},} arXiv e-prints, arXiv:2412.03653, \dodoi{10.48550/arXiv.2412.03653}

\bibitem[{E.~J.~D. {Jolley} \& Z. {Kuncic}(2008){Jolley} \& {Kuncic}}]{Jolley2008}
{Jolley}, E.~J.~D., \& {Kuncic}, Z. 2008, \bibinfo{title}{{Jet-enhanced accretion growth of supermassive black holes},} \mnras, 386, 989, \dodoi{10.1111/j.1365-2966.2008.13082.x}

\bibitem[{R. {Kale} \& C.~H. {Ishwara-Chandra}(2021){Kale} \& {Ishwara-Chandra}}]{Kale2021}
{Kale}, R., \& {Ishwara-Chandra}, C.~H. 2021, \bibinfo{title}{{CAPTURE: a continuum imaging pipeline for the uGMRT},} Experimental Astronomy, 51, 95, \dodoi{10.1007/s10686-020-09677-6}

\bibitem[{E. {Kara} {et~al.}(2017){Kara}, {Garc{\'\i}a}, {Lohfink}, {Fabian}, {Reynolds}, {Tombesi}, \& {Wilkins}}]{Kara2017}
{Kara}, E., {Garc{\'\i}a}, J.~A., {Lohfink}, A., {et~al.} 2017, \bibinfo{title}{{The high-Eddington NLS1 Ark 564 has the coolest corona},} \mnras, 468, 3489, \dodoi{10.1093/mnras/stx792}

\bibitem[{G.~A. {Khorunzhev} {et~al.}(2021){Khorunzhev}, {Meshcheryakov}, {Medvedev}, {Borisov}, {Burenin}, {Krivonos}, {Uklein}, {Shablovinskaya}, {Afanasiev}, {Dodonov}, {Sunyaev}, {Sazonov}, \& {Gilfanov}}]{Khorunzhev2021}
{Khorunzhev}, G.~A., {Meshcheryakov}, A.~V., {Medvedev}, P.~S., {et~al.} 2021, \bibinfo{title}{{Discovery of the Most X-ray Luminous Quasar SRGE J170245.3+130104 at Redshift \textbackslashboldsymbol\{z{\ensuremath{\approx}} 5.5\}},} Astronomy Letters, 47, 123, \dodoi{10.1134/S1063773721030026}

\bibitem[{A. {King}(2024){King}}]{King2024}
{King}, A. 2024, \bibinfo{title}{{The black hole masses of high-redshift QSOs},} \mnras, 531, 550, \dodoi{10.1093/mnras/stae1171}

\bibitem[{A. {Konigl}(1981){Konigl}}]{Konigl1981}
{Konigl}, A. 1981, \bibinfo{title}{{Relativistic jets as X-ray and gamma-ray sources.},} \apj, 243, 700, \dodoi{10.1086/158638}

\bibitem[{E. {Lambrides} {et~al.}(2024){Lambrides}, {Garofali}, {Larson}, {Ptak}, {Chiaberge}, {Long}, {Hutchison}, {Norman}, {McKinney}, {Akins}, {Berg}, {Chisholm}, {Civano}, {Cloonan}, {Endsley}, {Faisst}, {Gilli}, {Gillman}, {Hirschmann}, {Kartaltepe}, {Kocevski}, {Kokorev}, {Pacucci}, {Richardson}, {Stiavelli}, \& {Whalen}}]{Lambrides2024}
{Lambrides}, E., {Garofali}, K., {Larson}, R., {et~al.} 2024, \bibinfo{title}{{The Case for Super-Eddington Accretion: Connecting Weak X-ray and UV Line Emission in JWST Broad-Line AGN During the First Gyr of Cosmic Time},} arXiv e-prints, arXiv:2409.13047, \dodoi{10.48550/arXiv.2409.13047}

\bibitem[{M.~A. {Latif} {et~al.}(2022){Latif}, {Whalen}, {Khochfar}, {Herrington}, \& {Woods}}]{Latif2022}
{Latif}, M.~A., {Whalen}, D.~J., {Khochfar}, S., {Herrington}, N.~P., \& {Woods}, T.~E. 2022, \bibinfo{title}{{Turbulent cold flows gave birth to the first quasars},} \nat, 607, 48, \dodoi{10.1038/s41586-022-04813-y}

\bibitem[{M. {Laurenti} {et~al.}(2022){Laurenti}, {Piconcelli}, {Zappacosta}, {Tombesi}, {Vignali}, {Bianchi}, {Marziani}, {Vagnetti}, {Bongiorno}, {Bischetti}, {del Olmo}, {Lanzuisi}, {Luminari}, {Middei}, {Perri}, {Ricci}, \& {Vietri}}]{Laurenti2022}
{Laurenti}, M., {Piconcelli}, E., {Zappacosta}, L., {et~al.} 2022, \bibinfo{title}{{X-ray spectroscopic survey of highly accreting AGN},} \aap, 657, A57, \dodoi{10.1051/0004-6361/202141829}

\bibitem[{J.-T. {Li} {et~al.}(2021){Li}, {Wang}, {Yang}, {Zhang}, {Fu}, {Bian}, {Bregman}, {Fan}, {Li}, {Wu}, \& {Yu}}]{Li2021}
{Li}, J.-T., {Wang}, F., {Yang}, J., {et~al.} 2021, \bibinfo{title}{{Chandra Detection of Three X-Ray Bright Quasars at z > 5},} \apj, 906, 135, \dodoi{10.3847/1538-4357/abc750}

\bibitem[{H. {Liu} {et~al.}(2021){Liu}, {Luo}, {Brandt}, {Brotherton}, {Gallagher}, {Ni}, {Shemmer}, \& {Timlin}}]{Liu2021}
{Liu}, H., {Luo}, B., {Brandt}, W.~N., {et~al.} 2021, \bibinfo{title}{{On the Observational Difference between the Accretion Disk-Corona Connections among Super- and Sub-Eddington Accreting Active Galactic Nuclei},} \apj, 910, 103, \dodoi{10.3847/1538-4357/abe37f}

\bibitem[{E. {Liuzzo} {et~al.}(2013){Liuzzo}, {Giroletti}, {Giovannini}, {Boccardi}, {Tamburri}, {Taylor}, {Casadio}, {Kadler}, {Tosti}, \& {Mignano}}]{Liuzzo2013}
{Liuzzo}, E., {Giroletti}, M., {Giovannini}, G., {et~al.} 2013, \bibinfo{title}{{Exploring the bulk of the BL Lacertae object population. I. Parsec-scale radio structures},} \aap, 560, A23, \dodoi{10.1051/0004-6361/201322144}

\bibitem[{A. {Lupi} {et~al.}(2021){Lupi}, {Haiman}, \& {Volonteri}}]{Lupi2021}
{Lupi}, A., {Haiman}, Z., \& {Volonteri}, M. 2021, \bibinfo{title}{{Forming massive seed black holes in high-redshift quasar host progenitors},} \mnras, 503, 5046, \dodoi{10.1093/mnras/stab692}

\bibitem[{A. {Lupi} {et~al.}(2024{\natexlab{a}}){Lupi}, {Quadri}, {Volonteri}, {Colpi}, \& {Regan}}]{Lupi2024}
{Lupi}, A., {Quadri}, G., {Volonteri}, M., {Colpi}, M., \& {Regan}, J.~A. 2024{\natexlab{a}}, \bibinfo{title}{{Sustained super-Eddington accretion in high-redshift quasars},} \aap, 686, A256, \dodoi{10.1051/0004-6361/202348788}

\bibitem[{A. {Lupi} {et~al.}(2024{\natexlab{b}}){Lupi}, {Trinca}, {Volonteri}, {Dotti}, \& {Mazzucchelli}}]{Lupi2024b}
{Lupi}, A., {Trinca}, A., {Volonteri}, M., {Dotti}, M., \& {Mazzucchelli}, C. 2024{\natexlab{b}}, \bibinfo{title}{{Size matters: are we witnessing super-Eddington accretion in high-redshift black holes from JWST?},} \aap, 689, A128, \dodoi{10.1051/0004-6361/202451249}

\bibitem[{E. {Lusso} \& G. {Risaliti}(2016){Lusso} \& {Risaliti}}]{Lusso2016}
{Lusso}, E., \& {Risaliti}, G. 2016, \bibinfo{title}{{The Tight Relation between X-Ray and Ultraviolet Luminosity of Quasars},} \apj, 819, 154, \dodoi{10.3847/0004-637X/819/2/154}

\bibitem[{P. {Madau}(2025){Madau}}]{Madau2025}
{Madau}, P. 2025, \bibinfo{title}{{Chasing the Light: Shadowing, Collimation, and the Super-Eddington Growth of Infant Black Holes in JWST-Discovered AGNs},} arXiv e-prints, arXiv:2501.09854, \dodoi{10.48550/arXiv.2501.09854}

\bibitem[{P. {Madau} \& F. {Haardt}(2024){Madau} \& {Haardt}}]{Madau2024}
{Madau}, P., \& {Haardt}, F. 2024, \bibinfo{title}{{X-Ray Weak Active Galactic Nuclei from Super-Eddington Accretion onto Infant Black Holes},} \apjl, 976, L24, \dodoi{10.3847/2041-8213/ad90e1}

\bibitem[{R. {Maiolino} {et~al.}(2024){Maiolino}, {Scholtz}, {Witstok}, {Carniani}, {D'Eugenio}, {de Graaff}, {{\"U}bler}, {Tacchella}, {Curtis-Lake}, {Arribas}, {Bunker}, {Charlot}, {Chevallard}, {Curti}, {Looser}, {Maseda}, {Rawle}, {Rodr{\'\i}guez del Pino}, {Willott}, {Egami}, {Eisenstein}, {Hainline}, {Robertson}, {Williams}, {Willmer}, {Baker}, {Boyett}, {DeCoursey}, {Fabian}, {Helton}, {Ji}, {Jones}, {Kumari}, {Laporte}, {Nelson}, {Perna}, {Sandles}, {Shivaei}, \& {Sun}}]{Maiolino2024b}
{Maiolino}, R., {Scholtz}, J., {Witstok}, J., {et~al.} 2024, \bibinfo{title}{{A small and vigorous black hole in the early Universe},} \nat, 627, 59, \dodoi{10.1038/s41586-024-07052-5}

\bibitem[{R. {Maiolino} {et~al.}(2025){Maiolino}, {Risaliti}, {Signorini}, {Trefoloni}, {Juod{\v{z}}balis}, {Scholtz}, {{\"U}bler}, {D'Eugenio}, {Carniani}, {Fabian}, {Ji}, {Mazzolari}, {Bertola}, {Brusa}, {Bunker}, {Charlot}, {Comastri}, {Cresci}, {DeCoursey}, {Egami}, {Fiore}, {Gilli}, {Perna}, {Tacchella}, \& {Venturi}}]{Maiolino2024}
{Maiolino}, R., {Risaliti}, G., {Signorini}, M., {et~al.} 2025, \bibinfo{title}{{JWST meets Chandra: a large population of Compton thick, feedback-free, and intrinsically X-ray weak AGN, with a sprinkle of SNe},} \mnras, 538, 1921, \dodoi{10.1093/mnras/staf359}

\bibitem[{L. {Marcotulli} {et~al.}(2025){Marcotulli}, {Connor}, {Ba{\~n}ados}, {Boorman}, {Migliori}, {Grefenstette}, {Momjian}, {Siemiginowska}, {Stern}, {Belladitta}, {Cheung}, {Fabian}, {Khusanova}, {Mazzucchelli}, {Rojas-Ruiz}, \& {Urry}}]{Marcotulli2025}
{Marcotulli}, L., {Connor}, T., {Ba{\~n}ados}, E., {et~al.} 2025, \bibinfo{title}{{NuSTAR Observations of a Varying-flux Quasar in the Epoch of Reionization},} \apjl, 979, L6, \dodoi{10.3847/2041-8213/ad94ee}

\bibitem[{E. {Massaro} {et~al.}(2004){Massaro}, {Perri}, {Giommi}, \& {Nesci}}]{MassaroE2004}
{Massaro}, E., {Perri}, M., {Giommi}, P., \& {Nesci}, R. 2004, \bibinfo{title}{{Log-parabolic spectra and particle acceleration in the BL Lac object Mkn 421: Spectral analysis of the complete BeppoSAX wide band X-ray data set},} \aap, 413, 489, \dodoi{10.1051/0004-6361:20031558}

\bibitem[{W. {Massonneau} {et~al.}(2023){Massonneau}, {Volonteri}, {Dubois}, \& {Beckmann}}]{Massonneau2023}
{Massonneau}, W., {Volonteri}, M., {Dubois}, Y., \& {Beckmann}, R.~S. 2023, \bibinfo{title}{{How the super-Eddington regime regulates black hole growth in high-redshift galaxies},} \aap, 670, A180, \dodoi{10.1051/0004-6361/202243170}

\bibitem[{Y. {Matsuoka} {et~al.}(2018){Matsuoka}, {Iwasawa}, {Onoue}, {Kashikawa}, {Strauss}, {Lee}, {Imanishi}, {Nagao}, {Akiyama}, {Asami}, {Bosch}, {Furusawa}, {Goto}, {Gunn}, {Harikane}, {Ikeda}, {Izumi}, {Kawaguchi}, {Kato}, {Kikuta}, {Kohno}, {Komiyama}, {Lupton}, {Minezaki}, {Miyazaki}, {Morokuma}, {Murayama}, {Niida}, {Nishizawa}, {Oguri}, {Ono}, {Ouchi}, {Price}, {Sameshima}, {Schulze}, {Shirakata}, {Silverman}, {Sugiyama}, {Tait}, {Takada}, {Takata}, {Tanaka}, {Tang}, {Toba}, {Utsumi}, {Wang}, \& {Yamashita}}]{Matsuoka2018a}
{Matsuoka}, Y., {Iwasawa}, K., {Onoue}, M., {et~al.} 2018, \bibinfo{title}{{Subaru High-z Exploration of Low-luminosity Quasars (SHELLQs). IV. Discovery of 41 Quasars and Luminous Galaxies at 5.7 {\ensuremath{\leq}} z {\ensuremath{\leq}} 6.9},} \apjs, 237, 5, \dodoi{10.3847/1538-4365/aac724}

\bibitem[{C. {Mazzucchelli} {et~al.}(2023){Mazzucchelli}, {Bischetti}, {D'Odorico}, {Feruglio}, {Schindler}, {Onoue}, {Ba{\~n}ados}, {Becker}, {Bian}, {Carniani}, {Decarli}, {Eilers}, {Farina}, {Gallerani}, {Lai}, {Meyer}, {Rojas-Ruiz}, {Satyavolu}, {Venemans}, {Wang}, {Yang}, \& {Zhu}}]{Mazzucchelli2023}
{Mazzucchelli}, C., {Bischetti}, M., {D'Odorico}, V., {et~al.} 2023, \bibinfo{title}{{XQR-30: Black hole masses and accretion rates of 42 z {\ensuremath{\gtrsim}} 6 quasars},} \aap, 676, A71, \dodoi{10.1051/0004-6361/202346317}

\bibitem[{D. {McConnell} {et~al.}(2020){McConnell}, {Hale}, {Lenc}, {Banfield}, {Heald}, {Hotan}, {Leung}, {Moss}, {Murphy}, {O'Brien}, {Pritchard}, {Raja}, {Sadler}, {Stewart}, {Thomson}, {Whiting}, {Allison}, {Amy}, {Anderson}, {Ball}, {Bannister}, {Bell}, {Bock}, {Bolton}, {Bunton}, {Chippendale}, {Collier}, {Cooray}, {Cornwell}, {Diamond}, {Edwards}, {Gupta}, {Hayman}, {Heywood}, {Jackson}, {Koribalski}, {Lee-Waddell}, {McClure-Griffiths}, {Ng}, {Norris}, {Phillips}, {Reynolds}, {Roxby}, {Schinckel}, {Shields}, {Tremblay}, {Tzioumis}, {Voronkov}, \& {Westmeier}}]{McConnell2020}
{McConnell}, D., {Hale}, C.~L., {Lenc}, E., {et~al.} 2020, \bibinfo{title}{{The Rapid ASKAP Continuum Survey I: Design and first results},} \pasa, 37, e048, \dodoi{10.1017/pasa.2020.41}

\bibitem[{J.~P. {McMullin} {et~al.}(2007){McMullin}, {Waters}, {Schiebel}, {Young}, \& {Golap}}]{Mcmullin2007}
{McMullin}, J.~P., {Waters}, B., {Schiebel}, D., {Young}, W., \& {Golap}, K. 2007, \bibinfo{title}{{CASA Architecture and Applications},} ADASS XVI, ASP Conf. Series, 376, 127

\bibitem[{P. {Medvedev} {et~al.}(2021){Medvedev}, {Gilfanov}, {Sazonov}, {Schartel}, \& {Sunyaev}}]{Medvedev2021}
{Medvedev}, P., {Gilfanov}, M., {Sazonov}, S., {Schartel}, N., \& {Sunyaev}, R. 2021, \bibinfo{title}{{XMM-Newton observations of the extremely X-ray luminous quasar CFHQS J142952+544717=SRGE J142952.1 + 544716 at redshift z = 6.18},} \mnras, 504, 576, \dodoi{10.1093/mnras/stab773}

\bibitem[{P. {Medvedev} {et~al.}(2020){Medvedev}, {Sazonov}, {Gilfanov}, {Burenin}, {Khorunzhev}, {Meshcheryakov}, {Sunyaev}, {Bikmaev}, \& {Irtuganov}}]{Medvedev2020}
{Medvedev}, P., {Sazonov}, S., {Gilfanov}, M., {et~al.} 2020, \bibinfo{title}{{SRG/eROSITA uncovers the most X-ray luminous quasar at z $>$ 6},} \mnras, 497, 1842, \dodoi{10.1093/mnras/staa2051}

\bibitem[{D.~L. {Meier}(2002){Meier}}]{Meier2002}
{Meier}, D.~L. 2002, \bibinfo{title}{{Grand unification of AGN and the accretion and spin paradigms},} \nar, 46, 247, \dodoi{10.1016/S1387-6473(01)00189-0}

\bibitem[{R. {Middei} {et~al.}(2022){Middei}, {Giommi}, {Perri}, {Turriziani}, {Sahakyan}, {Chang}, {Leto}, \& {Verrecchia}}]{Middei2022}
{Middei}, R., {Giommi}, P., {Perri}, M., {et~al.} 2022, \bibinfo{title}{{The first hard X-ray spectral catalogue of Blazars observed by NuSTAR},} \mnras, 514, 3179, \dodoi{10.1093/mnras/stac1185}

\bibitem[{G. {Migliori} {et~al.}(2023){Migliori}, {Siemiginowska}, {Sobolewska}, {Cheung}, {Stawarz}, {Schwartz}, {Snios}, {Saxena}, \& {Kashyap}}]{Migliori2023}
{Migliori}, G., {Siemiginowska}, A., {Sobolewska}, M., {et~al.} 2023, \bibinfo{title}{{The extremely X-ray luminous radio-loud quasar CFHQS J142952 + 544717 at z = 6.18 under Chandra high-angular resolution lens},} \mnras, 524, 1087, \dodoi{10.1093/mnras/stad1959}

\bibitem[{A. {Moretti} {et~al.}(2014){Moretti}, {Ballo}, {Braito}, {Caccianiga}, {Della Ceca}, {Gilli}, {Salvaterra}, {Severgnini}, \& {Vignali}}]{Moretti2014}
{Moretti}, A., {Ballo}, L., {Braito}, V., {et~al.} 2014, \bibinfo{title}{{X-ray observation of ULAS J1120+0641, the most distant quasar at z = 7.08},} \aap, 563, A46, \dodoi{10.1051/0004-6361/201323051}

\bibitem[{A. {Moretti} {et~al.}(2021){Moretti}, {Ghisellini}, {Caccianiga}, {Belladitta}, {Della Ceca}, {Ighina}, {Sbarrato}, {Severgnini}, \& {Spingola}}]{Moretti2021}
{Moretti}, A., {Ghisellini}, G., {Caccianiga}, A., {et~al.} 2021, \bibinfo{title}{{Minute-timescale Variability in the X-ray Emission of the Highest Redshift Blazar},} \apj, 920, 15, \dodoi{10.3847/1538-4357/ac167a}

\bibitem[{T. {Murphy} {et~al.}(2013){Murphy}, {Chatterjee}, {Kaplan}, {Banyer}, {Bell}, {Bignall}, {Bower}, {Cameron}, {Coward}, {Cordes}, {Croft}, {Curran}, {Djorgovski}, {Farrell}, {Frail}, {Gaensler}, {Galloway}, {Gendre}, {Green}, {Hancock}, {Johnston}, {Kamble}, {Law}, {Lazio}, {Lo}, {Macquart}, {Rea}, {Rebbapragada}, {Reynolds}, {Ryder}, {Schmidt}, {Soria}, {Stairs}, {Tingay}, {Torkelsson}, {Wagstaff}, {Walker}, {Wayth}, \& {Williams}}]{Murphy2013}
{Murphy}, T., {Chatterjee}, S., {Kaplan}, D.~L., {et~al.} 2013, \bibinfo{title}{{VAST: An ASKAP Survey for Variables and Slow Transients},} \pasa, 30, e006, \dodoi{10.1017/pasa.2012.006}

\bibitem[{T. {Murphy} {et~al.}(2021){Murphy}, {Kaplan}, {Stewart}, {O'Brien}, {Lenc}, {Pintaldi}, {Pritchard}, {Dobie}, {Fox}, {Leung}, {An}, {Bell}, {Broderick}, {Chatterjee}, {Dai}, {d'Antonio}, {Doyle}, {Gaensler}, {Heald}, {Horesh}, {Jones}, {McConnell}, {Moss}, {Raja}, {Ramsay}, {Ryder}, {Sadler}, {Sivakoff}, {Wang}, {Wang}, {Wheatland}, {Whiting}, {Allison}, {Anderson}, {Ball}, {Bannister}, {Bock}, {Bolton}, {Bunton}, {Chekkala}, {Chippendale}, {Cooray}, {Gupta}, {Hayman}, {Jeganathan}, {Koribalski}, {Lee-Waddell}, {Mahony}, {Marvil}, {McClure-Griffiths}, {Mirtschin}, {Ng}, {Pearce}, {Phillips}, \& {Voronkov}}]{Murphy2021}
{Murphy}, T., {Kaplan}, D.~L., {Stewart}, A.~J., {et~al.} 2021, \bibinfo{title}{{The ASKAP Variables and Slow Transients (VAST) Pilot Survey},} \pasa, 38, e054, \dodoi{10.1017/pasa.2021.44}

\bibitem[{R. {Nanni} {et~al.}(2017){Nanni}, {Vignali}, {Gilli}, {Moretti}, \& {Brandt}}]{Nanni2017}
{Nanni}, R., {Vignali}, C., {Gilli}, R., {Moretti}, A., \& {Brandt}, W.~N. 2017, \bibinfo{title}{{The X-ray properties of z 6 luminous quasars},} \aap, 603, A128, \dodoi{10.1051/0004-6361/201730484}

\bibitem[{F. {Pacucci} \& R. {Narayan}(2024){Pacucci} \& {Narayan}}]{Pacucci2024}
{Pacucci}, F., \& {Narayan}, R. 2024, \bibinfo{title}{{Mildly Super-Eddington Accretion onto Slowly Spinning Black Holes Explains the X-Ray Weakness of the Little Red Dots},} \apj, 976, 96, \dodoi{10.3847/1538-4357/ad84f7}

\bibitem[{V.~S. {Paliya} {et~al.}(2020{\natexlab{a}}){Paliya}, {Ajello}, {Cao}, {Giroletti}, {Kaur}, {Madejski}, {Lott}, \& {Hartmann}}]{Paliya2020}
{Paliya}, V.~S., {Ajello}, M., {Cao}, H.~M., {et~al.} 2020{\natexlab{a}}, \bibinfo{title}{{Blazars at the Cosmic Dawn},} \apj, 897, 177, \dodoi{10.3847/1538-4357/ab9c1a}

\bibitem[{V.~S. {Paliya} {et~al.}(2020{\natexlab{b}}){Paliya}, {Dom{\'\i}nguez}, {Cabello}, {Cardiel}, {Gallego}, {Siana}, {Ajello}, {Hartmann}, {Gil de Paz}, \& {Stalin}}]{Paliya2020b}
{Paliya}, V.~S., {Dom{\'\i}nguez}, A., {Cabello}, C., {et~al.} 2020{\natexlab{b}}, \bibinfo{title}{{The First Gamma-Ray Emitting BL Lacertae Object at the Cosmic Dawn},} \apjl, 903, L8, \dodoi{10.3847/2041-8213/abbc06}

\bibitem[{B.~G. {Piner} {et~al.}(2010){Piner}, {Pant}, \& {Edwards}}]{Piner2010}
{Piner}, B.~G., {Pant}, N., \& {Edwards}, P.~G. 2010, \bibinfo{title}{{The Jets of TeV Blazars at Higher Resolution: 43 GHz and Polarimetric VLBA Observations from 2005 to 2009},} \apj, 723, 1150, \dodoi{10.1088/0004-637X/723/2/1150}

\bibitem[{Q. {Pognan} {et~al.}(2020){Pognan}, {Trakhtenbrot}, {Sbarrato}, {Schawinski}, \& {Bertemes}}]{Pognan2020}
{Pognan}, Q., {Trakhtenbrot}, B., {Sbarrato}, T., {Schawinski}, K., \& {Bertemes}, C. 2020, \bibinfo{title}{{Searching for super-Eddington quasars using a photon trapping accretion disc model},} \mnras, 492, 4058, \dodoi{10.1093/mnras/staa078}

\bibitem[{M. {Polletta} {et~al.}(2007){Polletta}, {Tajer}, {Maraschi}, {Trinchieri}, {Lonsdale}, {Chiappetti}, {Andreon}, {Pierre}, {Le F{\`e}vre}, {Zamorani}, {Maccagni}, {Garcet}, {Surdej}, {Franceschini}, {Alloin}, {Shupe}, {Surace}, {Fang}, {Rowan-Robinson}, {Smith}, \& {Tresse}}]{Polletta2007}
{Polletta}, M., {Tajer}, M., {Maraschi}, L., {et~al.} 2007, \bibinfo{title}{{Spectral Energy Distributions of Hard X-Ray Selected Active Galactic Nuclei in the XMM-Newton Medium Deep Survey},} \apj, 663, 81, \dodoi{10.1086/518113}

\bibitem[{S.~H. {Reddy} {et~al.}(2017){Reddy}, {Kudale}, {Gokhale}, {Halagalli}, {Raskar}, {de}, {Gnanaraj}, {Ajith Kumar}, \& {Gupta}}]{Reddy2017}
{Reddy}, S.~H., {Kudale}, S., {Gokhale}, U., {et~al.} 2017, \bibinfo{title}{{A Wideband Digital Back-End for the Upgraded GMRT},} Journal of Astronomical Instrumentation, 6, 1641011, \dodoi{10.1142/S2251171716410117}

\bibitem[{J. {Reynolds}(1994){Reynolds}}]{Reynolds1994}
{Reynolds}, J. 1994, \bibinfo{title}{{The Australia Telescope Compact Array Broad-band Backend: description and first results},} ATNF Technical Memos, AT/39.3/040

\bibitem[{A. {Ricarte} {et~al.}(2023){Ricarte}, {Narayan}, \& {Curd}}]{Ricarte2023}
{Ricarte}, A., {Narayan}, R., \& {Curd}, B. 2023, \bibinfo{title}{{Recipes for Jet Feedback and Spin Evolution of Black Holes with Strongly Magnetized Super-Eddington Accretion Disks},} \apjl, 954, L22, \dodoi{10.3847/2041-8213/aceda5}

\bibitem[{F. {Rigamonti} {et~al.}(2025){Rigamonti}, {Severgnini}, {Sottocorno}, {Dotti}, {Covino}, {Landoni}, {Bertassi}, {Braito}, {Cicone}, {Cupani}, {De Rosa}, {Della Ceca}, {Ighina}, {Singh}, \& {Vignali}}]{Rigamonti2025}
{Rigamonti}, F., {Severgnini}, P., {Sottocorno}, E., {et~al.} 2025, \bibinfo{title}{{ESPRESSO reveals a single but perturbed broad-line region in the supermassive black hole binary candidate PG 1302{\textendash}102},} \aap, 693, A117, \dodoi{10.1051/0004-6361/202452830}

\bibitem[{G. {Risaliti} {et~al.}(2009){Risaliti}, {Young}, \& {Elvis}}]{Risaliti2009}
{Risaliti}, G., {Young}, M., \& {Elvis}, M. 2009, \bibinfo{title}{{The Sloan Digital Sky Survey/XMM-Newton Quasar Survey: Correlation Between X-Ray Spectral Slope and Eddington Ratio},} \apjl, 700, L6, \dodoi{10.1088/0004-637X/700/1/L6}

\bibitem[{K. {Ross} {et~al.}(2024){Ross}, {Hurley-Walker}, {Galvin}, {Venville}, {Duchesne}, {Morgan}, {An}, {G{\"u}rkan}, {Hancock}, {Heald}, {Johnston-Hollitt}, \& {White}}]{Ross2024}
{Ross}, K., {Hurley-Walker}, N., {Galvin}, T.~J., {et~al.} 2024, \bibinfo{title}{{GaLactic and Extragalactic All-sky Murchison Widefield Array eXtended (GLEAM-X) survey II: Second Data Release},} \pasa, 41, e054, \dodoi{10.1017/pasa.2024.57}

\bibitem[{R.~J. {Sault} {et~al.}(1995){Sault}, {Teuben}, \& {Wright}}]{Sault1995}
{Sault}, R.~J., {Teuben}, P.~J., \& {Wright}, M.~C.~H. 1995, \bibinfo{title}{{A Retrospective View of MIRIAD},} ADASS IV, ASP Conf. Series, 77, 433, \dodoi{10.48550/arXiv.astro-ph/0612759}

\bibitem[{D.~V. {Savi{\'c}} {et~al.}(2024){Savi{\'c}}, {Hutsem{\'e}kers}, \& {Sluse}}]{Savic2024}
{Savi{\'c}}, D.~V., {Hutsem{\'e}kers}, D., \& {Sluse}, D. 2024, \bibinfo{title}{{Probing the broad line region geometry and size of the gravitationally lensed quasar Q2237+0305 with microlensing time series},} \aap, 687, A114, \dodoi{10.1051/0004-6361/202347953}

\bibitem[{A.~T.~P. {Schauer} {et~al.}(2017){Schauer}, {Regan}, {Glover}, \& {Klessen}}]{Schauer2017}
{Schauer}, A. T.~P., {Regan}, J., {Glover}, S. C.~O., \& {Klessen}, R.~S. 2017, \bibinfo{title}{{The formation of direct collapse black holes under the influence of streaming velocities},} \mnras, 471, 4878, \dodoi{10.1093/mnras/stx1915}

\bibitem[{F. {Shaban} {et~al.}(2022){Shaban}, {Siemiginowska}, {Suleiman}, {El-Nawawy}, \& {Ali}}]{Shaban2022}
{Shaban}, F., {Siemiginowska}, A., {Suleiman}, R.~M., {El-Nawawy}, M.~S., \& {Ali}, A. 2022, \bibinfo{title}{{X-ray properties of high-redshift Radio Loud and Radio Quiet Quasars observed by Chandra},} Journal of High Energy Astrophysics, 36, 152, \dodoi{10.1016/j.jheap.2022.10.002}

\bibitem[{O. {Shemmer} {et~al.}(2008){Shemmer}, {Brandt}, {Netzer}, {Maiolino}, \& {Kaspi}}]{Shemmer2008}
{Shemmer}, O., {Brandt}, W.~N., {Netzer}, H., {Maiolino}, R., \& {Kaspi}, S. 2008, \bibinfo{title}{{The Hard X-Ray Spectrum as a Probe for Black Hole Growth in Radio-Quiet Active Galactic Nuclei},} \apj, 682, 81, \dodoi{10.1086/588776}

\bibitem[{O. {Shemmer} {et~al.}(2005){Shemmer}, {Brandt}, {Vignali}, {Schneider}, {Fan}, {Richards}, \& {Strauss}}]{Shemmer2005}
{Shemmer}, O., {Brandt}, W.~N., {Vignali}, C., {et~al.} 2005, \bibinfo{title}{{The X-Ray Spectral Properties and Variability of Luminous High-Redshift Active Galactic Nuclei},} \apj, 630, 729, \dodoi{10.1086/432050}

\bibitem[{O. {Shemmer} {et~al.}(2014){Shemmer}, {Brandt}, {Paolillo}, {Kaspi}, {Vignali}, {Stein}, {Lira}, {Schneider}, \& {Gibson}}]{Shemmer2014}
{Shemmer}, O., {Brandt}, W.~N., {Paolillo}, M., {et~al.} 2014, \bibinfo{title}{{Exploratory X-Ray Monitoring of Luminous Radio-quiet Quasars at High Redshift: Initial Results},} \apj, 783, 116, \dodoi{10.1088/0004-637X/783/2/116}

\bibitem[{X. {Shen} {et~al.}(2020){Shen}, {Hopkins}, {Faucher-Gigu{\`e}re}, {Alexander}, {Richards}, {Ross}, \& {Hickox}}]{Shen2020}
{Shen}, X., {Hopkins}, P.~F., {Faucher-Gigu{\`e}re}, C.-A., {et~al.} 2020, \bibinfo{title}{{The bolometric quasar luminosity function at z = 0-7},} \mnras, 495, 3252, \dodoi{10.1093/mnras/staa1381}

\bibitem[{Y. {Shen} {et~al.}(2019){Shen}, {Wu}, {Jiang}, {Ba{\~n}ados}, {Fan}, {Ho}, {Riechers}, {Strauss}, {Venemans}, {Vestergaard}, {Walter}, {Wang}, {Willott}, {Wu}, \& {Yang}}]{Shen2019}
{Shen}, Y., {Wu}, J., {Jiang}, L., {et~al.} 2019, \bibinfo{title}{{Gemini GNIRS Near-infrared Spectroscopy of 50 Quasars at z {\ensuremath{\gtrsim}} 5.7},} \apj, 873, 35, \dodoi{10.3847/1538-4357/ab03d9}

\bibitem[{C. {Spingola} {et~al.}(2020){Spingola}, {Dallacasa}, {Belladitta}, {Caccianiga}, {Giroletti}, {Moretti}, \& {Orienti}}]{Spingola2020}
{Spingola}, C., {Dallacasa}, D., {Belladitta}, S., {et~al.} 2020, \bibinfo{title}{{Parsec-scale properties of the radio brightest jetted AGN at z $>$ 6},} \aap, 643, L12, \dodoi{10.1051/0004-6361/202039458}

\bibitem[{H. {Suh} {et~al.}(2025){Suh}, {Scharw{\"a}chter}, {Farina}, {Loiacono}, {Lanzuisi}, {Hasinger}, {Marchesi}, {Mezcua}, {Decarli}, {Lemaux}, {Volonteri}, {Civano}, {Yi}, {Han}, {Rawlings}, \& {Hung}}]{Suh2024}
{Suh}, H., {Scharw{\"a}chter}, J., {Farina}, E.~P., {et~al.} 2025, \bibinfo{title}{{A super-Eddington-accreting black hole \raisebox{-0.5ex}\textasciitilde1.5 Gyr after the Big Bang observed with JWST},} Nature Astronomy, 9, 271, \dodoi{10.1038/s41550-024-02402-9}

\bibitem[{A. {Tortosa} {et~al.}(2022){Tortosa}, {Ricci}, {Tombesi}, {Ho}, {Du}, {Inayoshi}, {Wang}, {Shangguan}, \& {Li}}]{Tortosa2022}
{Tortosa}, A., {Ricci}, C., {Tombesi}, F., {et~al.} 2022, \bibinfo{title}{{The extreme properties of the nearby hyper-Eddington accreting active galactic nucleus in IRAS 04416+1215},} \mnras, 509, 3599, \dodoi{10.1093/mnras/stab3152}

\bibitem[{A. {Tortosa} {et~al.}(2024){Tortosa}, {Zappacosta}, {Piconcelli}, {Bischetti}, {Done}, {Miniutti}, {Saccheo}, {Vietri}, {Bongiorno}, {Brusa}, {Carniani}, {Chilingarian}, {Civano}, {Cristiani}, {D'Odorico}, {Elvis}, {Fan}, {Feruglio}, {Fiore}, {Gallerani}, {Giallongo}, {Gilli}, {Grazian}, {Guainazzi}, {Haardt}, {Luminari}, {Maiolino}, {Menci}, {Nicastro}, {Petrucci}, {Puccetti}, {Salvestrini}, {Schneider}, {Testa}, {Tombesi}, {Tripodi}, {Valiante}, {Vallini}, {Vanzella}, {Vasylenko}, {Vignali}, {Vito}, {Volonteri}, \& {La Franca}}]{Tortosa2024}
{Tortosa}, A., {Zappacosta}, L., {Piconcelli}, E., {et~al.} 2024, \bibinfo{title}{{HYPERION. Shedding light on the first luminous quasars: A correlation between UV disc winds and X-ray continuum},} \aap, 691, A235, \dodoi{10.1051/0004-6361/202449662}

\bibitem[{C. {Vignali} {et~al.}(2005){Vignali}, {Brandt}, {Schneider}, \& {Kaspi}}]{Vignali2005}
{Vignali}, C., {Brandt}, W.~N., {Schneider}, D.~P., \& {Kaspi}, S. 2005, \bibinfo{title}{{X-Ray Lighthouses of the High-Redshift Universe. II. Further Snapshot Observations of the Most Luminous z$>$\raisebox{-0.5ex}\textasciitilde4 Quasars with Chandra},} \aj, 129, 2519, \dodoi{10.1086/430217}

\bibitem[{F. {Vito} {et~al.}(2019){Vito}, {Brandt}, {Bauer}, {Calura}, {Gilli}, {Luo}, {Shemmer}, {Vignali}, {Zamorani}, {Brusa}, {Civano}, {Comastri}, \& {Nanni}}]{Vito2019}
{Vito}, F., {Brandt}, W.~N., {Bauer}, F.~E., {et~al.} 2019, \bibinfo{title}{{The X-ray properties of z $>$ 6 quasars: no evident evolution of accretion physics in the first Gyr of the Universe},} \aap, 630, A118, \dodoi{10.1051/0004-6361/201936217}

\bibitem[{M. {Volonteri} \& M.~C. {Begelman}(2010){Volonteri} \& {Begelman}}]{Volonteri2010}
{Volonteri}, M., \& {Begelman}, M.~C. 2010, \bibinfo{title}{{Quasi-stars and the cosmic evolution of massive black holes},} \mnras, 409, 1022, \dodoi{10.1111/j.1365-2966.2010.17359.x}

\bibitem[{M. {Volonteri} {et~al.}(2021){Volonteri}, {Habouzit}, \& {Colpi}}]{Volonteri2021}
{Volonteri}, M., {Habouzit}, M., \& {Colpi}, M. 2021, \bibinfo{title}{{The origins of massive black holes},} Nature Reviews Physics, 3, 732, \dodoi{10.1038/s42254-021-00364-9}

\bibitem[{K. {Wachter} {et~al.}(1979){Wachter}, {Leach}, \& {Kellogg}}]{Wachter1979}
{Wachter}, K., {Leach}, R., \& {Kellogg}, E. 1979, \bibinfo{title}{{Parameter estimation in X-ray astronomy using maximum likelihood.},} \apj, 230, 274, \dodoi{10.1086/157084}

\bibitem[{F. {Wang} {et~al.}(2021{\natexlab{a}}){Wang}, {Yang}, {Fan}, {Hennawi}, {Barth}, {Banados}, {Bian}, {Boutsia}, {Connor}, {Davies}, {Decarli}, {Eilers}, {Farina}, {Green}, {Jiang}, {Li}, {Mazzucchelli}, {Nanni}, {Schindler}, {Venemans}, {Walter}, {Wu}, \& {Yue}}]{Wang2021}
{Wang}, F., {Yang}, J., {Fan}, X., {et~al.} 2021{\natexlab{a}}, \bibinfo{title}{{A Luminous Quasar at Redshift 7.642},} \apjl, 907, L1, \dodoi{10.3847/2041-8213/abd8c6}

\bibitem[{F. {Wang} {et~al.}(2021{\natexlab{b}}){Wang}, {Fan}, {Yang}, {Mazzucchelli}, {Wu}, {Li}, {Ba{\~n}ados}, {Farina}, {Nanni}, {Ai}, {Bian}, {Davies}, {Decarli}, {Hennawi}, {Schindler}, {Venemans}, \& {Walter}}]{Wang2021b}
{Wang}, F., {Fan}, X., {Yang}, J., {et~al.} 2021{\natexlab{b}}, \bibinfo{title}{{Revealing the Accretion Physics of Supermassive Black Holes at Redshift z {\ensuremath{\sim}} 7 with Chandra and Infrared Observations},} \apj, 908, 53, \dodoi{10.3847/1538-4357/abcc5e}

\bibitem[{D.~C. {Wells}(1985){Wells}}]{Wells1985}
{Wells}, D.~C. 1985, in Data Analysis in Astronomy, 195

\bibitem[{A. {Wierzcholska} \& S. {Wagner}(2025){Wierzcholska} \& {Wagner}}]{Wierzcholska2025}
{Wierzcholska}, A., \& {Wagner}, S. 2025, \bibinfo{title}{{Exceptional X-ray activity in BL Lacertae},} \aap, 693, A299, \dodoi{10.1051/0004-6361/202451349}

\bibitem[{C.~J. {Willott} {et~al.}(2010){Willott}, {Delorme}, {Reyl{\'e}}, {Albert}, {Bergeron}, {Crampton}, {Delfosse}, {Forveille}, {Hutchings}, {McLure}, {Omont}, \& {Schade}}]{Willott2010}
{Willott}, C.~J., {Delorme}, P., {Reyl{\'e}}, C., {et~al.} 2010, \bibinfo{title}{{The Canada-France High-z Quasar Survey: Nine New Quasars and the Luminosity Function at Redshift 6},} \aj, 139, 906, \dodoi{10.1088/0004-6256/139/3/906}

\bibitem[{W.~E. {Wilson} {et~al.}(2011){Wilson}, {Ferris}, {Axtens}, {Brown}, {Davis}, {Hampson}, {Leach}, {Roberts}, {Saunders}, {Koribalski}, {Caswell}, {Lenc}, {Stevens}, {Voronkov}, {Wieringa}, {Brooks}, {Edwards}, {Ekers}, {Emonts}, {Hindson}, {Johnston}, {Maddison}, {Mahony}, {Malu}, {Massardi}, {Mao}, {McConnell}, {Norris}, {Schnitzeler}, {Subrahmanyan}, {Urquhart}, {Thompson}, \& {Wark}}]{Wilson2011}
{Wilson}, W.~E., {Ferris}, R.~H., {Axtens}, P., {et~al.} 2011, \bibinfo{title}{{The Australia Telescope Compact Array Broad-band Backend: description and first results},} \mnras, 416, 832, \dodoi{10.1111/j.1365-2966.2011.19054.x}

\bibitem[{J. {Wolf} {et~al.}(2023){Wolf}, {Nandra}, {Salvato}, {Buchner}, {Onoue}, {Liu}, {Arcodia}, {Merloni}, {Ciroi}, {Di Mille}, {Burwitz}, {Brusa}, {Ishimoto}, {Kashikawa}, {Matsuoka}, {Urrutia}, \& {Waddell}}]{Wolf2023}
{Wolf}, J., {Nandra}, K., {Salvato}, M., {et~al.} 2023, \bibinfo{title}{{X-ray emission from a rapidly accreting narrow-line Seyfert 1 galaxy at z = 6.56},} \aap, 669, A127, \dodoi{10.1051/0004-6361/202244688}

\bibitem[{X.-B. {Wu} {et~al.}(2015){Wu}, {Wang}, {Fan}, {Yi}, {Zuo}, {Bian}, {Jiang}, {McGreer}, {Wang}, {Yang}, {Yang}, {Thompson}, \& {Beletsky}}]{Wu2015}
{Wu}, X.-B., {Wang}, F., {Fan}, X., {et~al.} 2015, \bibinfo{title}{{An ultraluminous quasar with a twelve-billion-solar-mass black hole at redshift 6.30},} \nat, 518, 512, \dodoi{10.1038/nature14241}

\bibitem[{Z. {Wu} {et~al.}(2007){Wu}, {Jiang}, {Gu}, \& {Liu}}]{Wu2007}
{Wu}, Z., {Jiang}, D.~R., {Gu}, M., \& {Liu}, Y. 2007, \bibinfo{title}{{VLBI observations of seven BL Lacertae objects from RGB sample},} \aap, 466, 63, \dodoi{10.1051/0004-6361:20066754}

\bibitem[{J. {Yang} {et~al.}(2021){Yang}, {Wang}, {Fan}, {Barth}, {Hennawi}, {Nanni}, {Bian}, {Davies}, {Farina}, {Schindler}, {Ba{\~n}ados}, {Decarli}, {Eilers}, {Green}, {Guo}, {Jiang}, {Li}, {Venemans}, {Walter}, {Wu}, \& {Yue}}]{Yang2021}
{Yang}, J., {Wang}, F., {Fan}, X., {et~al.} 2021, \bibinfo{title}{{Probing Early Supermassive Black Hole Growth and Quasar Evolution with Near-infrared Spectroscopy of 37 Reionization-era Quasars at 6.3 < z {\ensuremath{\leq}} 7.64},} \apj, 923, 262, \dodoi{10.3847/1538-4357/ac2b32}

\bibitem[{M. {Yue} {et~al.}(2024){Yue}, {Eilers}, {Ananna}, {Panagiotou}, {Kara}, \& {Miyaji}}]{Yue_2024_LRD}
{Yue}, M., {Eilers}, A.-C., {Ananna}, T.~T., {et~al.} 2024, \bibinfo{title}{{Stacking X-Ray Observations of ``Little Red Dots'': Implications for Their Active Galactic Nucleus Properties},} \apjl, 974, L26, \dodoi{10.3847/2041-8213/ad7eba}

\bibitem[{L. {Zappacosta} {et~al.}(2023){Zappacosta}, {Piconcelli}, {Fiore}, {Saccheo}, {Valiante}, {Vignali}, {Vito}, {Volonteri}, {Bischetti}, {Comastri}, {Done}, {Elvis}, {Giallongo}, {La Franca}, {Lanzuisi}, {Laurenti}, {Miniutti}, {Bongiorno}, {Brusa}, {Civano}, {Carniani}, {D'Odorico}, {Feruglio}, {Gallerani}, {Gilli}, {Grazian}, {Guainazzi}, {Marinucci}, {Menci}, {Middei}, {Nicastro}, {Puccetti}, {Tombesi}, {Tortosa}, {Testa}, {Vietri}, {Cristiani}, {Haardt}, {Maiolino}, {Schneider}, {Tripodi}, {Vallini}, \& {Vanzella}}]{Zappacosta2023}
{Zappacosta}, L., {Piconcelli}, E., {Fiore}, F., {et~al.} 2023, \bibinfo{title}{{HYPerluminous quasars at the Epoch of ReionizatION (HYPERION): A new regime for the X-ray nuclear properties of the first quasars},} \aap, 678, A201, \dodoi{10.1051/0004-6361/202346795}

\bibitem[{Y.-H. {Zhang} \& J.-C. {Li}(2017){Zhang} \& {Li}}]{Zhang2017}
{Zhang}, Y.-H., \& {Li}, J.-C. 2017, \bibinfo{title}{{Optical variability of the high synchrotron energy peaked blazar 1ES 1959+650 on various time-scales},} \mnras, 469, 1682, \dodoi{10.1093/mnras/stx942}

\bibitem[{Z. {Zuo} {et~al.}(2024){Zuo}, {Zhu}, {Brandt}, {Garmire}, {Vito}, {Wu}, \& {Xue}}]{Zuo2024}
{Zuo}, Z., {Zhu}, S., {Brandt}, W.~N., {et~al.} 2024, \bibinfo{title}{{The X-ray enhancements of radio-loud quasars at high redshift: new results at z = 4-7},} \mnras, 530, 360, \dodoi{10.1093/mnras/stae656}

\end{thebibliography}
\bibliographystyle{aasjournalv7}



\end{document}